\pdfoutput=1
\documentclass[aps,rmp,reprint,a4paper,amsmath,amssymb,longbibliography,tightenlines,floatfix]{revtex4-2}

\usepackage[T1]{fontenc}
\usepackage[utf8]{inputenc}
\usepackage[most]{tcolorbox}
\usepackage{lmodern,microtype,changepage,amsmath,amssymb,amsthm,mathtools,bm,tabularx,graphicx,enumitem,booktabs,tikz,xcolor,tabularray}
\definecolor{darkblue}{rgb}{0.1,0.1,.7}
\usepackage[colorlinks, linkcolor=darkblue, citecolor=darkblue, urlcolor=darkblue,
linktocpage, breaklinks, bookmarksnumbered,
bookmarksdepth=2]{hyperref}
\usepackage{orcidlink}

\makeatletter
\bibpunct{[}{]}{,}{n}{}{,}
\def\NAT@cmprs{\@ne}\def\NAT@sort{\@ne}
\AtBeginDocument{%
	\bibpunct{[}{]}{,}{n}{}{,}%
	\def\NAT@cmprs{\@ne}\def\NAT@sort{\@ne}%
	\NAT@set@cites}
\makeatother

\makeatletter
\let\rtx@orig@makecaption\@makecaption
\long\def\@makecaption#1#2{%
	\begingroup
	\pretolerance=-1 \hyphenpenalty=50 \emergencystretch=3em
	\rtx@orig@makecaption{#1}{#2}%
	\endgroup}
\makeatother

\usetikzlibrary{decorations.pathmorphing,arrows.meta,positioning,calc,fit,backgrounds}
\tcbuselibrary{listings}
\definecolor{inkcol}{RGB}{27,27,27}
\definecolor{slatecol}{RGB}{62,92,118}
\definecolor{slatesoft}{RGB}{122,145,168}
\definecolor{slatefill}{RGB}{233,238,243}
\definecolor{slatedeep}{RGB}{40,60,80}
\definecolor{linkdark}{RGB}{34,45,73}
\tikzset{>=Stealth, every picture/.append style={line cap=round,line join=round},
	wbndry/.style={draw=slatecol,line width=1pt},
	wdot/.style={circle,fill=inkcol,inner sep=0pt,minimum size=3pt},
	wodot/.style={circle,draw=slatecol,fill=white,inner sep=0pt,minimum size=4pt,line width=0.6pt},
	wscalar/.style={draw=inkcol,line width=0.55pt},
	wleg/.style={draw=inkcol,line width=0.55pt},
	wgluon/.style={draw=slatecol,line width=0.7pt,decorate,
		decoration={coil,aspect=0.5,segment length=4.4pt,amplitude=1.6pt}},
	wgrav/.style={draw=slatecol,line width=0.5pt,double,double distance=1.4pt},
	wtube/.style={draw=slatecol,dashed,line width=0.5pt},
	wbig/.style={draw=slatecol,line width=0.8pt,rounded corners},
	wcut/.style={draw=slatecol,dashed,line width=0.7pt},
	warr/.style={draw=slatecol,line width=0.6pt}}
	
\UseTblrLibrary{booktabs} 
\allowdisplaybreaks[1]
\theoremstyle{remark}
\newtheorem{lesson}{Lesson}[section]
\newcommand{\AdS}{\mathrm{AdS}}
\newcommand{\dS}{\mathrm{dS}}

\newcommand{\kT}{k_T}
\newcommand{\Kbb}{\mathcal{K}}
\newcommand{\Gbb}{\mathcal{G}}
\newcommand{\dd}{\mathrm{d}}
\newcommand{\ii}{\mathrm{i}}
\newcommand{\eps}{\epsilon}
\newcommand{\psiwf}{\psi}
\newcommand{\Ps}{\mathcal{P}}
\newcommand{\kkk}{\mathsf{KKK}}
\newcommand{\kkj}{\mathsf{KKJ}}
\newcommand{\kjj}{\mathsf{KJJ}}
\newcommand{\ul}[1]{\underline{#1}}
\DeclareMathOperator*{\Res}{Res}

\DeclareMathOperator{\Rez}{Re}
\newcommand{\half}{\tfrac12}
\newcommand{\vk}{\bm k}
\newcommand{\order}[1]{\mathcal{O}(#1)}

\begin{document}

\title{Cosmological Correlators from Scattering Amplitudes: A Review%
  \texorpdfstring{\\[2pt]\large\normalfont}{. }%
  Momentum-Space Methods in de~Sitter and Anti--de~Sitter Space}

\author{Soner Albayrak\,\orcidlink{0000-0002-9223-0115}}
\affiliation{\mbox{Department of Physics, Middle East Technical University, Ankara, Türkiye}}
\author{Savan Kharel\,\orcidlink{0000-0001-9455-4183}}
\affiliation{\mbox{Department of Physics, University of Chicago, Chicago, IL, USA}}
\date{\today}

\begin{abstract}
\noindent
The primordial statistics underlying the CMB and large-scale structure are encoded in late-time cosmological correlation functions (constructed from the coefficients of the \emph{wavefunction of the universe}). They are the closest analogue cosmology has to an $S$-matrix, yet the standard $S$-matrix toolkit does not directly apply. This review explains how the modern scattering-amplitudes program (spinor helicity, on-shell recursion, the double copy, positive geometry, and generalized unitarity) has nevertheless been extended to cosmology and, through a single analytic continuation, to Euclidean anti--de~Sitter space, where the same objects appear as holographic boundary correlators and are often cleanest to compute. The organizing framework is momentum space on the late-time boundary, where the primordial statistics are simplest and where the total energy of the external legs, the sum of their momentum magnitudes, controls the analytic structure.  At the total-energy singularity, the flat-space scattering amplitude emerges. We develop this toolkit starting from the wave equation and use it to compute tree-level scalar, gauge-field, and graviton correlators, exposing relations across spins and dimensions. We then show how bulk integrals can be bypassed using cosmological polytopes, on-shell recursion, and curved-space versions of the double copy, before extending the discussion to loops, soft limits, transition amplitudes, and (A)dS cutting rules. Finally, we connect these structures to inflationary observables and to three allied programs: the cosmological optical theorem, the cosmological bootstrap, and the cosmological collider. We conclude with a map of the current frontier and a set of open and interesting problems.
\end{abstract}

\maketitle

\tableofcontents{}

\section{Introduction}
\label{sec:intro}

Cosmology is a forensic science. Like Aragorn in \emph{The Lord of
	the Rings} reading a trail from footprints, broken branches, and traces almost
erased by time, cosmologists arrive long after the decisive events have passed.
They cannot rerun the history of the universe; they must reconstruct it
from what remains. In the inflationary picture, an early period of accelerated
expansion stretched quantum fluctuations to cosmological scales, leaving primordial
perturbations whose imprint survives in the anisotropies of the cosmic microwave
background (CMB) and, after
gravitational growth, in galaxies and the large-scale structure of the universe. 

To an excellent approximation, these primordial perturbations are Gaussian. For a
Gaussian random field, all statistical information is contained in the two-point
function, or equivalently the power spectrum. Small departures from Gaussianity
reveal themselves in the higher correlations, whose momentum dependence can
distinguish interactions, additional fields and initial states that would leave
nearly indistinguishable power spectra.
The \emph{shapes} of these correlations encode faint but potentially important traces of the physics of
the early universe: the particle content and interactions of the inflationary
era, at energies far exceeding anything the Large Hadron Collider can reach. These
correlators are the boundary data of a quantum evolution, fixed by the late-time
\emph{wavefunction of the universe} $\Psi[\phi]$ \cite{Hartle:1983ai}, a functional
of the field configuration $\phi$ on the late-time slice.

The wavefunction coefficients $\psiwf_n(\vk_1,\dots,\vk_n)$, defined by expanding
$\log\Psi$, play the role for cosmology that scattering amplitudes play in flat
space; equal-time correlators are obtained only after forming $|\Psi|^2$ and
integrating over the boundary field, and we keep ``wavefunction coefficient'' and
``correlator'' distinct throughout. Like amplitudes, the coefficients are
controlled by a small number of physical singularities. The sharpest of these is
vanishing total energy, $\kT \equiv \sum_i k_i \to 0$, reached by continuing away
from the physical region $k_i>0$, where $\psiwf_n$ develops a singularity whose
leading coefficient is the flat-space scattering amplitude, up to a normalization
and a power of $\kT$ fixed by the interactions
\cite{Raju:2012zr,Maldacena:2011nz}; singularities in the channel energies encode
factorization into lower-point processes.

In flat space, scattering amplitudes can be dramatically simpler than the
diagrams that compute them. The canonical example is the Parke--Taylor formula
for the maximally helicity violating (MHV) $n$-gluon amplitude, which collapses
the $220$ diagrams of the six-gluon process into a single ratio of spinor
variables \cite{Parke:1986gb}.\footnote{The count of $220$ refers to Feynman
	gauge with three- and four-point vertices; individual diagrams are not separately
	gauge invariant.} The claim here is not raw efficiency: off-shell recursion
already computes $n$-gluon trees in polynomial time \cite{Berends:1987me}, and
MHV is the cleanest helicity configuration, not the general case. The claim is
that the \emph{answer} has structure the Feynman expansion hides, and the lesson
of the modern $S$-matrix program \cite{Elvang:2013cua} is that this structure
follows from locality and unitarity made manifest through on-shell data.

The choice of variables is half the battle: spinor helicity trades redundant
polarization vectors for spinors. BCFW recursion
\cite{Britto:2004ap,Britto:2005fq} builds tree amplitudes from lower-point
on-shell amplitudes via a complex momentum deformation; color--kinematics
duality and the double copy \cite{Bern:2008qj,Bern:2010ue,Kawai:1985xq} express
gravity amplitudes as squares of gauge-theory ones, a tree-level theorem and a
conjecture with overwhelming evidence beyond. These are not merely aesthetic
gains. Generalized unitarity is production technology at collider precision
\cite{Bern:1994zx}, and the double copy has restructured post-Minkowskian
gravitational-wave calculations \cite{Bern:2019nnu}: on-shell structure, once
seen, can be computed with.

Can this technology be carried to curved spacetime? The obstruction
is that the amplitudes program presupposes an
$S$-matrix: asymptotic in- and out-states. Two of the most important settings in
theoretical physics have neither. In \emph{anti--de~Sitter space} (AdS), bulk scattering is replaced by correlators
of a boundary conformal field theory; there are no asymptotic states, only local
operators inserted on the timelike conformal boundary
\cite{Maldacena:1997re,Gubser:1998bc,Witten:1998qj}. In \emph{inflation}, the
out-state is replaced by the wavefunction of the universe on a quasi--de~Sitter
background, and the observables are equal-time correlators on a single spacelike late-time
slice \cite{Maldacena:2002vr,Maldacena:2011nz}. The aim of this review is therefore not to
transplant the $S$-matrix unchanged, but to identify which of its organizing
principles survive.

These correlators can be studied in three languages. In \emph{position
	space}, the kinematics of conformal symmetry are carried by cross-ratios; this is
the home of the mature field of the conformal bootstrap
\cite{Poland:2018epd,Simmons-Duffin:2016gjk}. In \emph{Mellin space}
\cite{Penedones:2010ue,Fitzpatrick:2011ia} a Witten diagram mirrors
a flat-space amplitude: contact terms become constants and exchanges become
sums of poles in the Mellin variables (the analogue of Mandelstam invariants). In \emph{momentum space}, the
Fourier transform along the $d$ boundary directions, the correlator becomes a
function of the boundary momenta \cite{Bzowski:2013sza,Gillioz:2022yze}. We work in
momentum space throughout, for three reasons. First, polarization sums, on-shell
recursion, color ordering, the double copy and unitarity cuts are all
momentum-space constructions, and a curved-space Witten diagram differs from the
corresponding flat Feynman diagram only by the radial factor of its external and
internal lines. Everything one knows about tree- and loop-level gauge theory and
gravity is therefore available.

Second, cosmological observables are naturally functions of comoving momenta, and their most
informative features are statements about momentum-space kinematics that a
position-space or Mellin representation would obscure.

Third, the Poincar\'e patch of (A)dS retains an exact $\mathbb R^d$ of boundary
translations, so momentum is conserved at every vertex (the radial integral carries the curvature). The price
is a sometimes unwieldy radial integral per bulk point. The gain is a correlator
whose singularities, the total energy $\kT$ and the channel energies, are
physical; the leading one again carries the flat-space
$S$-matrix.\footnote{There is something almost archaeological about this limit:
	strip away enough curved-space structure and one uncovers the flat-space amplitude
	buried underneath.}

\subsection{What we compute and what it means}
Four boundary objects recur in this review, and their relations should not be
collapsed into an identity. A Euclidean-AdS Witten diagram, the AdS analogue of a
Feynman diagram, computes a boundary correlator of single-trace operators dual to
bulk fields (for example, conserved currents $J$ dual to bulk gauge fields, the
stress tensor $T$ dual to the graviton, or scalar operators dual to bulk scalars).
For the Bunch--Davies state, the same bulk integral, continued in the radial
coordinate and couplings, determines a coefficient $\psiwf_n$ of the late-time
wavefunction $\Psi[\phi]$ introduced above (Section~\ref{sec:inflation}); the
continuation can carry phases and local counterterm ambiguities, which must be
tracked. The Born rule then combines $\psiwf_n$ with lower-order coefficients to
produce an equal-time correlator (Section~\ref{sec:wavefunction}). Finally, the
leading total-energy singularity of any of these determines a flat-space
amplitude.

The two-point function defines the power spectrum $\mathcal P(k)$, observed to be
nearly scale-invariant. The three-point function defines the bispectrum
$B(\vk_1,\vk_2,\vk_3)$, whose amplitude, for a given template, is conventionally
parametrized by $f_{\rm NL}$ and whose shape helps discriminate among mechanisms:
a \emph{local} shape, peaking in the squeezed limit, can arise naturally from
superhorizon evolution in models with multiple light fields; an \emph{equilateral} shape is often generated by
derivative self-interactions, including those responsible for a reduced sound
speed; and a \emph{folded} shape,
enhanced in flattened configurations, can signal a departure from the
Bunch--Davies initial state. Particles with masses of order the Hubble scale, produced during inflation, can
instead generate distinctive nonanalytic, often oscillatory, signals in the
squeezed limit, which are targets of the \emph{cosmological collider} program
\cite{Arkani-Hamed:2015bza,Lee:2016vti}.

\subsection{Organization and reader's guide}

The review is built in three parts: the targets (Section~\ref{sec:wavefunction}),
the anti--de~Sitter engine (Sections~\ref{sec:toolkit}--\ref{sec:loops}), and the
cosmological payoff (Sections~\ref{sec:inflation}--\ref{sec:outlook}). The stage is
set by momentum-space CFT technology
\cite{Bzowski:2013sza,Bzowski:2015pba,Raju:2010by,Raju:2011mp,Raju:2012zr,Raju:2012zs};
on this rests the computational program the review follows.\footnote{Higher-point gluon \cite{Albayrak:2018tam}, graviton
	\cite{Albayrak:2019yve} and scalar \cite{Albayrak:2020isk} correlators; a web
	relating spins and dimensions \cite{Albayrak:2023kfk}; the extension to loops
	\cite{Albayrak:2020bso}; a spinning-exchange relation \cite{Albayrak:2019asr}
	linking the algebra to cosmological polytopes \cite{Arkani-Hamed:2017fdk};
	curved-space recursion from three-dimensional spinor helicity
	\cite{Albayrak:2023jzl}; color--kinematics duality and a curvature-corrected
	double copy in (A)dS \cite{Albayrak:2020fyp}; soft limits, transition amplitudes
	and unitarity \cite{Albayrak:2024ddg,Albayrak:2023hie}.} Feeding on the same singularities, the
\emph{cosmological bootstrap}
\cite{Arkani-Hamed:2018kmz,Baumann:2020dch,Goodhew:2020hob,Baumann:2022jpr}
instead constrains inflationary correlators from symmetry, singularities,
locality and unitarity; the AdS--dS continuation \cite{Harlow:2011ke} ties the
two together.\footnote{An older holographic root is
	dS/CFT \cite{Strominger:2001pn,Witten:2001kn}, where asymptotic symmetries of the
	expanding patch act conformally on the very slice the correlators live on, made
	quantitative for inflation in \cite{McFadden:2009fg}.} This is a \emph{methods} review; for
inflationary model building, CMB data analysis and history, see
\cite{Baumann:2009ds,Baumann:2022mni,Baumann:2022jpr}.

\begin{figure*}[t]
	\centering
	\begin{tikzpicture}[scale=1.0,every node/.style={font=\footnotesize},
		sat/.style={wbig,draw=inkcol,line width=0.7pt,text=inkcol,align=center,
			minimum height=0.9cm},
		blkarr/.style={draw=inkcol,line width=0.6pt,-{Stealth[length=2.2mm]},
			shorten <=2pt,shorten >=2pt}]
		\node[wbig,draw=inkcol,line width=1pt,text=inkcol,
		align=center,minimum width=3.7cm,minimum height=1.0cm]
		(core) at (0,0) {momentum-space correlator\\[1pt] singular in $\kT$ and the channel energies $k_S{+}k_{\ul S}$};
		\node[sat,minimum width=2.9cm] (flat) at (-5.3,1.7)
		{\textbf{flat space}\\ $S$-matrix $A_n$};
		\node[sat,minimum width=2.9cm] (ads) at (0,2.5)
		{\textbf{anti--de~Sitter}\\ boundary correlator};
		\node[sat,minimum width=2.9cm] (ds) at (5.3,1.7)
		{\textbf{de~Sitter}\\ wavefunction $\psiwf_n$};
		\node[sat,minimum width=3.0cm] (data) at (5.3,-2.0)
		{\textbf{the sky}\\ power spectrum, bispectrum};
		\node[sat,minimum width=3.4cm] (meth) at (-5.3,-2.0)
		{\textbf{the toolkit}\\ residues, $\mathcal D$-ops, polytopes,\\ recursion, double copy, cuts};
		\draw[blkarr] (core)  -- (flat) node[midway,above,sloped]{$\kT\to0$};
		\draw[blkarr,{Stealth[length=2.2mm]}-{Stealth[length=2.2mm]}] (ads) -- (core);
		\draw[blkarr] (ds)    -- (core) node[midway,above,sloped]{$\eta=\ii z$};
		\draw[blkarr] (meth)  -- (core);
		\draw[blkarr] (core)  -- (data) node[midway,below,sloped]{Born rule};
	\end{tikzpicture}

	\caption{The map of the review. A single momentum-space object, an (A)dS boundary
		correlator (equivalently a de~Sitter wavefunction coefficient under the continuation
		$\eta=\ii z$), is computed by the amplitude-inspired toolkit, returns the flat-space
		$S$-matrix on its total-energy pole $\kT\to0$, and yields the observed cosmological
		statistics through the Born rule. The arrows denote maps, with
		convention-dependent normalizations, not literal equalities.}
	\label{fig:roadmap}
\end{figure*}
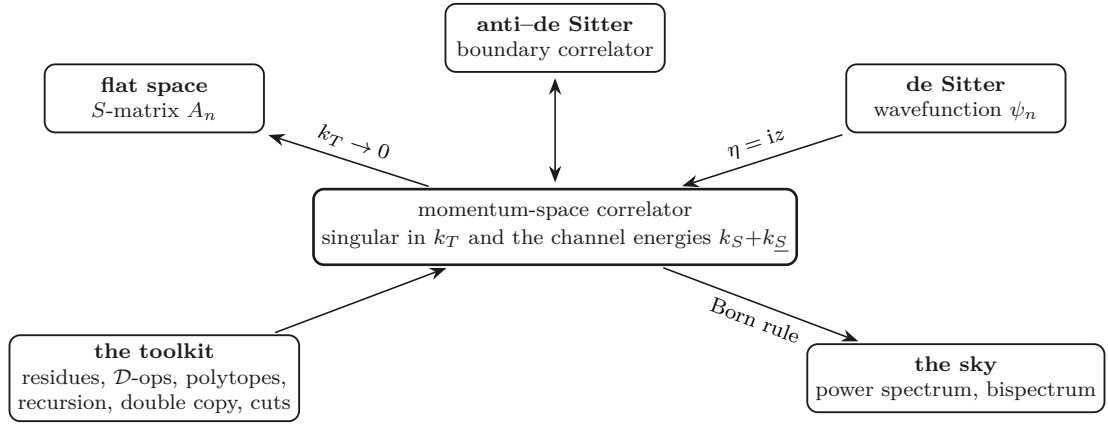

Figure~\ref{fig:roadmap} is the map. Section~\ref{sec:wavefunction} defines the
objects; Section~\ref{sec:toolkit} builds the toolkit and the AdS--dS dictionary;
Sections~\ref{sec:tree}--\ref{sec:loops} compute: trees, the web of theories,
the bulk integral bypassed, then loops. Section~\ref{sec:inflation} cashes the
results in inflation, Section~\ref{sec:frontier} maps the frontier,
Section~\ref{sec:outlook} closes with the dictionary and ten open problems, and
Appendix~\ref{app:conv} fixes conventions. For the cosmology alone, read
Sections~\ref{sec:wavefunction} and~\ref{sec:inflation}; for the technology
alone, start at Section~\ref{sec:toolkit}.

\paragraph{Conventions.} We work in the Poincar\'e patch with the boundary at $z\to0$;
$\vk$ is a boundary (spatial) momentum and $k\equiv|\vk|$ its magnitude, the ``energy''
of a leg. The total energy is $\kT=\sum_a k_a$; a partial energy is a sum of magnitudes
over a subset, $k_S=\sum_{a\in S}k_a$; and an underline denotes the magnitude of a
vector sum, $k_{\ul{12}}\equiv|\vk_1+\vk_2|$, the energy flowing on an internal line.
The combination $k_S+k_{\ul S}$, the \emph{channel energy}, is the quantity that
vanishes at a factorization singularity.
Keeping the two apart ($k_{12}=k_1+k_2$ versus $k_{\ul{12}}=|\vk_1+\vk_2|$) is the
whole art of reading these results. Boundary correlators carry a stripped
momentum-conserving delta function, denoted by the prime $\langle\cdots\rangle'$. Full
conventions are in Appendix~\ref{app:conv}. 

\section{Cosmological correlators and the wavefunction of the universe}
\label{sec:wavefunction}
This section fixes the objective. We briefly define the primordial fluctuation, the
wavefunction of the universe that carries its statistics, and the observables
built from both. The rest of the review is about computing them. The
distinction between the scalar proxy $\phi$ used in the calculations and the
observed curvature perturbation $\zeta$ is kept explicit throughout.

\subsection{Inflation, the fluctuation, and the Bunch--Davies mode}

Inflation (in the simplest realization) is a phase of quasi-exponential
expansion driven by a scalar field rolling slowly down a nearly flat potential.
To an excellent first approximation the background is dS space, with
\begin{equation}
	a(\eta)\simeq -\frac{1}{H\eta},
\end{equation}
in conformal time $\eta\in(-\infty,0)$ and with the Hubble rate $H$ varying only slowly.
In the dS approximation, we represent the end of inflation by a
late-time slice $\eta_0\to0^-$. In canonical single-field inflation the
physical scalar degree of freedom may be described by the gauge-invariant
curvature perturbation. Throughout much of this review we instead use
a minimally coupled massless scalar $\phi$ as a proxy for this fluctuation. The
proxy captures the late-time and analytic structure of the problem, but its
normalization is not the observed scalar amplitude (see Eq.~\eqref{eq:zeta-power}
below). Its free evolution
follows from the quadratic action of a massless scalar in de~Sitter. The
positive-frequency mode selected by requiring that sufficiently short
wavelengths approach the Minkowski vacuum is
\begin{equation}
	\label{eq:bd-mode}
	\begin{gathered}
		\phi_k(\eta)=\frac{H}{\sqrt{2k^3}}\,(1+\ii k\eta)\,
		e^{-\ii k\eta},\\[2pt]
		a(\eta)\phi_k(\eta)
		\xrightarrow{\ \eta\to-\infty\ }
		\frac{-\ii\,e^{-\ii k\eta}}{\sqrt{2k}}.
	\end{gathered}
\end{equation}
This is the familiar \emph{Bunch--Davies} condition
\cite{Bunch:1978yq}.

In perturbation theory the corresponding interacting vacuum is selected by
the usual $\ii\eps$ prescription. The mode \eqref{eq:bd-mode} oscillates as
$e^{-\ii k\eta}$; the bulk-to-boundary kernel of the wavefunction path integral
is instead normalized at the late-time slice and built from the conjugate
solution, so it behaves at early times as $K_k(\eta)\sim e^{+\ii k\eta}$. The
contour is taken to begin at $\eta=-\infty(1-\ii\eps)$, making this kernel
exponentially damped; keeping the two conventions distinct avoids an apparent
sign clash between the field mode and the wavefunction kernel.
This Lorentzian prescription is the counterpart of regularity in Euclidean
AdS, where the interior-regular solution is $K_\nu$ rather than the growing
$I_\nu$ (Section~\ref{sec:toolkit}). Under the analytic continuation
\begin{equation}
	\eta=\ii\,z,\qquad R_{\dS}=-\ii\,R_{\AdS},
\end{equation}
the two prescriptions map into one another, providing the first indication
of the close relation between the de~Sitter wavefunction and Euclidean-AdS
Witten diagrams. The continuation also acts on the radius, the couplings, and
the boundary counterterms; Section~\ref{sec:tk-dict} supplies the working
dictionary, and Section~\ref{sec:inflation} applies it to inflationary
quantities.
\subsection{The wavefunction of the universe}

Since the future boundary of de~Sitter is spacelike, the natural boundary
object is not an $S$-matrix but the late-time wavefunction. The wavefunction of
the universe is the bulk path integral over field histories that satisfy the
Bunch--Davies condition \cite{Bunch:1978yq} in the far past and approach a fixed
profile $\phi(\vk)$ on a late-time slice $\eta_0$,\footnote{This is the
	perturbative Poincar\'e-patch realization of the Hartle--Hawking, or Euclidean,
	vacuum prescription \cite{Hartle:1983ai}. We use it only perturbatively: to the
	order considered here, $\Psi$ is defined by its saddle point together with a
	finite number of loop corrections, so we require only its order-by-order
	perturbative definition.}
\begin{equation}
	\Psi_{\eta_0}[\phi]\equiv\langle\phi,\eta_0\,|\,{\rm BD}\rangle
	=\!\!\int\limits_{\substack{\Phi(\eta_0)=\phi\\[1pt]
			\Phi\,\sim\,{\rm BD}\ {\rm as}\ \eta\to-\infty}}\!\!
	\mathcal D\Phi\;e^{\,\ii S[\Phi]}\,.
\end{equation}

Expanding the logarithm of $\Psi$ in powers of the boundary field defines the
\emph{wavefunction coefficients} $\psiwf_n$,\footnote{It is the logarithm, rather
	than $\Psi$ itself, that is expanded: this isolates the connected contributions,
	while exponentiating restores the disconnected ones. The sum starts at $n=2$
	because the $n=0$ term is a field-independent normalization, which cancels between
	numerator and denominator in the Born rule \eqref{eq:born}, while the $n=1$ term is
	absent when expanding about the background solution.}
\begin{equation}
	\label{eq:Psi}
	\begin{split}
		\Psi[\phi]=\exp\bigg[\,\sum_{n\ge2}\frac1{n!}\!\int\!
		&\Big(\prod_{a=1}^n\frac{\dd^d k_a}{(2\pi)^d}\,\phi(\vk_a)\Big)(2\pi)^d
		\,\delta^d\!\Big(\textstyle\sum_a\vk_a\Big)\\[-2pt]
		&\times\psiwf_n(\vk_1,\dots,\vk_n)\bigg].
	\end{split}
\end{equation}
The momentum-conserving delta function expresses the unbroken boundary translation
invariance. Our convention absorbs all factors of $\ii$, coupling constants, and
external-leg normalizations into $\psiwf_n$; it is this convention that fixes the
signs in the Born-rule formulas below.

For the purposes of this review, $\psiwf_n$ are roughly the cosmological analogue of scattering amplitudes. They are
computed perturbatively from the bulk-to-boundary and
bulk-to-bulk propagators of the de~Sitter field, with every interaction vertex
on a single wavefunction branch, in contrast to the doubled contour of the
in-in formalism below.

\subsection{The Born rule: from wavefunction to correlators}

The wavefunction is not itself observable; the equal-time correlators are. They are
the moments of the probability distribution $|\Psi[\phi]|^2$,
\begin{equation}
	\langle\phi(\vk_1)\cdots\phi(\vk_n)\rangle
	=\frac{\int\mathcal D\phi\,|\Psi[\phi]|^2\,\phi(\vk_1)\cdots\phi(\vk_n)}
	{\int\mathcal D\phi\,|\Psi[\phi]|^2}\,.
\end{equation}
For the parity-even scalar correlators considered here, this is a
Gaussian-plus-interactions integral controlled by $2\Rez\,\psiwf_n$. In the
conventions of Eq.~\eqref{eq:Psi}, the first two nontrivial correlators are\footnote{A prime
	denotes the correlator with the momentum-conserving delta function stripped,
	$\langle\cdots\rangle=(2\pi)^d\delta^d(\sum_a\vk_a)\,\langle\cdots\rangle'$.}
\begin{equation}
	\label{eq:born}
	\begin{gathered}
		\langle\phi(\vk)\phi(-\vk)\rangle'=\frac{-1}{2\,\Rez\,\psiwf_2(k)},\\[4pt]
		\langle\phi(\vk_1)\phi(\vk_2)\phi(\vk_3)\rangle'
		=\frac{-2\,\Rez\,\psiwf_3(\vk_1,\vk_2,\vk_3)}
		{\displaystyle\prod_{a=1}^{3}2\,\Rez\,\psiwf_2(k_a)}\,.
	\end{gathered}
\end{equation}
At higher orders the Born rule requires the full connected expansion of the
boundary integral, in which higher $\psiwf_n$ mix with contractions generated
by lower coefficients; Eq.~\eqref{eq:born} is the leading-order dictionary,
not an all-orders identity.

One lesson motivates the strategy of computing $\psiwf_n$ rather than the
correlator directly: the wavefunction is the more transparent object. Its
coefficients expose the analytic structure of bulk time evolution, sharpest at
the \emph{total-energy singularity} $\kT\to0$ and, at higher points, at the
partial-energy singularities of its exchange channels. Squaring $\Psi$ and integrating
over the late-time fields erases phases and local pieces of that structure;
what the phase itself encodes is analyzed in \cite{Thavanesan:2025wvb}. The
wavefunction is to the correlator what an amplitude is to a cross section, and
this is why the review targets $\Psi$, reading off correlators at the
end.\footnote{On the relation between the wavefunction and in-in routes, and what
	each makes manifest, see \cite{Palma:2026qgn}; for the wavefunction's role in
	questions of state preparation and semiclassicality, \cite{Cotler:2025gui}.}

\subsection{The in-in route}

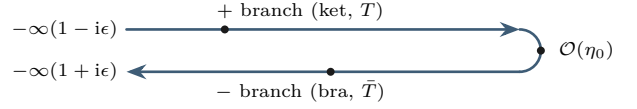
\begin{figure}[t]
	\centering
	\begin{tikzpicture}[scale=1.0]
		\draw[wbndry,->] (-2.9,0.28)--(2.3,0.28);
		\draw[wbndry] (2.3,0.28) arc (90:-90:0.28);
		\draw[wbndry,->] (2.3,-0.28)--(-2.9,-0.28);
		\node[wdot] at (2.58,0){};
		\node[right,font=\scriptsize] at (2.7,0){$\mathcal O(\eta_0)$};
		\node[left,font=\scriptsize] at (-2.9,0.28){$-\infty(1-\ii\eps)$};
		\node[left,font=\scriptsize] at (-2.9,-0.28){$-\infty(1+\ii\eps)$};
		\node[above,font=\scriptsize] at (-0.6,0.28){$+$ branch (ket, $T$)};
		\node[below,font=\scriptsize] at (-0.6,-0.28){$-$ branch (bra, $\bar T$)};
		\node[wdot] at (-1.6,0.28){};
		\node[wdot] at (-0.2,-0.28){};
	\end{tikzpicture}
	\caption{The in-in (Schwinger--Keldysh) contour of \eqref{eq:inin}: the ket is
		evolved forward on the $+$ branch and the bra backward on the $-$ branch, meeting
		at the late-time slice where the operator $\mathcal O$ is inserted. The conjugate
		far-past tilts project onto the Bunch--Davies vacuum. Vertices on the two branches
		carry opposite signs.}
	\label{fig:sk}
\end{figure}

There is a second route to a cosmological correlator. Rather than build $\Psi$ and square it, one
computes the equal-time expectation value directly in the interacting vacuum. This formalism can look unfamiliar at first. An $S$-matrix element connects asymptotic
in and out states. Here we want a different object, the expectation value of an operator
at a single moment in a single state. We therefore evolve the ket forward to that
moment, insert the operator, and evolve the bra backward. The time contour closes
on itself,
\begin{equation}
	\label{eq:inin}
	\begin{split}
		\langle\mathcal O(t)\rangle=\Big\langle 0\Big|\,
		&\bar T\,e^{+\ii\int_{-\infty}^{t}\dd t'\,H_{\rm int}(t')}\ \mathcal O(t)\\
		&\times\,T\,e^{-\ii\int_{-\infty}^{t}\dd t'\,H_{\rm int}(t')}\,\Big|0\Big\rangle,
	\end{split}
\end{equation}
with the ket running forward and the bra running backward
(Figure~\ref{fig:sk}). This is the in-in (Schwinger--Keldysh) formalism
\cite{Schwinger:1960qe,Keldysh:1964ud}, brought to inflation in
\cite{Maldacena:2002vr,Weinberg:2005vy}. The two branches acquire conjugate
far-past prescriptions,
\begin{equation}
	\eta_+\to-\infty(1-\ii\eps),\qquad \eta_-\to-\infty(1+\ii\eps),
\end{equation}
which project onto the Bunch--Davies vacuum.

Expanding the two exponentials and Wick contracting gives Feynman rules on a
doubly branched contour. Every interaction vertex carries a label: $+$ if it sits
on the forward branch and $-$ if it sits on the backward branch, contributing
$-\ii H_{\rm int}$ and $+\ii H_{\rm int}$, respectively.\footnote{For interactions
	containing time derivatives, the interaction Hamiltonian is not in general obtained
	simply by setting $H_{\rm int}=-L_{\rm int}$; the distinction must be treated
	carefully.} An internal line has four Schwinger--Keldysh components: the Feynman
propagator $G_{++}$, the anti-Feynman propagator $G_{--}$, and the two Wightman
functions $G_{+-}$ and $G_{-+}$, all constructed from the same Bunch--Davies mode
functions \eqref{eq:bd-mode}.\footnote{The systematic diagrammatic rules in this
	$\pm$-vertex language, developed into an efficient calculus for inflationary
	correlators, are given in \cite{Chen:2017ryl}; a systematic treatment of the
	in-in expansion and its convergence is given in \cite{Pinol:2023oux}.} For the Bunch--Davies state, these
four real-time propagators can be viewed as different boundary values of the same
analytically continued Euclidean two-point function under $\eta=\ii z$
(Section~\ref{sec:inflation}).

The two formalisms return the same correlators; they are two organizations of
the same quantum time evolution. Unitarity of that evolution moreover imposes relations among the wavefunction
coefficients themselves; their perturbative expression is the
\emph{cosmological optical theorem}, to which we return in
Section~\ref{sec:inflation}.

\subsection{The power spectrum}

The two-point coefficient is fixed by the free mode \eqref{eq:bd-mode} alone; the
derivation from the on-shell action is given in Section~\ref{sec:inflation}. Its
real part is $\Rez\,\psiwf_2=-k^3/H^2$, negative, as normalizability of $|\Psi|^2$ requires, and the
Born rule \eqref{eq:born} gives the scale-invariant power spectrum
\begin{equation}
	\langle\phi(\vk)\phi(-\vk)\rangle'=\frac{H^2}{2k^3},\qquad
	\Delta^2(k)\equiv\frac{k^3}{2\pi^2}\langle\phi\phi\rangle'=\frac{H^2}{4\pi^2},
\end{equation}
independent of $k$: the defining prediction of inflation, confirmed by the CMB at the
percent level, with a small measured tilt $n_s-1\approx-0.035$ encoding the slow
departure from exact de~Sitter \cite{Planck:2019kim}. This normalization belongs
to the proxy field; for the curvature perturbation in canonical slow-roll
single-field inflation,
\begin{equation}
	\label{eq:zeta-power}
	\langle\zeta(\vk)\zeta(-\vk)\rangle'=\frac{H^2}{4\epsilon_H M_{\rm Pl}^2 k^3},
	\qquad
	\Delta_\zeta^2=\frac{H^2}{8\pi^2\epsilon_H M_{\rm Pl}^2},
\end{equation}
at leading order, with $\epsilon_H=-\dot H/H^2$. In either normalization, scale
invariance is the statement that
$\psiwf_2\propto k^3$; a mass or a slow-roll correction shifts that exponent and
tilts the spectrum. All of the non-Gaussianity lives in $\psiwf_{n\ge3}$.

\subsection{The effective field theory of inflation}

Which interactions should we be computing? The toolkit of
Sections~\ref{sec:toolkit}--\ref{sec:loops} will evaluate any vertex it is handed,
so it is worth asking which vertices nature supplies. A monotonic rolling
background $\bar\phi(t)$ provides a physical clock: surfaces of constant field
value define surfaces of constant time. The underlying theory is invariant under
time diffeomorphisms, while the background is not, so time diffeomorphisms are
spontaneously broken and the fluctuation is described by the associated Goldstone
mode $\pi(t,\bm x)$ (a local error in the clock) with $\zeta=-H\pi$ at leading
order \cite{Cheung:2007st}.

In the unitary-gauge EFT one writes every operator allowed by the surviving
symmetries; the coefficient $M_2^4$ shifts the \emph{time}-kinetic term but not
the ordinary gradient term.\footnote{In the conventions of \cite{Cheung:2007st}, $M_2$ and $M_{\rm PL}$ are among the coupling coefficients in the unitary gauge EFT action\linebreak $\displaystyle\mathcal{S}=\int \dd^4x\sqrt{g}\left(\frac{1}{2}M_{\rm PL}^2R+\frac{1}{2}M_2(t)^4(g^{00}+1)^2+\dots\right)$.} In the decoupling regime, and neglecting higher-derivative corrections to the
dispersion relation, the fluctuation therefore propagates with sound speed
\begin{equation}
	\label{eq:eft-pi}
	c_s^{-2}=1-\frac{2M_2^4}{M_{\rm Pl}^2\dot H}\,.
\end{equation}
The \emph{same} $M_2^4$ necessarily switches on the cubic interaction
$\dot\pi(\partial_i\pi)^2$. Propagation and self-interaction are therefore tied
together: a small speed of sound forces large derivative self-interactions and,
barring cancellations with independent EFT operators, generically produces
$|f_{\rm NL}|\sim(1-c_s^2)/c_s^2$, with an equilateral-type shape. For
$\dot H<0$, the subluminal regime $c_s<1$ corresponds to $M_2^4>0$; negative
$M_2^4$, when compatible with stability, instead gives $c_s>1$.

These are precisely the kinds of local polynomial and derivative couplings of a
light scalar on a quasi--de~Sitter background that
Sections~\ref{sec:toolkit}--\ref{sec:loops} are built to handle. Their preferred
time direction reflects the inflationary clock, which spontaneously breaks
de~Sitter boosts; this makes the \emph{boostless} bootstrap of
Section~\ref{sec:inflation} \cite{Pajer:2020wnj} the natural setting.

Three limits calibrate the map from operators to shapes. DBI inflation
\cite{Alishahiha:2004eh} provides the canonical string-theoretic realization of
the $c_s\ll1$ regime. Multifield inflation \cite{Senatore:2010wk} introduces
additional light fields whose superhorizon conversion into $\zeta$ can generate
local-type non-Gaussianity. Quasi-single-field inflation interpolates between
these two shape limits \cite{Chen:2009zp,Chen:2009we}: a field with $m\sim H$
coupled to the inflaton sector leaves a characteristic nonanalytic momentum
dependence, computed from first principles in Section~\ref{sec:inflation}. This provides a
foundational example of the physics later systematized in the
cosmological-collider program.

\subsection{The bispectrum and its shapes}

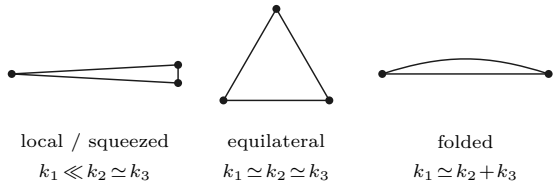
\begin{figure}[t]
	\centering
	\begin{tikzpicture}[scale=1.0,every node/.style={font=\scriptsize}]
		\begin{scope}[xshift=-3.35cm]
			\coordinate (a) at (0,0); \coordinate (b) at (2.2,0.12); \coordinate (c) at (2.2,-0.12);
			\draw[wleg] (a)--(b)--(c)--cycle;
			\node[wdot] at (a){}; \node[wdot] at (b){}; \node[wdot] at (c){};
			\node at (1.1,-0.9){local / squeezed};
			\node at (1.1,-1.3){$k_1\!\ll\!k_2\!\simeq\!k_3$};
		\end{scope}
		\begin{scope}[xshift=-0.55cm]
			\coordinate (a) at (0,-0.35); \coordinate (b) at (1.4,-0.35); \coordinate (c) at (0.7,0.86);
			\draw[wleg] (a)--(b)--(c)--cycle;
			\node[wdot] at (a){}; \node[wdot] at (b){}; \node[wdot] at (c){};
			\node at (0.7,-0.9){equilateral};
			\node at (0.7,-1.3){$k_1\!\simeq\!k_2\!\simeq\!k_3$};
		\end{scope}
		\begin{scope}[xshift=1.55cm]
			\coordinate (a) at (0,0); \coordinate (b) at (2.2,0); \coordinate (m) at (1.1,0.05);
			\draw[wleg] (a)--(b); \draw[wleg] (a) to[bend left=18] (b);
			\node[wdot] at (a){}; \node[wdot] at (b){};
			\node at (1.1,-0.9){folded};
			\node at (1.1,-1.3){$k_1\!\simeq\!k_2\!+\!k_3$};
		\end{scope}
	\end{tikzpicture}
	\caption{Three bispectrum shapes, drawn as the momentum triangle they peak on. The
		squeezed (local) limit is a soft limit and, in single-field inflation, is fixed by
		the single-field consistency relation; the equilateral limit responds to derivative
		self-interactions; the folded limit probes the
		initial state: departures from Bunch--Davies move singular support onto this
		physical boundary. A small oscillation superposed on the squeezed limit is the
		cosmological-collider signature of a massive particle.}
	\label{fig:shapes}
\end{figure}

The three-point correlator of the curvature perturbation, or bispectrum
$B_\zeta(\vk_1,\vk_2,\vk_3)=\langle\zeta\zeta\zeta\rangle'$, is the leading non-Gaussian
observable. Because the fluctuations are so nearly Gaussian, it is small; its
overall size is quoted as a number $f_{\rm NL}$, defined only once a template
shape and a normalization convention are chosen. Its \emph{shape}
(the dependence on the triangle that momentum conservation constrains to close) is
the useful diagnostic of inflationary physics (Figure~\ref{fig:shapes}). Shape analysis was introduced in \cite{Babich:2004gb}. Four shapes organize the discussion.
\begin{itemize}
	\item The \emph{local} shape peaks in the \emph{squeezed} configuration
	$k_1\ll k_2\simeq k_3$ and signals additional light fields. On an attractor,
	single-clock background in the Bunch--Davies state, a long-wavelength mode is
	locally indistinguishable from a rescaling
	of the spatial coordinates, so it cannot correlate with short-scale physics
	except through the tilt of the power spectrum; the resulting consistency
	relation \cite{Maldacena:2002vr,Creminelli:2004yq} fixes the squeezed limit to
	$B\to(1-n_s)\,\mathcal P(k_1)\mathcal P(k_3)$, a floor of order the slow-roll
	parameters. A detection of local non-Gaussianity above this floor would rule out
	this entire class: evading it requires giving up one of the assumptions,
	through an extra field, non-attractor evolution, or an excited initial state.
	\item The \emph{equilateral} shape peaks at $k_1\simeq k_2\simeq k_3$ and signals
	higher-derivative self-interactions of a single field; the shapes of general
	single-field inflation, which this template captures, were established in
	\cite{Chen:2006nt}.
	\item The \emph{folded} shape peaks when the triangle degenerates, $k_1\simeq k_2+k_3$,
	and probes the initial state (a departure from Bunch--Davies), which enhances
	precisely these flattened configurations \cite{Holman:2007na}.
	This physical boundary of the triangle should not be identified with the
	partial-energy singularities of an exchange diagram, which in the
	Bunch--Davies state sit at analytically continued kinematics
	(Section~\ref{sec:tree}); it is precisely a departure from Bunch--Davies that
	moves singular support onto the folded configuration, so the absence of
	folded singularities is a diagnostic of the Bunch--Davies condition.
	\item A small oscillatory modulation of the squeezed limit,
	$\propto(k_1/k_3)^{3/2}\cos[\mu\log\tfrac{k_1}{k_3}+\varphi]$, with
	$\mu=\sqrt{m^2/H^2-9/4}$ for a principal-series scalar in dS$_4$, is the
	\emph{cosmological collider} signal
	\cite{Arkani-Hamed:2015bza,Lee:2016vti} of a massive particle
	produced during inflation; the frequency encodes the mass, the angular
	dependence the spin, and the calculable phase $\varphi$ carries further mass
	and coupling information (Section~\ref{sec:inflation}).
\end{itemize}
A fifth template, the \emph{orthogonal} shape, is the combination of
equilateral-type EFT operators orthogonal to the equilateral template itself.
Introduced in \cite{Senatore:2009gt}, it is the third shape constrained by Planck
in Table~\ref{tab:obs} below.

Under the same attractor, single-clock, Bunch--Davies assumptions, the
consistency relation is not an isolated identity but the first member of an
infinite family: the soft modes of $\zeta$ (and of the graviton) are adiabatic
modes, large-gauge transformations that survive as physical states, and each one
implies a Ward identity relating a soft limit of an $(n+1)$-point function to
the $n$-point function.\footnote{This tower was derived systematically in
	\cite{Hinterbichler:2013dpa}, building on the identification of the conformal
	symmetries the adiabatic modes realize \cite{Hinterbichler:2012nm}; the companion
	conformal consistency relations were derived in \cite{Creminelli:2012ed}. Which
	soft-limit statements are robust theorems of single-field inflation, and how the
	squeezed limit separates single-field from multifield dynamics, was established
	in \cite{Assassi:2012zq}; how far they survive once the initial state is excited
	was charted in \cite{Flauger:2013hra}.} These identities are the cosmological
siblings of the soft theorems of scattering amplitudes, and like them they are
statements about residues at a soft singularity of $\psiwf_n$.

The link to the method is immediate. Each shape is a kinematic limit of $\psiwf_3$ (or
of a higher coefficient), and each limit is governed by a singularity: the total
energy $\kT\to0$ carries the flat-space interaction; the partial energies $k_S\to0$
and internal energies $k_{\ul S}\to0$ carry factorization onto lower-point processes.
These singularities sit at analytically continued kinematics; the physical folded
and squeezed limits, and the nonanalyticity of massive exchange, are governed by
them without coinciding with them.

\subsection{Observational status and targets}

How well are these shapes measured? The state of the art is the Planck 2018
analysis of temperature and polarization \cite{Planck:2019kim}, which reports (at
68\% CL, T+E)
\begin{equation}
    \begin{gathered}
        f_{\rm NL}^{\rm local}=-0.9\pm5.1,\qquad
        f_{\rm NL}^{\rm equil}=-26\pm47,\\
        f_{\rm NL}^{\rm ortho}=-38\pm24,
    \end{gathered}
\end{equation}
all consistent with zero. These are remarkable numbers: the fluctuations are
Gaussian to a part in $\sim10^4$. The same analysis, run directly on the EFT
parameters through the relation $f_{\rm NL}\sim1/c_s^2$ that follows from
\eqref{eq:eft-pi}, gives Planck's own bound $c_s\ge0.021$ at 95\% CL (T+E)
\cite{Planck:2019kim}: a collider-style exclusion limit on the derivative
interactions of the inflaton, within the operator assumptions of the Planck
EFT analysis.

But the numbers are far from the physically critical thresholds. The Simons
Observatory \cite{SimonsObservatory:2018koc},now observing, forecasts roughly
a factor-of-two tightening of the equilateral and orthogonal bounds, with a
corresponding sharpening of the $c_s$ exclusion. Combined with galaxy-survey
data, kinetic Sunyaev--Zel'dovich tomography could reach
$\sigma(f_{\rm NL}^{\rm local})\simeq1$ under the optimistic assumptions of
the Simons Observatory goal forecast. For a Stage-4 survey, the reference
design projected $\sigma(f_{\rm NL}^{\rm equil})\simeq21$,
$\sigma(f_{\rm NL}^{\rm ortho})\simeq9$, and
$\sigma(f_{\rm NL}^{\rm local})\simeq1.8$ with low-$\ell$ Planck data
\cite{Abazajian:2019eic}. Although CMB-S4 itself was discontinued in 2025,
these sensitivities provide a useful benchmark for future ground-based
surveys.

For the local shape the leverage shifts to large-scale structure, where the
number of modes is not bounded by the two-dimensional last-scattering
surface: the SPHEREx all-sky spectral survey \cite{SPHEREx:2014bgr},
launched in 2025, forecasts $\sigma(f_{\rm NL}^{\rm local})\simeq0.87$
under its survey and systematic-error model. The community studies of
primordial non-Gaussianity \cite{Meerburg:2019qqi} and of inflation across
upcoming facilities \cite{Achucarro:2022qrl} identify
$\sigma(f_{\rm NL}^{\rm local})\lesssim1$ as \emph{the} milestone: the
single-field consistency floor sits far below, at the slow-roll parameters,
so an order-unity detection is a discovery of physics beyond the attractor
single-clock framework, and a null result cuts deeply into multifield
model space. Table~\ref{tab:obs} summarizes the situation.

\begin{table*}[t]
    \centering\small
    \renewcommand{\arraystretch}{1.15}
    \caption{Bispectrum shapes, the Planck 2018 bounds
        \cite{Planck:2019kim}, historical design forecasts and projected
        survey sensitivities
        \cite{SimonsObservatory:2018koc,Abazajian:2019eic,SPHEREx:2014bgr,Meerburg:2019qqi,Achucarro:2022qrl},
        and the physics each shape isolates. The local, equilateral and
        folded templates are those of
        \cite{Babich:2004gb,Chen:2006nt,Holman:2007na} and the orthogonal
        template that of \cite{Senatore:2009gt}. Planck entries are
        measurements; the remaining entries are forecasts under the
        cited studies' assumptions. The Stage-4 figures are historical
        design sensitivities rather than those of a funded survey, and
        the Simons Observatory local-shape sensitivity refers to its
        goal forecast combined with galaxy-survey data. The crossing
        of $\sigma(f_{\rm NL}^{\rm local})\sim1$ is the
        single-field/multifield threshold.}
    \begin{tblr}{
        width=\linewidth,
        colspec={X[14,l] X[31,l] X[25,l] X[30,l]},
        colsep=4pt
    }
        \toprule
        \textbf{shape}
        & \textbf{Planck 2018 (68\% CL, T+E)}
        & \textbf{forecast sensitivity}
        & \textbf{what it measures} \\
        \midrule
        local
        & $f_{\rm NL}=-0.9\pm5.1$
        & $\sigma\simeq0.87$ (SPHEREx),
          $1$ (SO goal, kSZ+LSS),
          $1.8$ (Stage-4$+$Planck)
        & extra light fields \\
        equilateral
        & $f_{\rm NL}=-26\pm47$
        & $\sigma\simeq21$ (Stage-4)
        & $c_s<1$, $\dot\pi(\partial_i\pi)^2$, $\dot\pi^3$ \\
        orthogonal
        & $f_{\rm NL}=-38\pm24$
        & $\sigma\simeq9$ (Stage-4)
        & independent EFT combination \\
        folded
        & (in equil./ortho.\ basis)
        & --
        & initial state beyond Bunch--Davies \\
        collider
        & unconstrained at $f_{\rm NL}\sim1$
        & LSS, 21\,cm
        & masses and spins $\sim H$ \\
        \bottomrule
    \end{tblr}
    \label{tab:obs}
\end{table*}

The shapes being hunted are functions on the triangle, and the precision of a
template search is set by the precision of the theory that generates the
templates. The efficient computation of
$\psiwf_n$ (for massive exchange, for higher points, at loop level) is what
turns these experimental sensitivities into statements about the inflationary
Lagrangian.\footnote{For the model-by-model taxonomy of shapes and their
microphysical origins, the reviews \cite{Chen:2010xka} and
\cite{Wang:2013zva} are the standard references and complement the
amplitude-centric route taken here; for the effective-theory perspective
on primordial observables, see also \cite{Green:2022bre}.}

These are the targets. We now turn to the tools.

\section{The momentum-space toolkit}
\label{sec:toolkit}

Flat-space perturbation theory rests on full translation invariance: energy and
momentum are conserved at every vertex, and each vertex integral collapses onto
a delta function. (A)dS keeps the isometry group $SO(d{+}1,1)$ but breaks
translations along the holographic direction. Everything below exploits that one
asymmetry: boundary momenta are conserved, the radial ``energy'' is not.

From the wave equation we build the scalar propagators and their spinning
counterparts in the radial axial gauge, then the Witten-diagram Feynman rules
and the elementary $\AdS_4$ blocks, checking each against the conformal Ward
identities. The exposition follows the momentum-space
program of
\cite{Albayrak:2018tam,Albayrak:2019yve,Albayrak:2020isk,Albayrak:2023kfk}, the
spectral recursion of \cite{Raju:2010by,Raju:2011mp}, and the momentum-space
conformal technology of \cite{Bzowski:2013sza,Bzowski:2015pba}.

\subsection{Geometry of the Poincar\'e patch and the broken energy}
\label{sec:tk-geom}
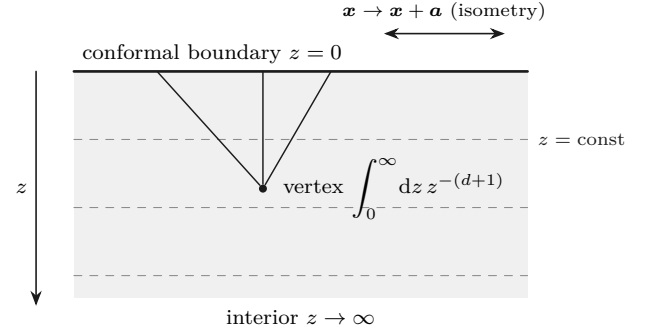
\begin{figure}[t]
	\centering
	\begin{tikzpicture}[scale=1.0]
		\fill[black!6] (0,0) rectangle (6,-3.0);
		\draw[wbndry,draw=inkcol] (0,0)--(6,0);
		\node[anchor=south west,font=\footnotesize] at (0,0){conformal boundary $z=0$};
		\draw[warr,draw=inkcol,<->] (4.1,0.5)--(5.7,0.5);
		\node[anchor=south,font=\scriptsize] at (4.9,0.55){$\bm x\to\bm x+\bm a$ (isometry)};
		\foreach \y in {-0.9,-1.8,-2.7} \draw[dashed,black!45,line width=0.4pt] (0,\y)--(6,\y);
		\node[anchor=west,font=\scriptsize,inkcol] at (6.02,-0.9){$z=\text{const}$};
		\draw[warr,draw=inkcol,->] (-0.5,0)--(-0.5,-3.1);
		\node[anchor=east,font=\footnotesize] at (-0.5,-1.55){$z$};
		\node[anchor=north,font=\footnotesize] at (3,-3.05){interior $z\to\infty$};
		\node[wdot] (v) at (2.5,-1.55){};
		\node[anchor=west,font=\footnotesize] at (2.65,-1.55){vertex $\displaystyle\int_0^\infty\!\dd z\,z^{-(d+1)}$};
		\draw[wleg] (v)--(1.1,0);
		\draw[wleg] (v)--(2.5,0);
		\draw[wleg] (v)--(3.4,0);
	\end{tikzpicture}
	\caption{The Poincar\'e patch of $\AdS_{d+1}$. Boundary translations
		$\bm x\to\bm x+\bm a$ are exact isometries, so boundary momentum is conserved;
		dashed constant-$z$ slices foliate the bulk, and the warp factor $1/z^2$ singles
		out the radial direction, whose translations are broken. External legs
		(bulk-to-boundary propagators) descend from $z=0$ to a bulk vertex integrated
		with the measure $\int_0^\infty\dd z\,z^{-(d+1)}$.}
	\label{fig:poincare}
\end{figure}

We start with the chart that makes the surviving symmetry manifest: the flat
slicing, or Poincar\'e patch. In Euclidean signature (radius set to unity)
\begin{equation}
	\dd s^2 \;=\; \frac{1}{z^2}\big(\dd z^2 + \delta_{\mu\nu}\,\dd x^\mu\dd x^\nu\big),
	\qquad \mu=1,\dots,d,\quad z>0,
	\label{eq:tk-metric}
\end{equation}
with the conformal boundary at $z\to0$ and the deep interior at $z\to\infty$
(Figure~\ref{fig:poincare}). The $d$ boundary translations $\bm x\to\bm x+\bm a$
are exact isometries of \eqref{eq:tk-metric}, since the metric depends on $\bm x$
only through $\dd\bm x$; the warp factor $1/z^2$ breaks
$z$-translations. We therefore Fourier transform along the boundary and keep $z$
in position space,
\begin{equation}
	\phi(z,\bm x)=\int\!\frac{\dd^d k}{(2\pi)^d}\,e^{\,\ii\vk\cdot\bm x}\,\phi(z,\vk),
	\label{eq:tk-fourier}
\end{equation}
so that the boundary momentum $\vk_a$ is a good label while its conjugate
radial ``energy'' is not.\footnote{``Energy'' is in quotation marks because no timelike Killing vector is
	involved: $k_a=|\vk_a|$ controls the radial dependence $e^{-k_a z}$ of the
	simplest modes and becomes a frequency after the continuation to de~Sitter,
	playing the role energy plays in flat space; it is not conserved, and its
	non-conservation is where all the physics of this review lives.} We write $k_a=|\vk_a|$ for
the magnitude (the ``energy'' of leg $a$), $\kT=\sum_a k_a$ for the total energy,
$k_{12}=k_1+k_2$ for a partial energy (a \emph{sum of magnitudes}), and
$k_{\ul{12}}=|\vk_1+\vk_2|$ for the internal energy flowing through an exchanged
line (the \emph{magnitude of a sum}). More generally, for a subset $S$ of the
external legs, $k_S=\sum_{a\in S}k_a$ and $k_{\ul S}=|\sum_{a\in S}\vk_a|$; the
combination $k_S+k_{\ul S}$, the total energy entering the sub-process of a
factorization channel, will also be called the \emph{partial energy} of that
channel, context distinguishing it from the plain sum $k_S$.

A bulk field of mass $m$ dual to a boundary operator of dimension $\Delta$ obeys
the radial wave equation (Bessel type); the two
near-boundary falloffs $z^{\Delta_\pm}$ are fixed by the equation
$\Delta(\Delta-d)=m^2$, i.e.
\begin{equation}
	\Delta_\pm=\frac d2\pm\nu,\qquad \nu=\sqrt{\tfrac{d^2}{4}+m^2}\,,
	\label{eq:nu}
\end{equation}
with the source falloff $\Delta_-=d-\Delta_+$ and the response falloff $\Delta_+$
(in the standard quantization used throughout). The order of the Bessel function
is precisely this $\nu$. In the radial axial gauge used below, the transverse
components of massless spinning fields obey the same equation with the order
shifted by the spin: a bulk gauge field or graviton dual to a conserved
spin-$\ell$ current of dimension $\Delta=d-2+\ell$ carries
\begin{equation}
	\nu_\ell=\ell-2+\tfrac d2 ,
	\label{eq:nu-ell}
\end{equation}
so $\nu_1=d/2-1$ for a conserved current (dual to a gauge field, $\Delta=d-1$)
and $\nu_2=d/2$ for the stress tensor (dual to the graviton, $\Delta=d$). (The
relation is specific to these massless, conserved sectors and this gauge; it is
not a general mass--dimension formula for arbitrary spinning fields.) This
uniform increase of the order by one unit per unit of spin is the seed of the
``web of theories'' of Section~\ref{sec:web}: a single master integral will serve
scalars, gluons, and gravitons alike \cite{Albayrak:2023kfk}.

The Euclidean AdS computation is also, up to the analytic continuation set out
at the end of this section (Section~\ref{sec:tk-dict}), the de~Sitter one: every
radial integral below doubles as a conformal-time integral of the wavefunction of
the universe.

\subsection{Propagators: bulk-to-boundary and the spectral bulk-to-bulk}
\label{sec:tk-prop}

The \emph{bulk-to-boundary propagator} $\Kbb_\nu$ is the radial solution that is
regular in the interior $z\to\infty$. Of the
two Bessel solutions, $I_\nu(kz)\sim e^{kz}$ blows up while $K_\nu(kz)\sim
e^{-kz}$ decays; interior regularity therefore selects $K_\nu$. Normalizing on a
boundary cutoff $z=\eps$ and stripping the divergent source factor
$\eps^{d/2-\nu}$,\footnote{We use the conventions of \cite{Bzowski:2013sza}; other normalizations rescale
	each external leg by a $k$-dependent factor, equivalent to a rescaling of the
	dual operator, and drop out of suitably normalized correlators.}
\begin{equation}
	\Kbb_\nu(k,z)=\frac{z^{d/2}K_\nu(kz)}{\eps^{d/2}K_\nu(k\eps)}
	\;\xrightarrow{\ \eps\to0\ }\;
	\frac{2^{1-\nu}}{\Gamma(\nu)}\,k^\nu\,z^{d/2}K_\nu(kz).
	\label{eq:btb}
\end{equation}
The selection of $K_\nu$ over $I_\nu$ is physical: it is the AdS face of the
infalling $\ii\eps$ prescription that picks the Bunch--Davies vacuum
(Section~\ref{sec:wavefunction}).

Internal lines carry the \emph{bulk-to-bulk propagator} $\Gbb_\nu$, the Green's
function of the wave operator, normalizable at the boundary and regular in the
interior.
The normalizable modes $z^{d/2}J_\nu(pz)$ are complete on the half-line, and resolving the source on them
gives the \emph{spectral (split) representation} \cite{Raju:2010by,Raju:2011mp}
\begin{equation}
	\Gbb_\nu(k;z,z')=\int_0^\infty\!\frac{p\,\dd p}{p^2+k^2}\,
	(zz')^{d/2}\,J_\nu(pz)\,J_\nu(pz')\,.
	\label{eq:btbb}
\end{equation}
The AdS internal line thus factorizes into two mode functions glued by a flat-space-like propagator,
which is the structural reason higher-point AdS diagrams so closely resemble
flat-space ones and why their $p$-integrals succumb to the residue theorem. For
$k_a>0$ the Euclidean integrand is nonsingular on the contour and needs no
$\ii\eps$; when the $p$-integral is evaluated by residues, or after continuation
to Lorentzian signature, the poles at $p=\pm\ii k$ are approached with the
$k^2+p^2+\ii\eps$ prescription used in Section~\ref{sec:loops}. The
pole at $p^2=-k^2$ is the on-shell condition for the internal line; its residue is
what the on-shell recursion of Section~\ref{sec:bypass} and the cutting rules of
Section~\ref{sec:loops} extract.

For spinning exchange one first fixes the gauge. The natural choice in
the Poincar\'e patch is the \emph{radial axial gauge} $A_z=0$ (gauge theory),
$h_{z\mu}=0$ (gravity): no Faddeev--Popov ghosts propagate, the surviving
boundary components obey the same Bessel equations, and the propagator numerator
is a sum over physical polarizations (for the complete momentum-space
bulk-to-bulk photon propagator see \cite{Moga:2025gdy}). The gauge propagator then decomposes on the
transverse and longitudinal projectors
\begin{equation}
	\vspace{-.3em}
	\Pi^{(1)}_{ij}=\delta_{ij}-\frac{\vk_i\vk_j}{k^2},\qquad
	\Pi^{(2)}_{ij}=\frac{\vk_i\vk_j}{k^2}\,,
	\label{eq:proj}
\end{equation}
separating a ``straight,'' transverse, scalar-like exchange ($\Pi^{(1)}$) from a
purely longitudinal ``crossed'' piece ($\Pi^{(2)}$); overall factors of $\ii$
belong to the propagator convention, not to the idempotent projectors
themselves. In the spectral representation
the gauge numerator reads $H_{ij}=\eta_{ij}+\vk_i\vk_j/p^2$, and its $1/p^2$ in the
\emph{spectral} momentum is a gauge artifact rather than a propagating
singularity: in gauge-invariant correlators it cancels among diagrams or against
transverse vertices.

\subsection{Feynman rules for Witten diagrams}
\label{sec:tk-feyn}

A boundary correlator is computed from \emph{Witten diagrams}, constructed exactly
as flat Feynman diagrams. An external state corresponds to a
bulk-to-boundary propagator $\Kbb$; an internal line to a bulk-to-bulk
propagator $\Gbb$; a vertex to the flat-space vertex read from the
bulk action, decorated by a power of $z$; and only the $d$ boundary momenta are
conserved. Concretely:
\begin{enumerate}[leftmargin=2.1em,itemsep=1pt,topsep=2pt]
	\item each external leg $a$ gets $\Kbb_{\nu_a}(k_a,z_v)$ ending on its vertex $v$;
	\item each internal line gets $\Gbb_\nu(k_{\ul{\cdots}};z_v,z_{v'})$ with its
	spectral integral $\int_0^\infty\!\dd p$, the internal energy $k_{\ul{\cdots}}$
	being the magnitude of the boundary momentum flowing through it;
	\item each vertex gets its flat-space tensor/coupling factor times a power of $z$;
	\item integrate every vertex with $\int_0^\infty\dd z_v\,z_v^{-(d+1)}$.
\end{enumerate}
Once the boundary momenta are imposed, what remains is a product of Bessel
functions integrated over the radial coordinates. These radial integrals are
the only genuinely curved-space feature, and the strategy is
to do them once and reuse them as \emph{building blocks}. The power of
$z$ at a vertex follows a simple counting rule derived in Section~\ref{sec:web}: a
vertex joining conserved currents of spins $\ell_1,\ell_2,\ell_3$ with $n$
derivatives carries $z^{\ell_1+\ell_2+\ell_3+n}$ \cite{Albayrak:2023kfk}. For the
Yang--Mills cubic vertex ($\ell_i=1$, $n=1$) this is $z^4$; for the quartic
($\ell_i=1$, $n=0$) again $z^4$; for the graviton cubic ($\ell_i=2$, $n=2$) it is
$z^8$.

\subsection{The \texorpdfstring{$d=3$}{d=3} building blocks}
\label{sec:tk-blocks}

Everything becomes explicit in $\AdS_4$ ($d=3$): the natural home of cosmology,
and an odd dimension, so the Bessel orders of massless and conformally
coupled fields are half-integers. Half-integer Bessels are elementary,
\begin{equation}
	\begin{gathered}
		K_{1/2}(kz)=\sqrt{\tfrac{\pi}{2kz}}\,e^{-kz},\qquad
		J_{1/2}(pz)=\sqrt{\tfrac{2}{\pi pz}}\,\sin(pz),\\
		J_{-1/2}(pz)=\sqrt{\tfrac{2}{\pi pz}}\,\cos(pz),
	\end{gathered}
	\label{eq:tk-halfint}
\end{equation}
so a $\nu=1/2$ external leg is a bare exponential $e^{-k_a z}$, an internal mode
is a $z^{-1/2}\sin(pz)$, and every vertex integral is a Laplace transform of
exponentials against sines. Fields with $\nu=3/2$, such as the minimally coupled
scalar and the graviton sector, dress the exponential with a polynomial in $kz$
but remain elementary; the seed blocks below are the $\nu=1/2$ case. Writing $K$ for an external leg, $J$ for an internal,
normalizable mode, and $E$ for the sum of the external energies entering the
vertex, the elementary blocks are
\begin{widetext}
	\begin{subequations}
		\label{eq:blocks}
		\begin{align}
			\kkk(k_1,k_2,k_3)&=\int_0^\infty\!\dd z\;e^{-Ez}
			&\hspace{-2.5em}&=\frac{1}{E}, &E&=k_{123},\\[2pt]
			\kkj(k_1,k_2;p)&=\int_0^\infty\!\dd z\;e^{-Ez}\sin(pz)
			&\hspace{-2.5em}&=\frac{p}{E^2+p^2}, &E&=k_{12},\\[2pt]
			\kjj(k_1;p,q)&=\int_0^\infty\!\dd z\;e^{-Ez}\sin(pz)\sin(qz)
			&\hspace{-2.5em}&=\frac{E}{2}\!\left[\frac{1}{E^2+(p{-}q)^2}-\frac{1}{E^2+(p{+}q)^2}\right],
			&E&=k_1,
		\end{align}
	\end{subequations}
\end{widetext}
\noindent where the $z$-powers of the $\AdS_4$ measure and vertices are absorbed
into the definitions so the answers come out clean,\footnote{In the computation of $\kkk$, the three
	$K_{1/2}$'s lead to $(\pi/2)^{3/2}(k_1k_2k_3)^{-1/2}z^{-3/2}e^{-k_{123}z}$; the $z^4$
	vertex-plus-measure weight cancels every constant and power, leaving
	$\int_0^\infty\dd z\,e^{-k_{123}z}$. The product-to-sum identity
	$2\sin pz\,\sin qz=\cos(p{-}q)z-\cos(p{+}q)z$ handles $\kjj$.} and a $J$-mode
carries the normalization $\sqrt{2/(\pi p)}$ of \eqref{eq:tk-halfint}, tracked
explicitly when gluing blocks. Adding external legs only lengthens $E$: the
four- and five-point blocks $\mathsf{KKKK}$ and $\mathsf{KKKJ}$ are
\eqref{eq:blocks} with $E=k_{1234}$ and $E=k_{123}$, and so on to any
multiplicity.

The denominators already display the two singularity families that organize the
review. The all-$K$ blocks carry the \emph{total-energy pole} $1/k_{123}$,
$1/k_{1234}$: it sits at $\kT\to0$, outside the physical region $k_a>0$, but its
residue survives, $\lim_{\kT\to0}\kT\,\kkk=1$ (in a full diagram the order and
coefficient of the leading total-energy singularity depend on the interaction
and external fields), the germ of the flat-space limit
\cite{Maldacena:2002vr,Maldacena:2011nz,Raju:2010by,Raju:2011mp}, made precise in
Section~\ref{sec:tree}. The
mixed $K/J$ blocks instead carry \emph{partial-energy} denominators such as
$(k_1{+}k_2)^2+p^2$; under the spectral integral
$\int p\,\dd p/(p^2+k_{\ul{12}}^2)$ of an internal line, the pole at
$p=\ii(k_1{+}k_2)$ converts them into poles in the partial energy
$k_1+k_2+k_{\ul{12}}$ of a sub-process: the factorization mechanism of Sections~\ref{sec:tree}
and~\ref{sec:bypass}.

These energies uplift to relativistic invariants. Assign each leg the null
$(d{+}1)$-momentum $\hat{\vk}_a=(\ii k_a,\vk_a)$, with
$\hat{\vk}_a^2=-k_a^2+|\vk_a|^2=0$: the external data sit on a lightcone one
dimension up, and the exchanged Mandelstam invariant reads
\begin{equation}
	s=(\hat{\vk}_1+\hat{\vk}_2)^2=k_{\ul{12}}^2-k_{12}^2,
	\label{eq:tk-mandelstam}
\end{equation}
vanishing on the flat mass shell and measuring, elsewhere, the off-shellness that
(A)dS supplies.

Away from $d=3$, or for minimally coupled fields, the elementary reductions of
\eqref{eq:blocks} fail, and the three-point seed is the general \emph{triple-$K$
	integral}
\begin{equation}
	I_{\alpha\{\nu_1\nu_2\nu_3\}}(k_1,k_2,k_3)
	=\int_0^\infty\!\dd z\;z^{\alpha}\prod_{a=1}^3 k_a^{\nu_a}K_{\nu_a}(k_a z),
	\label{eq:tripleK}
\end{equation}
the central special function of momentum-space CFT \cite{Bzowski:2013sza}. It
converges at $z\to\infty$ for any $k_a>0$ but at $z\to0$ only when
$\alpha+1>\sum_a|\nu_a|$; outside that window it is defined by shifting the parameters,
$\alpha\to\alpha+u\eps$ and $\nu_a\to\nu_a+v_a\eps$, and the resulting poles in
$\eps$ are physical: removed by local counterterms, they leave behind
scheme-dependent contact terms and the momentum-space face of conformal
anomalies and the renormalization of composite operators \cite{Bzowski:2015pba}. The blocks
\eqref{eq:blocks} are the simplest cases in which \eqref{eq:tripleK} degenerates
to elementary functions.

\subsection{Conformal Ward identities}
\label{sec:tk-ward}

The boundary correlator is constrained by the conformal group, whose generators,
after the Fourier transform \eqref{eq:tk-fourier}, act on the momenta as
differential operators \cite{Bzowski:2013sza}. Rotations and translations are manifest; the content sits in
dilatations and special conformal transformations. Acting on the reduced $n$-point
function (the momentum-conserving delta stripped, a function of $n-1$ independent
momenta), the dilatation Ward identity is first order,
\begin{equation}
	0=\Big[\,\sum_{a=1}^n\Delta_a-(n-1)d
	-\sum_{a=1}^{n-1}\vk_a\!\cdot\!\frac{\partial}{\partial\vk_a}\,\Big]
	\langle\!\langle\mathcal O_1\cdots\mathcal O_n\rangle\!\rangle,
	\label{eq:ward-dil}
\end{equation}
the statement that the correlator is homogeneous of definite degree, while the
special-conformal identity is genuinely dynamical and \emph{second order},
\begin{equation}
	\begin{split}
		0=\sum_{a=1}^{n-1}\Big[\,&2(\Delta_a-d)\frac{\partial}{\partial k_a^\kappa}
		-2\,k_a^{\lambda}\frac{\partial^2}{\partial k_a^{\lambda}\partial k_a^\kappa}\\
		&+k_{a\kappa}\frac{\partial^2}{\partial k_a^{\lambda}\partial k_{a\lambda}}\,\Big]
		\langle\!\langle\cdots\rangle\!\rangle.
	\end{split}
	\label{eq:ward-sct}
\end{equation}
This second-order character is the defining feature of momentum-space CFT: the
conformal constraint, algebraic in position space, becomes a differential equation
in momentum space, whose solutions are the special functions (triple-$K$
integrals, and the dilogarithms and Appell functions met later) that organize the
subject \cite{Bzowski:2013sza,Gillioz:2022yze}. On a scalar three-point function,
rotational invariance reduces the correlator to $f(k_1,k_2,k_3)$, and \eqref{eq:ward-sct} collapses to
$(\mathcal K_i-\mathcal K_j)f=0$ with the Bessel operator
$\mathcal K_a=\partial_{k_a}^2+\tfrac{1-2\nu_a}{k_a}\partial_{k_a}$. The triple-$K$
integral \eqref{eq:tripleK} solves this because each factor obeys
$\mathcal K_a\big[k_a^{\nu_a}K_{\nu_a}(k_a z)\big]=z^2\,k_a^{\nu_a}K_{\nu_a}(k_a z)$,
so $\mathcal K_a I$ is the \emph{same} object for every $a$ and the differences
annihilate it. This is the momentum-space avatar of conformal invariance and the reason
\eqref{eq:tripleK} is, for generic weights and up to the overall OPE
coefficient, the unique conformal three-point seed; degenerate cases require
the renormalization just described.\footnote{Momentum-space CFT is a subject in its own right: solutions of
	\eqref{eq:ward-sct} by one-loop master integrals and the anomaly program grown
	from them \cite{Coriano:2013jba,Coriano:2020ees}; the generalized hypergeometric
	structure of the Ward identities in $d>2$ \cite{Coriano:2020ccb}; conformal Ward
	identities for scalar and tensor four-point functions \cite{Coriano:2019nkw};
	renormalization of the
	tensorial sector \cite{Bzowski:2018fql}; simplex representations of $n$-point
	functions \cite{Bzowski:2019kwd,Bzowski:2020kfw}; general-$d$ handbooks
	\cite{Bzowski:2022rlz,Bzowski:2023jwt}; spinor-helicity methods for spinning
	correlators in $d=3$ \cite{Jain:2020rmw,Jain:2021vrv,Maldacena:2011jn};
	parity-odd three-point functions in $d=3$ \cite{Jain:2021wyn}; correlators from
	slightly broken higher-spin symmetry \cite{Jain:2020puw}; supersymmetric
	extensions via super spinor-helicity and Grassmann twistors \cite{Jain:2023idr};
	momentum-space conformal blocks and crossing
	\cite{Gillioz:2018mto,Isono:2018rrb,Isono:2019wex}; the Lorentzian continuation
	\cite{Bautista:2019qxj}; momentum-space correlators for lightcone Hamiltonian
	truncation \cite{Anand:2019lkt}. We need only the corner built on \eqref{eq:tripleK}.}

\subsection{From anti--de~Sitter to de~Sitter}
\label{sec:tk-dict}

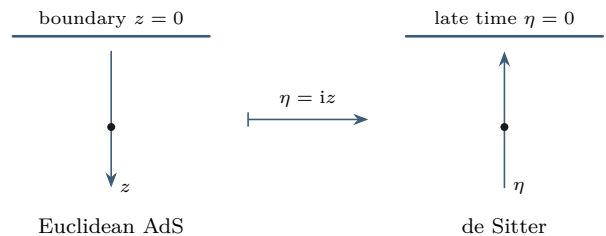
\begin{figure}[t]
	\centering
	\begin{tikzpicture}[scale=1.0]
		\draw[wbndry] (-3.9,1.1)--(-1.3,1.1); \node[above,font=\scriptsize] at (-2.6,1.1){boundary $z=0$};
		\draw[warr,->] (-2.6,0.9)--(-2.6,-0.9); \node[right,font=\scriptsize] at (-2.6,-0.9){$z$};
		\node[wdot] at (-2.6,-0.1){};
		\node[font=\footnotesize] at (-2.6,-1.4){Euclidean $\AdS$};
		\draw[warr,|->] (-0.8,0)--(0.8,0); \node[above,font=\scriptsize] at (0,0.05){$\eta=\ii z$};
		\draw[wbndry] (1.3,1.1)--(3.9,1.1); \node[above,font=\scriptsize] at (2.6,1.1){late time $\eta=0$};
		\draw[warr,->] (2.6,-0.9)--(2.6,0.9); \node[right,font=\scriptsize] at (2.6,-0.9){$\eta$};
		\node[wdot] at (2.6,-0.1){};
		\node[font=\footnotesize] at (2.6,-1.4){de~Sitter};
	\end{tikzpicture}
	\caption{The AdS--dS continuation \eqref{eq:cont}. A Euclidean AdS Witten diagram of
		radial depth $z$ becomes, under $\eta=\ii z$, a de~Sitter wavefunction coefficient on
		the late-time slice $\eta\to0$; the bulk-to-boundary propagator $e^{-kz}$ continues to
		the Bunch--Davies mode $e^{\ii k\eta}$, and the radial integrals of
		this section become the conformal-time integrals of the wavefunction.}
	\label{fig:adsds}
\end{figure}

Everything in this toolkit is Euclidean AdS; cosmology is one continuation away
(Figure~\ref{fig:adsds}),
\begin{equation}
	\label{eq:cont}
	R_{\dS}=-\ii\,R_{\AdS},\qquad \eta=\ii\,z,
\end{equation}
under which the radial coordinate becomes conformal time with the late-time
boundary at $\eta\to0$ \cite{Harlow:2011ke}, the bulk-to-boundary mode $e^{-kz}$
of \eqref{eq:btb} becomes the Bunch--Davies mode $e^{\,\ii k\eta}$
\cite{Bunch:1978yq}, and \emph{the radial integrals of this section become the
	conformal-time integrals of the wavefunction of the universe}
\cite{Maldacena:2002vr,Maldacena:2011nz,Weinberg:2005vy}. The continuation acts
on everything at once: radius, couplings, tensor structures, integration
contour, and boundary counterterms, so an AdS correlator and a de~Sitter
wavefunction coefficient share a bulk integral but need not share overall
phases or local terms. We compute almost everything in Euclidean AdS, where the analytic structure is cleanest, and read
off cosmology at the end; the care the continuation requires is the business of
Section~\ref{sec:inflation}.
\begin{lesson}
	The wavefunction of the universe is an anti--de~Sitter object. Under
	\eqref{eq:cont} the late-time Bunch--Davies wavefunction is the continuation of
	the Euclidean AdS partition function, and each coefficient $\psiwf_n$ is a
	continued EAdS boundary correlator (for massive fields, a finite sum of them).
	One object, two evaluations; the matching of couplings, phases, and local terms
	is the dictionary's fine print.
\end{lesson}

The toolkit is complete: propagators, Feynman rules, elementary blocks, the
conformal constraints they satisfy, and the continuation that reads every result
cosmologically. Section~\ref{sec:tree} assembles it into
explicit correlators, beginning with the simplest spinning object there is.

\section{Tree-level correlators, efficiently}
\label{sec:tree}

We can now compute. Our interest is cosmology. Yet we begin with gluons, which,
a careful reader will object, are not by themselves primordial observables. In four
spacetime dimensions the classical Yang--Mills action is Weyl invariant, so a
gauge field does not feel the expansion of a conformally flat universe. Its mode
function is the flat-space one in conformal coordinates, nothing is amplified at
horizon crossing, and no primordial gauge field is produced this way. But the same Weyl
invariance is precisely what makes the gluon a clean lab in
$(\mathrm A)\dS_4$: its Bessel order is $\nu=\tfrac12$, as discussed in Section~\ref{sec:tk-blocks}.

Three further reasons to consider gluons. The modern amplitudes program was built on gluons: the Parke--Taylor
formula \cite{Parke:1986gb}, BCFW recursion \cite{Britto:2005fq}, and
color--kinematics duality \cite{Bern:2008qj} were all discovered in Yang--Mills theory,
so gluons are the natural first place to test how directly
flat-space technology imports. The web of Section~\ref{sec:web} and
the double copy of Section~\ref{sec:bypass} then recycle the gluon computation
into the graviton one, and the graviton, unlike the gluon, is a primordial
observable in principle, through the tensor bispectrum
$\langle\gamma\gamma\gamma\rangle$ computed below \cite{Maldacena:2011nz}. And in AdS the gluon correlator is the conserved-current
correlator $\langle JJ\cdots J\rangle$ of the boundary CFT, of interest
independently of cosmology.

This section is the computational heart of the review: we compute the four-point
gluon correlator once, in full, extract the two structural principles that make
it efficient, and then treat gravitons and scalars by the same logic.
The focus is method rather than one calculation per theory: every spinning correlator
factors into a polarization part, which is pure index algebra, times a small set of
\emph{scalar} integrals, the \emph{masters}. Similar masters serve gluons,
gravitons, and scalars alike, and they are not new objects: they are the
integrals that build the wavefunction of the universe. Throughout, the toolkit of Section~\ref{sec:toolkit} is fixed:
flat-space Yang--Mills, Einstein, and scalar vertices, external Bessel-$K$ legs,
and bulk-to-bulk lines opened in the spectral representation \eqref{eq:btbb}. The
work is to assemble them, reduce the radial integrals to the $d=3$ building
blocks \eqref{eq:blocks}, and do the remaining spectral integrals by residues.

\subsection{Gluons: masters and residues}
\label{sec:tree-gluons}

Color is stripped before any kinematics. At tree level the connected $n$-point
current correlator decomposes into single traces,
$\langle J^{a_1}\!\cdots J^{a_n}\rangle'=\sum_{\sigma\in S_n/\mathbb Z_n}
\operatorname{Tr}(T^{a_{\sigma(1)}}\!\cdots T^{a_{\sigma(n)}})\,W_n(\sigma)$,
where the prime strips the momentum-conserving delta function. We call
$W_n(1\cdots n)$ a color-ordered boundary correlator, reserving $\mathcal M$
for the scalar master integrals below \cite{Albayrak:2018tam}.

\paragraph{Warm-up: the three-point correlator.} A single cubic vertex contracted
with three $\Kbb$ legs gives, in $d=3$,
\begin{equation}
	\begin{aligned}
		W_3^{ijk}={}&{}V^{ijk}(\vk_1,\vk_2,\vk_3)\;\kkk(k_1,k_2,k_3)\\
		={}&{}\frac{V^{ijk}(\vk_1,\vk_2,\vk_3)}{\kT},\qquad \kT=k_1+k_2+k_3,
	\end{aligned}
	\label{eq:3pt}
\end{equation}
with $V^{ijk}$ the familiar color-ordered flat Yang--Mills vertex,
$V^{ijk}=\eta^{ij}(k_1{-}k_2)^k+\eta^{jk}(k_2{-}k_3)^i+\eta^{ki}(k_3{-}k_1)^j$ up
to normalization: a purely kinematic flat-space numerator multiplying the
total-energy pole $1/\kT$, into which all of the AdS dynamics has collapsed.
Its residue $\lim_{\kT\to0}\kT\,W_3=V^{ijk}$ is exactly the color-ordered flat three-gluon
amplitude, and the full $d$-dependence of the same diagram is the triple-$K$
integral \eqref{eq:tripleK}, reproducing the unique conformal three-point
structure of \cite{Bzowski:2013sza}. One diagram thus already carries the flavor of the many ingredients of the
program: a flat-space vertex, a radial integral reduced to an
elementary block, a total-energy singularity whose residue is the amplitude, and
exact conformal consistency. Under the continuation $\eta=\ii z$, completed by
the dictionary of Section~\ref{sec:tk-dict} for couplings and normalizations,
it becomes the inflationary three-point coefficient $\psiwf_3$.

\begin{figure}[t]
	\centering
	\begin{tikzpicture}[scale=0.8,baseline=(current bounding box.south)]
		\draw[wbndry] (0,1.8)--(2.6,1.8);
		\node[wdot] (a) at (0.6,0.9){}; \node[wdot] (b) at (2.0,0.9){};
		\draw[wgluon] (0.2,1.8)--(a); \draw[wgluon] (0.9,1.8)--(a);
		\draw[wgluon] (1.7,1.8)--(b); \draw[wgluon] (2.4,1.8)--(b);
		\draw[wgluon] (a)--(b);
		\node[above,font=\scriptsize] at (0.2,1.8){$1$}; \node[above,font=\scriptsize] at (0.9,1.8){$2$};
		\node[above,font=\scriptsize] at (1.7,1.8){$3$}; \node[above,font=\scriptsize] at (2.4,1.8){$4$};
		\node[font=\footnotesize] at (1.3,0.1){(a) $s$};
	\end{tikzpicture}\hfill
	\begin{tikzpicture}[scale=0.8,baseline=(current bounding box.south)]
		\draw[wbndry] (0,1.8)--(2.6,1.8);
		\node[wdot] (a) at (1.3,1.25){}; \node[wdot] (b) at (1.3,0.55){};
		\draw[wgluon] (0.9,1.8)--(a); \draw[wgluon] (1.7,1.8)--(a);
		\draw[wgluon] (0.2,1.8)--(b); \draw[wgluon] (2.4,1.8)--(b);
		\draw[wgluon] (a)--(b);
		\node[above,font=\scriptsize] at (0.2,1.8){$1$}; \node[above,font=\scriptsize] at (0.9,1.8){$2$};
		\node[above,font=\scriptsize] at (1.7,1.8){$3$}; \node[above,font=\scriptsize] at (2.4,1.8){$4$};
		\node[font=\footnotesize] at (1.3,0.1){(b) $t$};
	\end{tikzpicture}\hfill
	\begin{tikzpicture}[scale=0.8,baseline=(current bounding box.south)]
		\draw[wbndry] (0,1.8)--(2.6,1.8);
		\node[wdot] (a) at (1.3,0.9){};
		\draw[wgluon] (0.2,1.8)--(a); \draw[wgluon] (0.9,1.8)--(a);
		\draw[wgluon] (1.7,1.8)--(a); \draw[wgluon] (2.4,1.8)--(a);
		\node[above,font=\scriptsize] at (0.2,1.8){$1$}; \node[above,font=\scriptsize] at (0.9,1.8){$2$};
		\node[above,font=\scriptsize] at (1.7,1.8){$3$}; \node[above,font=\scriptsize] at (2.4,1.8){$4$};
		\node[font=\footnotesize] at (1.3,0.1){(c) contact};
	\end{tikzpicture}
	\caption{The three planar contributions to the color-ordered four-gluon correlator;
		gluon lines are drawn curly, the boundary as a slate rule. The $u$-channel $(13)(24)$
		is non-planar and absent.}
	\label{fig:channels}
\end{figure}
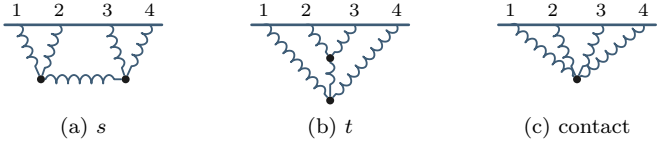

For the ordering $(1234)$ the planar four-gluon amplitude is
an $s$-channel exchange $(12)(34)$, a $t$-channel exchange $(23)(41)$, and a contact
term; the non-adjacent exchange $(13)(24)$ ($u$-channel) is absent. The three topologies are shown in
Figure~\ref{fig:channels}.

\paragraph{Gluing two blocks with the spectral integral.} The $s$-channel of
Figure~\ref{fig:channels}(a) joins legs $1,2$ at a bulk vertex $z$ and legs
$3,4$ at $z'$, linked by a gluon bulk-to-bulk line of boundary momentum
$\vk_{12}=-\vk_{34}$. The detailed construction can be found in \cite{Albayrak:2018tam}. Opening that line in
the spectral representation \eqref{eq:btbb} introduces a radial momentum $p$, a
flat-style propagator $1/(p^2+k_{\ul{12}}^2)$, the tensor numerator
$H_{mn}(\vk_{12},p)=\eta_{mn}+(\vk_{12})_m(\vk_{12})_n/p^2$, and two normalizable
Bessel $J$-modes ($J$-modes hereafter) attaching to the vertices. Each vertex then carries two external $K$-legs
and one internal $J$-mode, so that
\begin{equation}
	\label{eq:4s-setup}
	\begin{split}
		W_{4,s}=\int_0^\infty\!\dd p\,
		&\frac{-\ii\,p\,V^{12m}\,H_{mn}(\vk_{12},p)\,V^{34n}}{p^2+k_{\ul{12}}^2}\\
		&\times\mathcal V(k_1,k_2;p)\,\mathcal V(k_3,k_4;p),
	\end{split}
\end{equation}
where $\mathcal V(k_a,k_b;p)$ is the radial integral at a vertex: two external
legs, one normalizable mode, the vertex weight and the measure. 

In $d=3$ each
half-integer $K$-leg collapses to an exponential and the $J$-mode to
$\sqrt{2/(\pi p)}\,\sin(pz)$ times the absorbed $z$-powers, so by the $\kkj$ block of
\eqref{eq:blocks},
$\mathcal V(k_a,k_b;p)=\sqrt{2/(\pi p)}\;\kkj(k_a,k_b;p)=\sqrt{2/\pi}\,\sqrt p/(k_{ab}^2+p^2)$.
The product of the two vertex integrals contributes
$\tfrac2\pi\,p/[(k_{12}^2+p^2)(k_{34}^2+p^2)]$, leaving a
\emph{single} spectral integral,
\begin{multline}
	\label{eq:4s-omega}
	W_{4,s}=-\ii\,\tfrac2\pi\,V^{12m}V^{34n}
	\!\int_0^\infty\!\!\dd p\;
	\frac{p^2\,H_{mn}(\vk_{12},p)}
	{(p^2{+}k_{\ul{12}}^2)}\\
	\times\frac{1}{(p^2{+}k_{12}^2)(p^2{+}k_{34}^2)}.
\end{multline}
Inserting $H_{mn}$ splits this into two scalar integrands: the $\eta_{mn}$ piece
keeps the $p^2$ in the numerator, while the longitudinal piece has its $1/p^2$
cancel it. Writing
$W_{4,s}=-\ii\,V^{12m}V^{34n}\big[\eta_{mn}\,\mathcal M^{(1)}
+(\vk_{12})_m(\vk_{12})_n\,\mathcal M^{(2)}\big]$, the entire content of the
$s$-channel up to polarization algebra is the pair of \emph{master integrands}
\begin{equation}
	\label{eq:4s-masters}
	\boxed{\;
		\begin{gathered}
			\mathcal M^{(1)}=\frac{1}{k_{1234}\,k_{12\ul{12}}\,k_{34\ul{12}}}\,,\\[2pt]
			\mathcal M^{(2)}=\frac{k_{1234\ul{12}}}
			{k_{12}\,k_{34}\,k_{\ul{12}}\,k_{1234}\,k_{12\ul{12}}\,k_{34\ul{12}}}\;,
		\end{gathered}\;}
\end{equation}
where $k_{12\ul{12}}=k_1+k_2+k_{\ul{12}}$ is a partial energy. The transverse master
$\mathcal M^{(1)}$ is the \emph{conformally coupled scalar exchange
	wavefunction}: the two-site wavefunction \eqref{eq:psi2} and the seed of the web
\eqref{eq:seed}.
The longitudinal master $\mathcal M^{(2)}$ is the curvature's first fingerprint.

\paragraph{The residue engine.} Both spectral integrals belong to one family,
evaluated by residues. With $a,b,c>0$,
\begin{equation}
	\label{eq:masters-I}
	\begin{aligned}
		I_1&=\!\int_0^\infty\!\frac{p^2\,\dd p}{(p^2{+}a^2)(p^2{+}b^2)(p^2{+}c^2)}\\
		&=\frac{\pi/2}{(a{+}b)(b{+}c)(c{+}a)},\\[3pt]
		I_0&=\!\int_0^\infty\!\frac{\dd p}{(p^2{+}a^2)(p^2{+}b^2)(p^2{+}c^2)}\\
		&=\frac{\pi(a{+}b{+}c)/2}{abc\,(a{+}b)(b{+}c)(c{+}a)},
	\end{aligned}
\end{equation}
so that $\mathcal M^{(1)}=\tfrac2\pi I_1(k_{\ul{12}},k_{12},k_{34})$ and
$\mathcal M^{(2)}=\tfrac2\pi I_0(k_{\ul{12}},k_{12},k_{34})$, reproducing
\eqref{eq:4s-masters} on using $k_{12}+k_{34}=\kT$ and
$k_{\ul{12}}+k_{12}=k_{12\ul{12}}$.
Consider the evaluation of $I_1$. The integrand is even in $p$ and falls off as
$p^{-4}$, so $I_1$ equals half the contour integral closed in the upper half-plane,
picking up the simple poles at $p=\ii a,\ii b,\ii c$; the residue at $p=\ii a$ is
$(\ii a)^2/[2\ii a\,(b^2-a^2)(c^2-a^2)]=-a/[2\ii(b^2-a^2)(c^2-a^2)]$, and summing
the three,
\begin{equation}
	\label{eq:I1res}
	\begin{split}
		I_1=\frac\pi2\bigg[&\frac{-a}{(b^2-a^2)(c^2-a^2)}
		+\frac{-b}{(a^2-b^2)(c^2-b^2)}\\
		&+\frac{-c}{(a^2-c^2)(b^2-c^2)}\bigg].
	\end{split}
\end{equation}
Each term is an intricate rational function with spurious-looking denominators
$b^2-a^2$, but the sum collapses to a compact form by the partial-fraction
identity
\begin{equation}
	\label{eq:pf}
	\sum_{\mathrm{cyc}}\frac{-a}{(b^2-a^2)(c^2-a^2)}=\frac{1}{(a+b)(b+c)(c+a)},
\end{equation}
whose right-hand side has poles \emph{only} at the physical energy sums $a+b$,
$b+c$, $c+a$, never at the spurious differences. This is why the answer is a product
of total- and partial-energy factors, and it is the momentum-space shadow of unitarity: a
singularity occurs only where a physical sub-process can go on shell, a statement
made precise by the factorization and cutting rules of
Sections~\ref{sec:bypass} and~\ref{sec:loops}. At higher
points each internal line contributes one such integral, done one at a time by
exactly this method, so the entire tree-level $\AdS_4$ gluon sector is rational.

In summary: a spinning correlator is a fixed polarization sum acting on a small set of
\emph{scalar} master integrands. Compute the masters once; the spin is algebra: the gauge correlator is a
cosmological scalar skeleton dressed by Yang--Mills numerators.

\paragraph{Factorization is an AdS cut.} The partial-energy pole of
$\mathcal M^{(1)}$ at $k_{12\ul{12}}\to0$ factorizes,
\begin{equation}
	\Res_{k_{12\ul{12}}\to0}\mathcal M^{(1)}=\frac{1}{k_{1234}\,k_{34\ul{12}}}\,,
\end{equation}
into the left three-point sub-amplitude times the rest of the diagram glued across
the internal line: the (A)dS avatar of an $s$-channel unitarity cut, the line cut
by replacing its propagator with the on-shell exchange of the corresponding state.
For spinning exchanges, the polarization sum and normalization that complete this
scalar statement into a product of three-point objects are supplied by the full
wavefunction coefficient.
The same structure returns as the transition amplitudes sewn in the on-shell
recursion of Section~\ref{sec:bypass} and as the facets of the cosmological
polytope \cite{Arkani-Hamed:2017fdk}.

\paragraph{The total-energy residue is the flat-space amplitude.} The
\emph{total-energy} pole at $\kT=k_{1234}\to0$ carries the flat $S$-matrix. The
limit is reached by analytic continuation of the energies; setting
$k_{34}\to-k_{12}$ so that $\kT\to0$,
\begin{equation}
	\label{eq:flatlim}
	\boxed{\;
		\begin{aligned}
			\Res_{\kT\to0}\mathcal M^{(1)}
			&=\frac{1}{k_{12\ul{12}}\,k_{34\ul{12}}}\bigg|_{k_{34}\to-k_{12}}\\
			&=\frac{1}{(k_{\ul{12}}{+}k_{12})(k_{\ul{12}}{-}k_{12})}\\
			&=\frac{1}{k_{\ul{12}}^2-k_{12}^2}=\frac1s\;,
		\end{aligned}\;}
\end{equation}
with $s=k_{\ul{12}}^2-k_{12}^2=(\hat{\vk}_1+\hat{\vk}_2)^2$ the flat Mandelstam
invariant of the null uplift \eqref{eq:tk-mandelstam}. The residue is exactly
the $s$-channel pole of the flat four-gluon amplitude: \emph{the AdS correlator,
	continued to vanishing total energy, knows about flat-space scattering}. This is
the flat-space limit \cite{Raju:2012zr,Maldacena:2002vr,Maldacena:2011nz},
treated systematically in \cite{Li:2021snj,Jain:2022ujj};
at three points it is the residue check of the warm-up \eqref{eq:3pt}. The limit
has since been sharpened into a systematic expansion around flat space
\cite{Lee:2023kno,Cespedes:2025dnq}.
In general the $(d{+}1)$-dimensional amplitude is the coefficient of the leading
$\kT\to0$ pole, of order $(n/2-1)(d-3)+1$ for gluons and $(n/2-1)(d-1)+1$ for
gravitons (the two orders differ by
$n-2$), and the statement survives to all loop orders in
dimensional regularization \cite{Raju:2012zr}. The longitudinal master
$\mathcal M^{(2)}$ carries a simple pole at $\kT\to0$ as well (its residue on the
flat locus is $-1/(k_{12}^2\,s)$), but it multiplies the longitudinal tensor
structure rather than the flat vertex contraction of \eqref{eq:flatlim}. The
split into transverse and longitudinal masters is a gauge-dependent organization,
and the flat amplitude emerges only once contractions, channel sums, and contact
terms combine; what is invariant is that the boundary correlator carries content
beyond the $S$-matrix, genuinely AdS data, not a correction to be discarded. The amplitude-to-correlator
dictionary can be systematized: $\AdS_4$ gluon correlators expand in flat-space
amplitude structures plus curvature corrections \cite{Gomez:2026yno}, and
cosmological dressing rules convert flat-space amplitudes into (A)dS
wavefunction coefficients and correlators \cite{Chowdhury:2025ohm}.
\begin{lesson}
	The total-energy pole is the correlator's memory of flat space: its residue is
	the $S$-matrix, and everything beyond the pole is intrinsically (A)dS.
\end{lesson}

\paragraph{Assembling the four-point function.} Collecting the $s$-channel, its
$t$-channel cyclic image $1\!\to\!2\!\to\!3\!\to\!4$, and the contact term
$\mathcal M_c=V_c^{1234}/k_{1234}$ (a single quartic vertex, one radial integral)
gives the complete color-ordered boundary correlator
\begin{equation}
	\begin{aligned}
		\label{eq:4pt-full}
		\hspace{-1em}
		W_4=&
		-\frac{\ii\,V^{12m}V^{34n}\big[\eta_{mn}+\tfrac{k_{1234\ul{12}}}{k_{12}k_{34}k_{\ul{12}}}(\vk_{12})_m(\vk_{12})_n\big]}
		{k_{1234}\,k_{12\ul{12}}\,k_{34\ul{12}}}
		\\&-\big(\,{}_{12}^{34}\!\to{}_{23}^{41}\,\big)
		+\frac{V_c^{1234}}{k_{1234}} .       
	\end{aligned}
\end{equation}
\noindent
Every denominator is a product of the total energy $k_{1234}$ and internal or
partial energies ($k_{\ul{12}}$, $k_{12\ul{12}}$, and cyclic); every numerator is a
polynomial built from the flat Yang--Mills vertices. The result is rational: no
transcendental functions appear at four points in $\AdS_4$. 

These are the first explicit higher-point
gluon correlators \cite{Albayrak:2018tam}; the first four-point MHV
gluon correlator in $\AdS_4$, already organized as a flat-space-pole term plus a
genuinely AdS remainder, was obtained in \cite{Raju:2012zs}. The same engine runs
at five and six points, one spectral integral per internal line, and the
results remain rational \cite{Albayrak:2018tam}, with one qualitatively new
feature: a doubly longitudinal master appears that \emph{vanishes} at $\kT\to0$
altogether, a structure the flat $S$-matrix cannot see. Five-point gluon
amplitudes in AdS have also been computed in Mellin space \cite{Alday:2022lkk}.

\subsection{Gravitons: stripping the tensor structure}
\label{sec:tree-gravitons}

Every graviton leg and internal line carries two indices, and the bulk-to-bulk
numerator, built quadratically from the gauge one,
$H_{ab,cd}=H_{ac}H_{bd}+H_{ad}H_{bc}-\tfrac{2}{d-1}H_{ab}H_{cd}$ with
$H_{ab}=\eta_{ab}+\vk_a\vk_b/p^2$, contains pieces scaling as $p^0,p^{-2},p^{-4}$.
Expanded naively, a single exchange splits into three radial integrals, and a diagram
with $n$ internal lines into as many as $3^n$: nine at five points, twenty-seven at
six. The remedy is never to expand \cite{Albayrak:2019yve}; relatedly, AdS gluon and
graviton amplitudes organize as differential operators dressing flat-space
structures \cite{Li:2022tby}.

\paragraph{The differential operator.} The whole $p$-dependence can be re-expanded
on a basis of projectors $\mathbb{P}^{(0)},\mathbb{P}^{(1)},\mathbb{P}^{(2)}$ that depend on the boundary
momentum $\vk$ but \emph{not} on the spectral momentum $p$,
\begin{equation}
	\label{eq:Hgrav}
	H_{ab,cd}(\vk,p)=\mathbb{P}^{(0)}_{ab,cd}
	+\frac{k^2+p^2}{p^2}\,\mathbb{P}^{(1)}_{ab,cd}
	+\Big(\frac{k^2+p^2}{p^2}\Big)^2\mathbb{P}^{(2)}_{ab,cd}.
\end{equation}
These are built from the transverse and longitudinal projectors
$T_{ab}=\eta_{ab}-\vk_a\vk_b/k^2$ and $L_{ab}=\vk_a\vk_b/k^2$ that underlie
\eqref{eq:proj}: the gauge numerator is
$H_{ab}=T_{ab}+\tfrac{k^2+p^2}{p^2}L_{ab}$, so \eqref{eq:Hgrav} collects powers
of $L$, the superscript counting the longitudinal factors. Multiplied by
the scalar propagator $1/(p^2+k^2)$ that accompanies them, the three coefficients
collapse to $1/(p^2+k^2)$, $1/p^2$, and $(k^2+p^2)/p^4$. Each is reproduced by
acting on a \emph{single} scalar factor $\mathcal M_n$ (numerator set to unity)
with an operation on the internal energy $k$: the identity, the soft limit
$\lim_{\vk\to0}$, or a derivative, the $\mathbb{P}^{(2)}$ channel needing the
last two combined. Assembling them defines the operator that restores the spin.
From here on $W_n$ denotes a correlator with its tensor structure attached and
$\mathcal M_n$ its stripped scalar factor, so that
\begin{equation}
	\label{eq:Dop}
	\boxed{\;
		\begin{gathered}
			W^{\,\mathrm{grav}}_n=\Big(\prod_{e}\mathcal D^{\vk_e}\Big)\,
			\mathcal M_n^{\,\mathrm{scalar}},\\[2pt]
			\mathcal D^{\vk}=\frac{\ii}{2}\Big[\mathbb{P}^{(0)}
			+\big(\mathbb{P}^{(1)}+\mathbb{P}^{(2)}\big)\lim_{\vk\to0}
			-\mathbb{P}^{(2)}\,k^2\lim_{\vk\to0}\partial_{k^2}\Big]\;,
		\end{gathered}\;}
\end{equation}
one operator per internal line $e$, acting on the single scalar factor
$\mathcal M_n$, a multiple radial integral of exactly the kind evaluated for
gluons. The $3^n$ tensor integrals have become one scalar integral plus a purely
algebraic differentiation.\footnote{Two features of
	$\mathcal D^{\vk}$ are important. It carries \emph{no} spectral momentum (the
	$p$-dependence was entirely absorbed into the scalar coefficients of
	\eqref{eq:Hgrav}), so it commutes through $\int\!\dd p$. And it carries \emph{no}
	bulk coordinate (it acts on the boundary kinematics through $\vk$ and
	$\partial_{k^2}$, not on $z$), so it commutes through every radial integral, and the
	operators for different lines commute with one another. The bulk integral is
	therefore done \emph{once}, for a scalar; the spin-2 structure is restored at the
	very end by an operator that passes through everything. Where the scalar skeleton
	was the first efficiency principle, this is the second.} Twistor-inspired
variables offer a complementary compact language for spinning (A)dS correlators
\cite{Baumann:2024ttn,Rost:2025uyj}.

\paragraph{Three points: primordial tensor non-Gaussianity.} With no internal line,
$\mathcal D$ does not act, and the stress-tensor three-point function is a single
contact diagram: the cubic Einstein vertex, with tensor structure
$\mathcal T_3(\eps_a,\vk_a)$, dressed by its scalar factor. In
$\AdS_4$, with $e_2=k_1k_2+k_2k_3+k_3k_1$,
\begin{equation}
	\label{eq:grav3}
	W^{\,\mathrm{grav}}_3=\mathcal T_3(\eps_a,\vk_a)\times
	\Big[\frac{k_1k_2k_3}{k_{123}^2}+\frac{e_2}{k_{123}}-k_{123}\Big],
\end{equation}
whose bracket, unlike the bare gluon pole, carries the subleading pieces
$e_2/k_{123}$ and $k_{123}$ alongside the $k_1k_2k_3/k_{123}^2$ singularity. This is
exactly the inflationary graviton non-Gaussianity of
\cite{Maldacena:2011nz}, there obtained from a bulk \emph{time} integral. Continued
to de Sitter it gives the cubic wavefunction coefficient $\psiwf_3$, and the Born
rule, quadratic kernels and normalization included, then delivers the primordial
tensor bispectrum $\langle\gamma\gamma\gamma\rangle$; the agreement is a check on
the entire momentum-space machinery.

\paragraph{Four points: the spin-2 imprint.} The $s$-channel scalar factor,
\begin{equation}
	\mathcal M_4=\int_0^\infty\!\frac{p\,\dd p}{k_{\ul{12}}^2+p^2}\,B(k_1,k_2;p)\,B(k_3,k_4;p),
\end{equation}
is a single spectral integral of two spin-2 blocks, the $\nu=\tfrac32$ analogues
of $\kkj$,\footnote{Explicitly, with the $\sqrt{2/(\pi p)}$ normalization
	stripped as in \eqref{eq:blocks},
	$B(k_1,k_2;p)=p^2\big(p^2+k_{12}^2+2k_1k_2\big)/\big(k_{12}^2+p^2\big)^2$:
	rational, with the squared partial-energy denominator.} whose
\emph{squared} denominators
produce, after the residue evaluation,
\begin{equation}
	\label{eq:grav4}
	\mathcal M_4\sim\frac{\text{(polynomial in energies)}}{\big(k_{12\ul{12}}\,k_{34\ul{12}}\big)^2k_T^3},
\end{equation}
a \emph{double} pole on each partial energy, in contrast to the simple poles of the
gluon master \eqref{eq:4s-masters}, with the total energy entering up to $1/\kT^3$.
The double partial-energy pole is the imprint of spin two in this two-derivative
massless exchange; derivative counting and external weights can shift pole orders
in other theories. The $s$-channel of
$W^{\,\mathrm{grav}}_4$ is assembled as
$V_{12}\,\mathcal D^{\vk_{12}}\mathcal M_4\,V_{34}$, with $V_{12},V_{34}$ the
vertex tensor factors and only one scalar factor ever integrated; the remaining
channels repeat the pattern, and the explicit polarization-dressed result is in
\cite{Albayrak:2019yve}. The first four-point stress-tensor correlator in
$\AdS_4$ (rational, with each three-point factor the square of a gluon one, a double
copy already at three points) was obtained in \cite{Raju:2012zs}.

\paragraph{Three routes to one trispectrum.} The graviton four-point function can
now be reached along three independent routes. Besides the differential-operator
route just described, the de Sitter graviton trispectrum has been assembled
from \emph{gluon} building blocks \cite{Armstrong:2023phb}, importing the
flat-space double-copy philosophy: the gauge-theory partial-energy structures of
\eqref{eq:4s-masters} are squared and recombined, with the double partial-energy
poles of \eqref{eq:grav4} emerging from the squaring rather than from any tensor
integral. Independently, the same object was determined by the de Sitter
bootstrap \cite{Bonifacio:2022vwa}, fixing
the answer from symmetry, singularity structure, and unitarity, the wavefunction
never touching a bulk integral. Where the constructions overlap they agree:
double poles, $1/\kT^3$ behavior and all. The pipeline extends beyond the
trispectrum: on-shell amplitudes in AdS pass directly to cosmological gluon and
graviton correlators \cite{Mei:2024sqz}.

\paragraph{Flat and collinear limits do not commute.} Because $\mathcal D$ commutes
with the bulk integration, both limits act on the scalar factor. The flat-space
limit is the coefficient of the leading total-energy pole: in the conventions of
the stripped scalar factor, and up to the vertex and external-state
normalizations fixed in \cite{Raju:2012zr},
\begin{equation}
	\label{eq:gravflat}
	\mathcal A_n^{\,\mathrm{flat}}=\lim_{\kT\to0}\frac{\kT^{\,n-1}}{\Gamma(n-1)\,\prod_i k_i}\,\mathcal M_n,
\end{equation}
the exponent $n-1$ being the $d=3$ value of the general power $(n/2-1)(d-1)+1$
\cite{Raju:2012zr}; at four points the $1/\kT^3$ pole of \eqref{eq:grav4} supplies it.
The collinear limit sends an internal momentum soft, $\vk\to0$, whereupon
$\mathcal D$ degenerates to the constant
de~Donder (harmonic-gauge) projector
$\lim_{\vk\to0}\mathcal D^{\vk}_{ab,cd}=\tfrac{\ii}{2}\big(\eta_{ac}\eta_{bd}+\eta_{ad}\eta_{bc}-\tfrac{2}{d-1}\eta_{ab}\eta_{cd}\big)$.
The two limits \emph{do not commute} \cite{Albayrak:2019yve}: the collinear limit sees the
longitudinal modes (the $\mathbb{P}^{(1)},\mathbb{P}^{(2)}$ pieces) that the flat-space limit
discards. The non-commutativity is a clean diagnostic separating the flat $S$-matrix
content of the correlator from its intrinsically curved-space content, the same
dichotomy met for gluons.

\subsection{Scalars: three tiers of transcendentality}
\label{sec:tree-scalars}

\begin{figure}[t]
	\centering
	\begin{tikzpicture}[every node/.style={font=\footnotesize},
		tier/.style={wbig,draw=inkcol,align=center,text width=6.9cm,
			inner sep=4pt,minimum height=0.8cm}]
		\node[tier] (a) at (0,0)
		{Tier I: rational\\ \;(conformal coupling $+$ conformal\\ interaction $=$ wavefunction)};
		\node[tier] (b) at (0,-1.65)
		{Tier II: dilogarithm \;($\phi^3$ in $\AdS_4$)};
		\node[tier] (c) at (0,-2.95)
		{Tier III: triple-$K$\\ \;(minimal coupling; logs, $\arctan$, $\gamma_E$)};
		\draw[warr,draw=inkcol,->] (a)--(b) node[midway,right=2pt,font=\scriptsize]{more transcendental};
		\draw[warr,draw=inkcol,->] (b)--(c);
	\end{tikzpicture}
	\caption{The three tiers of scalar Witten diagrams, set by distance from
		conformality: rational at the conformal point, a dilogarithm one step away, and
		genuine triple-$K$ transcendentals once the coupling is minimal.}
	\label{fig:tiers}
\end{figure}
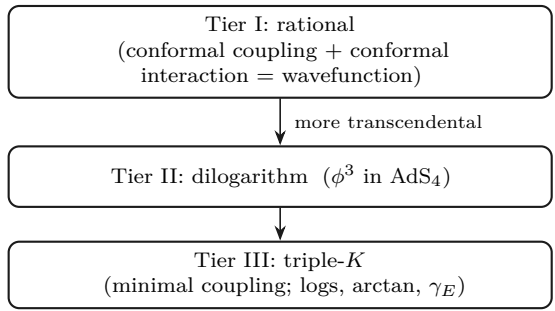

The scalar is the place to see the machinery at its simplest and across its full range of
difficulty. The correlators sort into three tiers, set by how far the
theory sits from conformality (Figure~\ref{fig:tiers}) \cite{Albayrak:2020isk}.
What sets the tier is the effective mass $\mu^2=m^2+\xi\mathcal R$: a conformally coupled scalar,
$\xi_c=(d-1)/4d$, has $\mu_c^2=(1-d^2)/4$ and hence Bessel order
\begin{equation}
	\nu=\tfrac12\sqrt{d^2+4\mu^2}=\tfrac12\qquad(\text{any }d),
\end{equation}
so its bulk-to-boundary leg is the bare $e^{-kz}$ in every dimension! Hence a Weyl
rescaling maps its action exactly onto flat half-space; the conformal interactions
$\phi^6,\phi^4,\phi^3$ then live in $\AdS_{3,4,6}$.

\begin{enumerate}[leftmargin=1.5em,itemsep=1pt]
	\item \emph{Rational}: conformal coupling \emph{and} a conformal interaction. The
	Weyl map is exact and the diagram is the cosmological wavefunction, a rational
	function of energies. The four-point exchange is precisely $\mathcal M^{(1)}$ of
	\eqref{eq:4s-masters}: the scalar
	skeleton and the wavefunction are the same object.
	\item \emph{Dilogarithm}: conformal coupling, non-conformal interaction. The
	canonical case is $\phi^3$ in $\AdS_4$, where $n=3\neq n_c=4$ and the Weyl map
	leaves a $z$-dependent coupling $\lambda_3(z)=\lambda_3 z^{-1}$. Writing it as a
	Laplace transform $z^{-1}=\int_0^\infty\!\dd\omega\,e^{-\omega z}$ promotes $\omega$
	to the energy of an extra exponential leg, turning the exchange into a rational seed
	with two shifted energies followed by two ordinary integrals,
	\begin{equation}
		\label{eq:sc-omega}
		\begin{split}
			W_{4,1}=\ii\lambda_3^2\!\int_0^\infty\!\!\dd\omega_1\dd\omega_2\;
			&\frac{1}{(k_{1234}{+}\omega_1{+}\omega_2)}\\
			&\times\frac{1}{(k_{12\ul{12}}{+}\omega_1)(k_{34\ul{12}}{+}\omega_2)}.
		\end{split}
	\end{equation}
	The $\omega_2$ integral gives a logarithm and the $\omega_1$ integral a
	dilogarithm,\footnote{For the systematic evaluation of such tree-level integrals
		and the transcendental functions they produce, see \cite{Raman:2025tsg}.}
	\begin{equation}
		\label{eq:dilog}
		\begin{split}
			W_{4,1}=-\frac{\ii\lambda_3^2}{4\,k_{\ul{12}}}\Big[&
			2\,\mathrm{Li}_2\!\big(\tfrac{k_{1234}}{k_{12}-k_{\ul{12}}}\big)
			-2\,\mathrm{Li}_2\!\big(\tfrac{k_{1234}}{k_{12\ul{12}}}\big)\\
			&-2\log k_{12\ul{12}}\log k_{34\ul{12}}
			+\cdots+\pi^2\Big],
		\end{split}
	\end{equation}
	the simplest genuinely transcendental AdS correlator \cite{Albayrak:2020isk}, with its
	polylogarithmic branch points sitting, as always, on the total- and partial-energy
	loci $k_{1234},k_{12\ul{12}},k_{34\ul{12}}$. Read cosmologically, this is the AdS
	sibling of the de~Sitter dilogarithm of the wavefunction (Section~\ref{sec:inflation}):
	rationality is a feature of the conformal locus, not of AdS correlators in general.
	\item \emph{Triple-$K$}: minimal coupling ($\xi=0$, $\nu=d/2$). Even the
	three-point function is a genuine triple-$K$ integral \eqref{eq:tripleK}, governed by
	the momentum-space CFT technology of \cite{Bzowski:2013sza,Bzowski:2015pba}.
	For a minimally coupled scalar in $\AdS_4$ ($\nu=\tfrac32$) the regulated seed
	already carries a logarithm and an Euler--Mascheroni constant: in one convenient
	subtraction scheme,
	$\mathsf{KKK}\sim\tfrac19\kT^3-k_1k_2k_3+\tfrac13(k_1^3+k_2^3+k_3^3)(1-\gamma_E-\log(\kT/\mu_R))$,
	with $\mu_R$ the renormalization scale. The polynomial terms and constants are
	scheme dependent; the nonlocal logarithm is not. The mixed block $\kkj$ carries an
	$\arctan$ of energy ratios: genuine transcendentality at three points, in sharp
	contrast to the rational $\nu=\tfrac12$ case; the $\gamma_E$ and $\mu_R$ are the
	fingerprint of the regularization triple-$K$ integrals require.
\end{enumerate}

The moral: half-integer $\nu$ alone does not make a diagram rational. It takes
three things at once: $\nu=\tfrac12$, bare exponential legs, and a conformal
vertex. The minimally coupled scalar in $\AdS_4$ has $\nu=\tfrac32$, still
half-integer, yet transcendental: the extra power of $z$ in the measure
$\dd z/z^{d+1}$ spoils the flat-space structure. The conformal exchange
$\mathcal M^{(1)}$ is the simplest seed of all; the web of Section~\ref{sec:web}
and the bootstrap of Section~\ref{sec:bypass} grow the whole tree-level
landscape from it.

\section{A web of theories: spins and dimensions}
\label{sec:web}

Section~\ref{sec:tree} reduced every spinning correlator to polarization algebra
acting on a few scalar masters. The converse is just as useful: theories that
share masters are, at the level of the scalar factor, \emph{one theory}.
Following \cite{Albayrak:2023kfk}, this section makes that exact for the broad
parity-even class of cubic vertices whose indices are contracted by boundary
metrics and momenta. Within that class, the radial integral of a cubic Witten
diagram sees the spins, the derivative count, and the boundary dimension only
through a single power of $z$ and a set of Bessel orders; change $(\ell_a,n,d)$
while holding those fixed and one theory's scalar factor becomes another's.
Gluons, gravitons, scalars, and higher-spin currents are woven into one web, and
the spinning wavefunction coefficients $\psiwf_n$ of de~Sitter are scalar
coefficients evaluated at shifted orders, once the tensor structures and contact
terms of Section~\ref{sec:tree} are restored.

\subsection{Index-counting the cubic vertex}

Each external field enters a diagram through its bulk-to-boundary profile, fixed by
its spin $\ell$ and the dimension $\Delta$ of the dual operator. A spin-$\ell$ field
dual to a symmetric traceless operator carries the universal radial function of the
toolkit, Eq.~\eqref{eq:btb}, dressed by a polarization tensor and a spin-dependent
prefactor,
\begin{equation}
	G_{i_1\cdots i_\ell}(k,z)\propto\eps_{i_1\cdots i_\ell}\,
	z^{\,d/2-\ell}\,k^{\nu}K_\nu(kz),\qquad
	\nu=\Delta-\tfrac d2 .
\end{equation}
For a \emph{conserved} current the dimension is protected, $\Delta=d-2+\ell$;
together with the unconstrained scalar case, the Bessel orders are
\begin{equation}
	\label{eq:nuspin}
	\nu_a=\ell_a-2+\tfrac d2\quad\text{(conserved)},\qquad
	\nu_a=\alpha_a\quad\text{(scalar)} .
\end{equation}
The spin enters the Bessel function \emph{only} through the order, while the
prefactor power takes the universal form $z^{d/2-\ell}$. This is the first hint that, from the radial point of
view, spin is close to a relabeling.

The interaction confirms it. Consider a cubic vertex coupling three
conserved currents of spins $\ell_1,\ell_2,\ell_3$ through $n$ boundary
derivatives, with indices contracted by boundary metrics and momenta. Its
$z$-dependence follows without any computation, by counting indices. The vertex must
supply $\ell_1+\ell_2+\ell_3$ boundary indices to meet the three legs; of these, $n$
are carried by the boundary momenta $\vk_a^i$ that the derivatives provide, and the
remaining $\ell_1+\ell_2+\ell_3-n$ are supplied in pairs by inverse boundary metrics.
Since each derivative momentum must \emph{also} have its index raised, the vertex
contains
\begin{equation}
	\underbrace{\tfrac{\ell_1+\ell_2+\ell_3-n}{2}}_{\text{contracting leg indices}}
	+\underbrace{\vphantom{\tfrac{a}{2}}\,n\,}_{\text{raising momenta}}
	=\tfrac{\ell_1+\ell_2+\ell_3+n}{2}
\end{equation}
inverse boundary metrics. In the Poincar\'e patch $g^{ij}=z^2\eta^{ij}$, so each
carries a factor $z^2$, and the vertex weight is $z^{\,\ell_1+\ell_2+\ell_3+n}$.
Assemble the diagram (three legs $z^{d/2-\ell_a}$, the vertex weight, and the bulk
measure $z^{-(d+1)}$) and the spins cancel out of the exponent entirely:
\begin{equation}
	\begin{split}
		\int_0^\infty\!&\frac{\dd z}{z^{d+1}}\,z^{\,\sum_a\ell_a+n}
		\prod_{a=1}^3 z^{\,d/2-\ell_a}E_{\nu_a}(k_a,z)\\
		&=\int_0^\infty\!\dd z\; z^{\,\frac{d-2}{2}+n}\prod_{a=1}^3 E_{\nu_a}(k_a,z).
	\end{split}
\end{equation}

Within this class, the scalar factor of the cubic Witten diagram is the generalized triple-$K$
integral
\begin{equation}
	\label{eq:master}
	\int_0^\infty\!\dd z\; z^{\frac{d-2}{2}+n}\,
	\prod_{a=1}^{3} E_{\nu_a}(k_a,z),
	\quad \nu_a=\ell_a-2+\tfrac d2 \ \text{or}\ \alpha_a,
\end{equation}
with $E_\nu=K_\nu$ on an external leg and $J_\nu$ on an internal one. Up to
external normalizations and the stripped tensor polynomial, it depends on the
theory only through the pair $\big(\tfrac{d-2}{2}+n,\ \{\nu_a\}\big)$. Vertices
with radial derivatives, explicit curvature tensors, parity-odd contractions, or
nonconserved fields require additional reduction before they take this form.

\noindent The vertex weight $z^{\sum_a\ell_a+n}$ is eaten by the leg prefactors
$z^{d/2-\ell_a}$; the spins survive only in the Bessel orders $\nu_a$ and the
derivative count $n$.
Equation~\eqref{eq:master} is the generalized triple-$K$ integral of
Eq.~\eqref{eq:tripleK}, a single object controlling every cubic correlator in the
review. Table~\ref{tab:web} collects the vertex weights for the theories at hand:
the Yang--Mills cubic ($\ell_i=1$, $n=1$) carries $z^4$, the
Einstein cubic ($\ell_i=2$, $n=2$) carries $z^8$, and the scalar--scalar--graviton
vertex ($\ell=2,0,0$, $n=2$) carries $z^4$, reproducing the weights used in
Section~\ref{sec:tree}.

\begin{table*}[t]
	\centering\small
	\renewcommand{\arraystretch}{1.25}
	\caption{Radial weights from index-counting the cubic vertex, assembled into the
		master integral \eqref{eq:master} with orders $\nu_a$ from \eqref{eq:nuspin}.
		Raising all three spins by one and adding a derivative raises the weight by
		$z^4$, the shift that drives the web.}
	\begin{tabular}{@{}l@{\hspace{2.2em}}l@{\hspace{2.2em}}l@{\hspace{2.2em}}l@{\hspace{2.2em}}l@{}}
		\toprule
		\textbf{theory} & \textbf{vertex} & \textbf{spins $(\ell_1\ell_2\ell_3)$, $n$} & \textbf{vertex weight $z^{\sum\ell+n}$} & \textbf{radial power}\\
		\midrule
		conformal scalar & $\phi^3$ & $(000)$, $n=0$ & $z^{0}$ & $\frac{d-2}{2}$\\
		Yang--Mills & $A^2\partial A$ & $(111)$, $n=1$ & $z^{4}$ & $\frac{d-2}{2}+1$\\
		Einstein gravity & $h^2\partial^2 h$ & $(222)$, $n=2$ & $z^{8}$ & $\frac{d-2}{2}+2$\\
		scalar--scalar--graviton & $\phi^2\partial^2 h$ & $(002)$, $n=2$ & $z^{4}$ & $\frac{d-2}{2}+2$\\
		\bottomrule
	\end{tabular}
	\label{tab:web}
\end{table*}

\subsection{Two shift relations}

Because the scalar factor \eqref{eq:master} sees this class of theories only through the power
$\tfrac{d-2}{2}+n$ and the orders $\nu_a=\ell_a-2+\tfrac d2$, any move on
$(\ell_a,n,d)$ that preserves this data is a symmetry of the correlator. Two moves
generate the entire web.

\paragraph{Spin/derivative/dimension shift.} Raising every spin by one, adding one
derivative, and lowering the boundary dimension by two leaves the power and all three
orders invariant,
\begin{multline}
	\big(\ell_1\ell_2\ell_3;\,n\big)\ \text{in}\ \AdS_{d+1}\;\Longleftrightarrow\;\\
	\big((\ell_1{+}1)(\ell_2{+}1)(\ell_3{+}1);\,n{+}1\big)\ \text{in}\ \AdS_{d-1},
\end{multline}
since $\nu_a\mapsto(\ell_a{+}1)-2+\tfrac{d-2}{2}=\nu_a$ and
$\tfrac{(d-2)-2}{2}+(n{+}1)=\tfrac{d-2}{2}+n$.

\paragraph{Spin-to-scalar shift.} A spinning interaction is equivalently a purely
scalar one in a higher dimension,
\begin{multline}
	\big(\ell_1\ell_2\ell_3;\,n\big)\ \text{in}\ \AdS_{d+1}\;\Longleftrightarrow\;\\
	\text{scalars of}\ \Delta_a=d-2+\ell_a+n\ \text{in}\ \AdS_{d+2n+1}.
\end{multline}
A scalar in boundary dimension $d+2n$ has
$\alpha_a=\Delta_a-\tfrac{d+2n}{2}=\ell_a-2+\tfrac d2=\nu_a$, and a non-derivative
scalar cubic carries $z^{(d+2n-2)/2}=z^{(d-2)/2+n}$: every spinning cubic scalar
factor is literally a scalar Witten diagram in a shifted dimension.

These are the momentum-space face of the weight-shifting operators of the conformal
and cosmological bootstraps: the same algebra that raises spins and dimensions in
the spinning bootstrap \cite{Baumann:2020dch} is realized here as an invariance of
the radial integrand. In position space the operators were built systematically from
finite-dimensional conformal representations,
generating arbitrary spinning conformal blocks from scalar ones \cite{Karateev:2017jgd};
in momentum space the analogous shift operators follow from the simplex
representation of CFT correlators \cite{Caloro:2022zuy}; their
cosmological incarnation reduces every boost-invariant de~Sitter correlator to a
handful of scalar seed exchanges \cite{Baumann:2019oyu}, and connects to
amplitude-style unifying relations among correlators \cite{Chen:2023xlt}; the
weight-shifting algebra itself has been extended and systematized further
\cite{De:2026stn,deKorte:2026sdu}. When the exchanged spinning
field is massive, its current is no longer conserved. The correlators and Ward
identities of non-conserved spinning operators were worked out
in \cite{Marotta:2022jrp}; they are exactly the kinematics that massive
spinning exchange, the cosmological collider of Section~\ref{sec:inflation},
demands. The web organizes what those operators generate, and supplies what they
consume: the scalar factors of \eqref{eq:master}, computed once, are the seed data
on which every weight-shifting chain acts.

\subsection{The web}
Composing the shifts identifies the kinematic factors of widely separated theories.
The cleanest instance ties the two gauge theories of the review to a scalar,
\begin{equation}
	\label{eq:web}
	\hspace{-.5em}
	\text{gluon}_{\,\AdS_{d+1}}\Longleftrightarrow\
	\text{graviton}_{\,\AdS_{d-1}} \Longleftrightarrow\ \text{scalar}_{\,\AdS_{d+3}},
\end{equation}
because the gluon ($\ell{=}1,n{=}1$) has $\nu_a=\tfrac d2-1$ at power $z^{d/2}$,
while the graviton ($\ell{=}2,n{=}2$) in dimension $d-2$ has
$\nu_a=\tfrac{d-2}{2}=\tfrac d2-1$ at the very same power. Neither shift is confined to
spins $0,1,2$: the counting applies kinematically to any conserved higher-spin
current, so higher-spin cubic factors are scalar diagrams in shifted dimensions,
a bridge to Vasiliev theory that we do not pursue (consistency of interacting
higher-spin theories imposes constraints that this radial identity does not
address). The practical payoff is immediate: a scalar factor computed once, for
the simplest member of a web, delivers the radial integral of every other member
for free; what remains is tensor algebra and theory-specific normalization. That
is exactly why the gluon and graviton computations of Section~\ref{sec:tree}
looked so alike.

\subsection{Seed equivalences}

The web makes a sharper prediction than similarity. To turn it into explicit
four-point correlators one resolves a spin-one \emph{internal} line on the
transverse and longitudinal projectors of Eq.~\eqref{eq:proj}, leaving two scalar
\emph{seed integrals}
\begin{equation}
	\label{eq:seeds}
	\begin{gathered}
		M_{\perp}=\int_0^\infty\!\frac{p\,\dd p}{k_{\ul{12}}^2+p^2}\,
		\kkj(k_1,k_2;p)\,\kkj(k_3,k_4;p),\\[2pt]
		M_{\parallel}=\int_0^\infty\!\frac{\dd p}{p}\,
		\kkj(k_1,k_2;p)\,\kkj(k_3,k_4;p),
	\end{gathered}
\end{equation}
where $\kkj$ is the bulk-point integral \eqref{eq:master} of two external $K$-legs
against one internal $J$-mode of spectral momentum $p$,\footnote{With the $J$-mode
	normalization included: $\kkj$ here denotes \eqref{eq:master} with one $J_\nu$,
	equal at $d=3$, $\nu=\tfrac12$ to $\sqrt{2/(\pi p)}$ times the stripped block of
	\eqref{eq:blocks}, exactly as in the gluing of Section~\ref{sec:tree}.} and
$k_{\ul{12}}=|\vk_1+\vk_2|$ is the internal energy. Exactly this
transverse/longitudinal decomposition was used at $d=3$ to
compute the four-point scalar correlator in slow-roll inflation from the late-time
wavefunction with AdS/CFT technology \cite{Ghosh:2014kba}; the results of \cite{Albayrak:2023kfk}
reproduce that decomposition and generalize it in $d$ and in spin.
The transverse and
longitudinal pieces of the spinning correlator pass the manifestly local test
of \cite{Jazayeri:2021fvk} \emph{separately} \cite{Albayrak:2023kfk}: nontrivial
evidence that the split is more than bookkeeping, though the individual pieces
are not separately gauge-invariant observables; only their dressed sum is.

At four points the seed integrals, once the spinning tensor structures are
stripped, coincide with those of specific scalars:
\begin{equation}
	\label{eq:seed}
	\langle JJJJ\rangle\big|_{\mathrm{seed}}=\langle\sigma\sigma\sigma\sigma\rangle_{A},
	\qquad
	\langle TTTT\rangle\big|_{\mathrm{seed}}=\langle\phi\phi\phi\phi\rangle_{G},
\end{equation}
where $\sigma_A$ is a ``vectorlike'' scalar of dimension $\Delta=d-1$, so its order
$\alpha=\tfrac d2-1=\nu_1$ equals the gluon's, and $\phi_G$ is a ``tensorlike'' scalar
of $\Delta=d$, so $\alpha=\tfrac d2=\nu_2$ equals the graviton's. Using the energy
variables of Section~\ref{sec:tree} (total energy $\kT$, partial energies
$k_{12\ul{12}}=k_{12}+k_{\ul{12}}$ and $k_{34\ul{12}}=k_{34}+k_{\ul{12}}$), the gluon
transverse seed in $\AdS_4$ is
\begin{equation}
	\label{eq:gluonseed}
	M_{\perp}\big|_{d=3}=\frac{1}{\kT\,k_{12\ul{12}}\,k_{34\ul{12}}}
	=\mathcal M^{(1)},
\end{equation}
\emph{exactly} the transverse master $\mathcal M^{(1)}$ of \eqref{eq:4s-masters}: the
conformally coupled scalar exchange wavefunction and the two-site cosmological
wavefunction $\psiwf_4$. Its flat-space limit is the residue of the total-energy pole
at $\kT\to0$: continuing $k_{34}\to-k_{12}$ collapses \eqref{eq:gluonseed} onto
$1/(k_{\ul{12}}^2-k_{12}^2)=1/s$, the flat $s$-channel pole of \eqref{eq:flatlim}. The
longitudinal seed is even simpler:
$M_{\parallel}\big|_{d=3}=1/(k_{12}\,k_{34}\,\kT)$. The
graviton (tensorlike) seed instead carries the \emph{squared} partial energies
$(k_{12\ul{12}}\,k_{34\ul{12}})^2$, the double-pole imprint of the two-derivative
vertex at each end: exactly the denominator of the four-point graviton scalar factor
of Section~\ref{sec:tree}. A single scalar function, evaluated at two sets of orders,
produced both spinning correlators.

\begin{figure*}[t]
	\centering
	\begin{tikzpicture}[>=stealth,every node/.style={font=\footnotesize}]
		\node[wbig,align=center] (s) at (0,0)
		{scalar seed\\[1pt]$\langle\sigma\sigma\sigma\sigma\rangle,\ \langle\phi\phi\phi\phi\rangle$};
		\node[wbig,align=center] (g) at (4.7,0)
		{gluon (spin 1)\\[1pt]$\langle JJJJ\rangle$};
		\node[wbig,align=center] (h) at (9.4,0)
		{graviton (spin 2)\\[1pt]$\langle TTTT\rangle$};
		\draw[warr,<->] (s)--(g); \draw[warr,<->] (g)--(h);
		\node[above] at (2.35,0.12) {\scriptsize shift $d$};
		\node[above] at (7.05,0.12) {\scriptsize shift $d$};
		\draw[wscalar] (-0.7,-1.15)--(0.7,-1.15); \node[below] at (0,-1.25){\scriptsize scalar};
		\draw[wgluon]  (4.0,-1.15)--(5.4,-1.15);  \node[below] at (4.7,-1.25){\scriptsize gluon};
		\draw[wgrav]   (8.7,-1.15)--(10.1,-1.15); \node[below] at (9.4,-1.25){\scriptsize graviton};
		\node[align=center] at (4.7,-2.15)
		{\scriptsize one master integral\;
			$\displaystyle\int\!\dd z\,z^{\frac{d-2}{2}+n}\!\prod_a E_{\nu_a}$
			:\; spin enters only through the orders $\nu_a$ and the count $n$};
	\end{tikzpicture}
	\caption{The web of theories. One master integral \eqref{eq:master} generates the
		scalar, gluon, and graviton correlators; raising the spin shifts the Bessel orders
		and the effective dimension, so the spin-1 and spin-2 correlators are the
		vectorlike and tensorlike scalar \emph{seeds} \eqref{eq:seed} dressed by tensor
		structure. Scalar lines are thin, gluon lines curly, graviton lines doubled.}
	\label{fig:web}
\end{figure*}
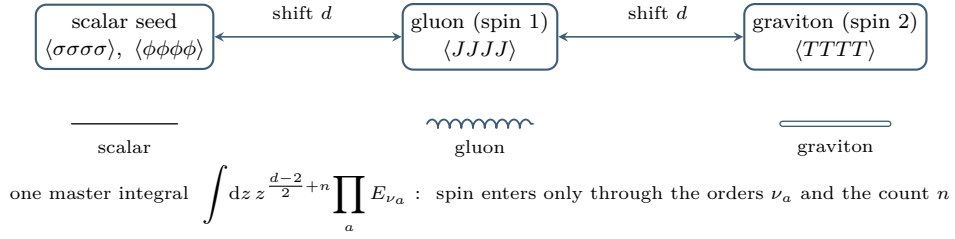

\begin{lesson}
	For the cubic vertex class of this section, the web is not an analogy but an
	identity: the gluon and graviton four-point seed integrals are the vectorlike
	and tensorlike scalar wavefunctions; the full correlators follow by dressing
	them with the spin-1 and spin-2 tensor structures, couplings, and contact terms.
\end{lesson}

\noindent This is the \emph{static} face of the double copy. The web fixes which
scalar seeds equal which spinning correlators by matching orders on a single line;
the double copy of Section~\ref{sec:bypass} reads the same relations
\emph{dynamically}, squaring numerators channel by channel. That the two viewpoints
agree wherever both apply is a demanding cross-check on the whole momentum-space
program. The $\AdS_4$
scalar--graviton correlator assembled from these seeds matches the in-in computation
of the graviton-exchange trispectrum \cite{Seery:2008ax} and the weight-shifting,
boostless, and double-copy determinations compared in \cite{Albayrak:2023kfk},
where the $\AdS_6$ result is new.
Cosmologically, the web means the inflationary wavefunction of a spinning field
in this class need never be computed from scratch: its scalar factor is a
wavefunction coefficient $\psiwf_n$ at shifted orders, and its manifestly local,
total-energy, and factorization singularities are inherited wholesale from the
scalar seed; the tensor structures and Ward identities are then restored as in
Section~\ref{sec:tree}.

\section{Bypassing the bulk: polytopes, recursion, and the double copy}
\label{sec:bypass}

Section~\ref{sec:web} collapsed every spin onto a handful of scalar seeds, but
each seed is still a bulk radial integral, however cleverly the tensor algebra
is stripped first. Three developments remove that
integral outright, replacing it by combinatorics, by on-shell sewing, or by the
double copy.\footnote{For a focused review of wavefunction and $S$-matrix
	techniques in de~Sitter, complementary to this section, see
	\cite{Benincasa:2022omn}.} Each admits two readings: as an efficient computation of an
$\AdS_{d+1}$ Witten diagram and, after the continuation $\eta=\ii z$
(Section~\ref{sec:inflation}), as a statement about the inflationary wavefunction
of the universe. The objects these methods produce, the wavefunction coefficients
$\psiwf_n$, assemble into the cosmological correlators through the Born rule
\eqref{eq:born}; the flat-space
scattering amplitude sits inside each as the residue of the total-energy pole
$\kT\to0$. We
take the three routes in turn, keeping the cosmological reading in view throughout.

\subsection{Cosmological polytopes and graph tubings}
\label{sec:bypass-poly}

In de~Sitter the basic intermediate object is the wavefunction $\Psi[\phi]$ of
Section~\ref{sec:wavefunction}, whose coefficients $\psiwf_n$ determine every
equal-time correlator through the Born rule \eqref{eq:born}. The construction is cleanest for a
conformally coupled scalar. In $\dS_4$ such a field is Weyl-equivalent to a
massless field in flat half-space: the scale factor drops out of the quadratic action
and the mode function is the bare exponential $K(k,\eta)=e^{\,\ii k\eta}$, which the
Wick rotation $\eta=\ii z$ carries to the $\AdS_4$ radial profile $e^{-kz}$ of
Section~\ref{sec:toolkit}. Every vertex integral runs only over the half-line
$z\in[0,\infty)$ (the past up to the boundary), and that truncation is the entire
source of the rational, ``curved-space'' structure we are after.

An internal line of energy $y=|\vk|$ is the radial Green's function with a Dirichlet
condition at the boundary,
\begin{equation}
	\big(-\partial_z^2+y^2\big)\,\Gbb(y;z,z')=\delta(z-z'),\qquad \Gbb(y;0,z')=0,
	\label{eq:bbG}
\end{equation}
whose unique solution is a free propagator minus its mirror image across $z=0$,
\begin{equation}
	\Gbb(y;z,z')=\frac{1}{2y}\Big[\,e^{-y|z-z'|}-e^{-y(z+z')}\,\Big].
	\label{eq:image}
\end{equation}
The first term is time-ordered; the second, the \emph{image term}, enforces the
Dirichlet condition. The total-energy pole arises instead from the common
translation of all vertices toward the deep interior; the image term, as we now
see, supplies the correct numerator.

\paragraph{The two-site wavefunction, by hand.} The four-point exchange has two
vertices of energy $x_L,x_R$ joined by one line of energy $y$; we write
$\psiwf(\mathfrak g_2)$ for the wavefunction of this two-site graph $\mathfrak g_2$
(with two conformally coupled legs per vertex it is the four-point coefficient
$\psiwf_4$, at $x_L=k_{12}$, $x_R=k_{34}$, $y=k_{\ul{12}}$). Inserting
\eqref{eq:image} into the nested radial integral,
\begin{equation}
	\begin{split}
		\psiwf(\mathfrak g_2)=\frac{1}{2y}\int_0^\infty\!\!\dd z_1\,\dd z_2\;
		&e^{-x_L z_1-x_R z_2}\\
		&\times\Big[e^{-y|z_1-z_2|}-e^{-y(z_1+z_2)}\Big]\,.
	\end{split}
\end{equation}
The image term factorizes at once into $-\tfrac{1}{2y}(x_L+y)^{-1}(x_R+y)^{-1}$. For
the time-ordered term, split the domain at $z_1=z_2$: on $z_1>z_2$ the integrand is
$e^{-(x_L+y)z_1}e^{-(x_R-y)z_2}$, and
\begin{equation}
	\begin{split}
		\int_0^\infty\!\!&\dd z_2\,e^{-(x_R-y)z_2}\!\int_{z_2}^\infty\!\!\dd z_1\,
		e^{-(x_L+y)z_1}\\
		&=\frac{1}{x_L{+}y}\int_0^\infty\!\!\dd z_2\,e^{-\kT z_2}
		=\frac{1}{(x_L{+}y)\,\kT}\,,
	\end{split}
\end{equation}
using $x_L+x_R=\kT$; the region $z_2>z_1$ gives $1/[(x_R+y)\kT]$. Collecting the three
pieces over the common denominator $(x_L+y)(x_R+y)$, the numerators are $\kT+2y$ (from
the time-ordered term, since $(x_L+y)+(x_R+y)=\kT+2y$) and $-\kT$ (from the image
term). The two combine to $(\kT+2y)-\kT=2y$, which cancels the prefactor $1/2y$
exactly, and
\begin{equation}
	\boxed{\;\psiwf(\mathfrak g_2)=\frac{1}{\kT\,(x_L+y)(x_R+y)}\,,\qquad \kT=x_L+x_R\,.}
	\label{eq:psi2}
\end{equation}
Equation~\eqref{eq:psi2} is nothing but the transverse gluon master
$\mathcal M^{(1)}$ of \eqref{eq:4s-masters}, and equally the
$d=3$ seed of the web of theories \eqref{eq:seed}. The ``scalar skeleton'' that kept
reappearing in Section~\ref{sec:tree} was the cosmological two-site wavefunction all
along; the radial integrals of the scalar sector were never doing independent work.

\begin{lesson}
	The image term in \eqref{eq:image} is the boundary condition made flesh: it cancels
	the spurious $1/2y$ and fixes the numerator, while the total-energy pole $\kT\to0$,
	whose residue is the flat-space amplitude, arises from the collective radial
	translation of all vertices. Both ingredients are needed for the rational structure
	of the wavefunction.
\end{lesson}

\paragraph{The cosmological polytope.} This structure was given a geometric home in
\cite{Arkani-Hamed:2017fdk}. To each graph $\mathfrak g$ one
associates a convex \emph{cosmological polytope} $\mathcal P_{\mathfrak g}$, built by
gluing one triangle per edge. Its \emph{canonical form} (the unique projective
form with logarithmic singularities on, and only on, its boundary) is exactly the
wavefunction coefficient, $\Omega(\mathcal P_{\mathfrak g})=\psiwf_{\mathfrak g}$. The
dictionary is precise: the facets of $\mathcal P_{\mathfrak g}$ are in one-to-one
correspondence with the connected subgraphs, each lying on the hyperplane
$E_{\mathfrak g}=\sum_{v\in\mathfrak g}x_v+\sum_{e\in\partial\mathfrak g}y_e\to0$, a
total- or partial-energy singularity. The recursive factorization on those poles
is the statement that a facet is itself a product of lower cosmological polytopes
\cite{Arkani-Hamed:2017fdk,Benincasa:2018ssx,Arkani-Hamed:2024jbp}. For the two-site chain the polytope is a triangle,
with dimension $D=2$ and $F=3$ facets; a simplicial polytope of dimension $D$ with $F$ facets has a
canonical form $\Omega=N/\prod_i\ell_i$ with $\deg N=F-D-1$, here a constant,
reproducing \eqref{eq:psi2}.

\paragraph{Positive geometry.} The polytope is an instance of a \emph{positive
	geometry} \cite{Arkani-Hamed:2017tmz}: a real semialgebraic set with a unique
canonical form in the above sense, every boundary component itself a positive
geometry carrying the residue as its own canonical form. The definition is recursive,
and so is the physics it encodes: taking a residue of
$\Omega(\mathcal P_{\mathfrak g})$ \emph{is} factorizing the wavefunction on an energy
singularity. Simplices, polygons, the amplituhedron, and the associahedron of
bi-adjoint $\phi^3$ theory all satisfy the axioms; the cosmological polytope enrolls
the cosmological wavefunction in the same family, and the triangulation technology
of \cite{Arkani-Hamed:2017tmz} is what converts its canonical form into the explicit
energy sums below. The combinatorics predates the physics: tubes and tubings are the language of the
\emph{graph associahedra} constructed from Coxeter-complex blow-ups
\cite{CarrDevadoss:2006}, although the compatibility rule used below, in which
disjoint tubes may be adjacent, is tailored to the wavefunction and differs from the
standard graph-associahedron face poset; the geometry of these wavefunction
polytopes continues to be developed \cite{Benincasa:2024leu,Li:2026gns}, and the
cosmohedra of Section~\ref{sec:frontier} supply a distinct geometry for the full
planar wavefunction. Topologists had named the combinatorics of the wavefunction of
the universe more than a decade before it was written down.

\paragraph{Physics from facets.} The facet on $\kT\to0$, the \emph{scattering
	facet}, has a remarkable rigidity: its
combinatorial geometry forces the residue
of the total-energy pole to be Lorentz invariant and unitary, with factorization on
poles and positive factorization coefficients, even though the polytope is assembled
from purely spatial boundary data with no Hilbert space in sight
\cite{Arkani-Hamed:2018bjr}. Within this class of scalar toy models, flat-space
S-matrix theory emerges as a corollary of the cosmological geometry, not an input
to it. The remaining
facets constrain the analytic structure away from $\kT=0$: no pair of partially
overlapping subgraphs bounds a common face, so the wavefunction cannot have
simultaneous singularities in partially overlapping channels. Its sequential
discontinuities there vanish, a direct analogue of the Steinmann relations obeyed by
flat-space amplitudes, established for the wavefunction in
\cite{Benincasa:2020aoj}. Nor is the construction confined to the conformally coupled
scalar: for light states in de~Sitter the wavefunction is generated from the
conformally coupled seed by iterated integrals in the vertex energies, one per power
of the scale factor at each vertex, with cosmological polytopes controlling every
term of the expansion \cite{Benincasa:2019vqr}. The primer \cite{Benincasa:2022gtd}
is the pedagogical entry point to this circle of ideas, from the half-line integrals
of this subsection to the frontier.

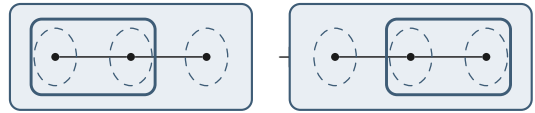
\begin{figure}[t]
	\centering
	\begin{tikzpicture}[scale=1.0]
		\begin{scope}
			\fill[slatefill,rounded corners] (-0.6,-0.7) rectangle (2.6,0.7);
			\node[wdot] (a) at (0,0){}; \node[wdot] (b) at (1.0,0){}; \node[wdot] (c) at (2.0,0){};
			\draw[wscalar] (a)--(b)--(c);
			\draw[wtube] (a) ellipse (0.28 and 0.38);
			\draw[wtube] (b) ellipse (0.28 and 0.38);
			\draw[wtube] (c) ellipse (0.28 and 0.38);
			\draw[slatecol,line width=1.1pt,rounded corners] (-0.32,-0.5) rectangle (1.32,0.5);
			\draw[wbig] (-0.6,-0.7) rectangle (2.6,0.7);
		\end{scope}
		\node at (3.1,0){\large $+$};
		\begin{scope}[xshift=3.7cm]
			\fill[slatefill,rounded corners] (-0.6,-0.7) rectangle (2.6,0.7);
			\node[wdot] (a) at (0,0){}; \node[wdot] (b) at (1.0,0){}; \node[wdot] (c) at (2.0,0){};
			\draw[wscalar] (a)--(b)--(c);
			\draw[wtube] (a) ellipse (0.28 and 0.38);
			\draw[wtube] (b) ellipse (0.28 and 0.38);
			\draw[wtube] (c) ellipse (0.28 and 0.38);
			\draw[slatecol,line width=1.1pt,rounded corners] (0.68,-0.5) rectangle (2.32,0.5);
			\draw[wbig] (-0.6,-0.7) rectangle (2.6,0.7);
		\end{scope}
	\end{tikzpicture}
	\caption{The two complete tubings of the three-site chain; the overlapping pairs
		$\{12\},\{23\}$ cannot coexist, so each tubing carries one of them. Their sum
		\eqref{eq:tubing} is $\psiwf(\mathfrak g_3)$ of \eqref{eq:allline}. Vertex tubes are dashed, the
		whole-graph tube solid, the active pair-tube highlighted.}
	\label{fig:tubing}
\end{figure}

\paragraph{Graph tubings.} The efficient, purely combinatorial face of the same
geometry is the \emph{tubing} rule. A \emph{tube} $t$ is a connected subgraph; its
energy is the sum of the external energies it encloses plus the energies of the lines
crossing its boundary,
\begin{equation}
	\varepsilon(t)=\sum_{v\in t}x_v+\!\!\sum_{e\,\text{crossing}\,\partial t}\!\!y_e\,,
	\qquad\varepsilon(\mathfrak g)=\kT\,;
\end{equation}
two tubes are compatible when one contains the other or they are disjoint, and
incompatible when they partially overlap; a \emph{complete tubing} $T$ is a maximal
set of pairwise-compatible tubes. Then, for tree graphs,
\begin{equation}
	\psiwf(\mathfrak g)=\sum_{\text{complete tubings }T}\ \prod_{t\in T}\frac{1}{\varepsilon(t)}\,.
	\label{eq:tubing}
\end{equation}
Complete tubings are the simplices of one triangulation of $\mathcal P_{\mathfrak g}$,
each inverse-energy product the simplex's share of the canonical form; other
triangulations give different termwise representations of the same answer. No bulk
integral is ever evaluated.
For two sites the three tubes are mutually compatible: the two vertex tubes
($\varepsilon=x_L+y,\,x_R+y$), each pierced by the internal line, and the whole-graph
tube ($\varepsilon=\kT$). The single complete tubing reproduces \eqref{eq:psi2}. For the
three-site chain $1$--$2$--$3$ with edges $y_a,y_b$, the adjacent-pair tubes $\{12\}$
and $\{23\}$ overlap and cannot coexist, so there are exactly two complete tubings,
sharing the four tubes $\{123\},\{1\},\{2\},\{3\}$ and differing only in the middle
tube (Figure~\ref{fig:tubing}):
\begin{equation}
	\begin{split}
		\psiwf(\mathfrak g_3)=\frac{1}{\kT\,(x_1{+}y_a)(x_2{+}y_a{+}y_b)(x_3{+}y_b)}\\
		\times\left[\frac{1}{x_1{+}x_2{+}y_b}+\frac{1}{x_2{+}x_3{+}y_a}\right].
	\end{split}
\end{equation}
Over a common denominator the bracket's numerator collapses to
$(x_1{+}x_2{+}y_b)+(x_2{+}x_3{+}y_a)=x_1+2x_2+x_3+y_a+y_b$, so
\begin{equation}
	\begin{split}
		\psiwf(\mathfrak g_3)=\;&\frac{x_1+2x_2+x_3+y_a+y_b}
		{\;\kT\,(x_1{+}y_a)(x_2{+}x_3{+}y_a)(x_2{+}y_a{+}y_b)\;}\\[2pt]
		&\times\frac{1}{(x_3{+}y_b)(x_1{+}x_2{+}y_b)}\,.
	\end{split}
	\label{eq:allline}
\end{equation}
The denominator is the product of the six subgraph (facet) energies; the numerator is
a \emph{linear} polynomial, and its degree was fixed before any computation: the
three-site polytope has $F=6$, $D=4$, so $\deg N=F-D-1=1$. The geometry predicts the
shape of the answer, and the combinatorics reproduces the bulk integral to the last
term.

\paragraph{Spinning fields and the ``new relation''.} Gauge polarizations re-enter the
scalar skeleton line by line. Each internal gluon line resolves on the transverse and
longitudinal projectors \eqref{eq:proj}: a \emph{straight} line (transverse) keeps its
energy $y_e$ in every tube it pierces and its scalar factor is exactly the conformally
coupled propagator \eqref{eq:image}; a \emph{crossed} line (longitudinal) carries the
extra weight $(k^2+p^2)/p$ in the spectral integrand, engineered to cancel
the propagator denominator. That cancellation is the whole story: the crossed integral
is the straight integral with its denominator removed, which is what a straight line
becomes as its momentum is sent soft. This is the ``new relation'' of
\cite{Albayrak:2019asr},
\begin{equation}
	\big[\text{crossed line, momentum }\vk\big]
	\;=\;\lim_{\vk\to0}\big[\text{straight line}\big]\,,
\end{equation}
so a crossed line contributes \emph{zero} energy to every tube it crosses:
one sets $y_e\to0$ in $\varepsilon(t)$ for that line. A truncated graph with $n$
internal lines would naively split into $2^n$ scalar graphs; the new relation collapses
them to a single straight-only master, a wavefunction coefficient computed by
\eqref{eq:tubing}, dressed by algebraic soft limits. For the four-gluon exchange the
straight channel is \eqref{eq:psi2} and the crossed channel its $y\to0$ limit
$1/[\kT\,x_L\,x_R]$; contracting with the cubic vertices reproduces the four-gluon
correlator of Section~\ref{sec:tree} with no bulk integral evaluated. The scalar
construction runs in any dimension \cite{Albayrak:2020isk}; away from the conformal
point the Laplace transform of Section~\ref{sec:tree-scalars} applies, producing
dilogarithms rather than rational functions. The lesson to carry forward: because
the conformally coupled skeleton is
literally the cosmological wavefunction, these Witten-diagram manipulations are, term
for term, computations of inflationary $\psiwf_n$; the tubing rule \eqref{eq:tubing}
is a bootstrap for wavefunction coefficients and, through the Born rule, for
cosmological correlators \cite{Arkani-Hamed:2017fdk,Benincasa:2018ssx}.

\subsection{On-shell recursion in (A)dS}
\label{sec:bypass-recursion}

The most powerful flat-space methods never draw a diagram. BCFW recursion deforms two
external momenta along a complex \emph{null} direction, $\vk_1\to\vk_1+z\bm q$,
$\vk_n\to\vk_n-z\bm q$ with $q^2=q\!\cdot\!k_1=q\!\cdot\!k_n=0$, keeping every momentum
on shell. The shifted amplitude $\mathcal A(z)$ is then meromorphic, and if it vanishes
at large $z$ Cauchy's theorem writes the physical amplitude as a sum over
factorization channels,
\begin{equation}
	\mathcal A(0)=\sum_I\mathcal A_L(z_I)\,\frac{1}{P_I^2}\,\mathcal A_R(z_I),
\end{equation}
each a product of lower-point on-shell amplitudes at the single value $z_I$ where the
channel goes on shell, built entirely from three-point seeds, with no contact vertex.

Three features of (A)dS force a modification \cite{Raju:2010by,Raju:2011mp}. Boundary momenta are not
null (they carry an energy $k=|\vk|$ as well as a direction), so there is no null
shift vector $\bm q$, not even a complex one in three dimensions \cite{Raju:2012zr};
radial momentum is not conserved; and meromorphy in the deformation is not automatic.
The resolution, introduced in \cite{Raju:2012zr} and carried to the all-plus
graviton in \cite{Albayrak:2023jzl}, is an \emph{all-line} shift (the
curved-space descendant of the flat-space all-line deformation \cite{Risager:2005vk})
that deforms every external momentum at once,
\begin{equation}
	\vk_n\to\vk_n+\alpha_n\,\eps_n\,\omega\,,\qquad \sum_n\alpha_n\,\eps_n=0\,,
	\label{eq:allshift}
\end{equation}
with the linear dependence of the polarizations, automatic in $d=3$ for $n\ge4$, ensuring
$\sum_n\vk_n(\omega)=\sum_n\vk_n$ so that momentum conservation survives for all
$\omega$. The objects sewn at the channels are not vacuum correlators but
\emph{transition amplitudes}: lower-point correlators with a normalizable Bessel-$J$
mode on the internal leg \cite{Raju:2010by,Raju:2011mp,Raju:2012zs}. The bulk-to-bulk propagator, written over
the full real $p$-line, then turns each channel integral into a pure residue sum.

\paragraph{Three-dimensional spinor helicity.} In Lorentzian boundary signature the
$\AdS_4$ boundary Lorentz group is $SO(2,1)$, with double cover $SL(2,\mathbb R)$;
the Euclidean correlators used elsewhere in this review are reached by
complexification. A momentum is promoted to the complex null four-vector
$\hat{\vk}=(\ii k,\vk)$ and factorized into spinors
$\lambda_\alpha,\bar\lambda_{\dot\alpha}$, a formalism developed systematically
for gauge potentials and for fields of any mass
\cite{Nagaraj:2019zmk,Basile:2024ydc}. The same three-dimensional
spinor-helicity variables organize the momentum-space correlators of conserved
higher-spin currents
\cite{Jain:2020rmw,Jain:2021vrv,Maldacena:2011jn,Caron-Huot:2021kjy} and lift
to twistor space \cite{Bala:2025gmz,CarrilloGonzalez:2025qjk}; the lectures
\cite{S:2025pmh} are a pedagogical entry point. Because the momentum is not null the
formalism needs one bracket more than flat space: alongside the angle brackets
$\langle ij\rangle,\langle\bar\imath\bar\jmath\rangle$ there is the mixed
$[\,i\,\bar\jmath\,]$, which carries the energy through
\begin{equation}
	k_j=-\tfrac{\ii}{2}\,[\,j\,\bar\jmath\,]\,,
\end{equation}
with a generalized Schouten identity supplying the analogue of momentum conservation.
The crucial difference is in the on-shell condition. Under \eqref{eq:allshift} only the
undotted spinors shift, and they shift \emph{linearly} in $\omega$, so the bracket
$\langle\pi_1(\omega)\pi_2(\omega)\rangle$ governing an internal line is
\emph{quadratic} in $\omega$: the on-shell condition is
$a\,\omega^2+b\,\omega+c=0$, and each channel contributes the \emph{two} residues
\begin{equation}
	\omega_\pm=\frac{-b\pm\sqrt{b^2-4ac}}{2a}\,,
	\label{eq:omega-roots}
\end{equation}
the curved-space doubling that distinguishes the (A)dS recursion from flat-space BCFW,
where the analogous condition is linear with a single root. Spurious poles introduced
by the deformation cancel between the two roots, and the boundary term at infinity is
fixed by Ward identities to remove the divergent part of the residue sum.

\paragraph{The all-plus graviton in \texorpdfstring{$\AdS_4$}{AdS4}.} Gluing two graviton transition
amplitudes over the three bipartitions $\{12\}\{34\},\{13\}\{24\},\{14\}\{23\}$,
all generated from one base channel by relabeling operators $\mathfrak d_i$,
yields the four-point all-plus graviton correlator in $\AdS_4$ with \emph{no contact
	vertex} \cite{Albayrak:2023jzl}. Each transition amplitude is the spinor-helicity
\emph{square} of a gluon one, a first sighting of the double copy. This is a clean curved-space diagnostic: the flat-space
four-point all-plus graviton amplitude vanishes at tree level and first appears at
one loop, so a nonzero tree-level $\AdS_4$ answer is purely a curvature effect,
computed here as a finite sum of residues \eqref{eq:omega-roots} and checked
numerically to high precision. Read in de~Sitter, the recursion therefore delivers
a genuinely new piece of the graviton wavefunction, not a re-derivation.

The on-shell program has since matured into a full amplitude bootstrap. In a
Mellin-momentum representation, these recursive techniques, combined with
double-copy input, yield the complete four-point graviton amplitude in (A)dS
momentum space \cite{Mei:2023jkb}. The bootstrap extends to $n$-point gluon and
graviton scattering, with factorization, the flat-space limit, and soft limits
replacing the Lagrangian as the defining data \cite{Mei:2024abu}. Complementary recursions are now available: tree-level AdS
correlators satisfy recursion relations derived from the cosmological
scattering equations \cite{Armstrong:2022mfr}, supergluon amplitudes in
$\AdS_5\times S^3$ are BCFW-constructible \cite{Cao:2023cwa}, and a BCFW
recursion for Yang--Mills theory exists in the de~Sitter static patch
\cite{Albrychiewicz:2021ndv}.

\subsection{Color--kinematics duality and the double copy}
\label{sec:bypass-dc}

In flat space gravity is the square of gauge theory
\cite{Bern:2022wqg,Adamo:2022dcm}. Write a tree gauge amplitude
over cubic graphs, $\mathcal A=\sum_{i}c_i n_i/D_i$. Color--kinematics (BCJ) duality
is the statement that the kinematic numerators can be gauged to satisfy the same Jacobi
identity as the color factors, $n_s+n_t+n_u=0$. The \emph{double copy}
$\mathcal M^{\mathrm{grav}}=\sum_i n_i\tilde n_i/D_i$ then trades color for a second
copy of kinematics; equivalently, the KLT relations \cite{Kawai:1985xq}
express the graviton amplitude as a quadratic form in color-ordered gauge amplitudes.
At four points everything rests on the single massless identity $s+t+u=0$. Its
(A)dS analogue fails,
in a controlled way \cite{Albayrak:2020fyp}. Several independent tracks have mapped
curved-space color--kinematics and double-copy structure;\footnote{Momentum-space
	color--kinematics in $\AdS_4$ \cite{Armstrong:2020woi}; the three-point (A)dS/CFT
	double copy \cite{Farrow:2018yni}; embedding-space BCJ relations with curvature
	corrections \cite{Diwakar:2021juk}; the differential representation, with the
	duality imposed at the integrand level \cite{Herderschee:2022ntr}; on-shell
	correlators in general curved symmetric spacetimes \cite{Cheung:2022pdk}; a
	Mellin-space double copy for the $\AdS_5\times S^5$ supergraviton
	\cite{Zhou:2021gnu}; differential double-copy relations for (A)dS three-point
	functions \cite{Lee:2022fgr}; double-copy structure in parity-violating
	three-dimensional CFT correlators \cite{Jain:2021qcl}; color--kinematics
	duality for correlators in generic spacetimes \cite{Sivaramakrishnan:2021srm};
	the differential representation extended to Mellin space \cite{Li:2023azu};
	BCJ relations for supergluons in $\AdS_5\times S^3$ \cite{Drummond:2022dxd};
	the self-dual sector in $\AdS_4$ and in cosmology
	\cite{Lipstein:2023pih,Chowdhury:2024dcy,CarrilloGonzalez:2024sto}; the
	convolutional double copy on homogeneous spaces and in (A)dS
	\cite{Borsten:2021zir,Liang:2023zxo}; a twistor-space double copy for $\AdS_4$
	boundary correlators \cite{Ansari:2025fvi}; stringy KLT relations on AdS
	\cite{Alday:2025bjp}; and a twisted de Rham approach to the string double copy
	in AdS \cite{Kakkad:2025klm}.} we follow
\cite{Albayrak:2020fyp}, which works directly in the momentum-space variables of
this review.

With the null uplift of \eqref{eq:tk-mandelstam}, the flat
invariant is $s=k_{\ul{12}}^2-k_{12}^2$, and the \emph{AdS} Mandelstam invariant is its
curvature deformation
\begin{equation}
	\tilde s_{12}=k_{12\ul{12}}\,k_{34\ul{12}}=\kT\,\big(k_{12}+k_{\ul{12}}\big)+s_{12}\,,
\end{equation}
the first form a product of the two partial energies meeting at the exchanged line,
the second the flat invariant plus a curvature term. The
decisive fact is that the three do not sum to zero,
\begin{equation}
	\boxed{\;\tilde s+\tilde t+\tilde u
		=\kT\sum_{(ab)}\big(k_{ab}+k_{\ul{ab}}\big)+(s+t+u)\;\neq\;0\;,}
	\label{eq:stu}
\end{equation}
the excess vanishing only in the flat-space limit $\kT\to0$: it is not an explicit
power of the curvature radius but registers, in these variables, the failure of
radial energy conservation. Almost every consequence follows from \eqref{eq:stu}.

\paragraph{Full-rank map, modified BCJ.} Stripping color leaves two independent
color-ordered correlators $A(1234),A(1324)$, related to the numerators by a
$2\times2$ matrix. In flat space that map is degenerate (its kernel is the BCJ
relation among color-ordered amplitudes), but in AdS it is \emph{full rank}, precisely
because $\tilde s+\tilde t+\tilde u\neq0$. The numerators are therefore uniquely
recovered from the integrated correlators, and there are \emph{no extra BCJ relations}:
the integrated four-point correlator carries \emph{two} independent functions where
flat space has one. The kinematic Jacobi identity, generically violated
($n_s+n_t+n_u\neq0$), is restored by a generalized gauge transformation
$n_i\to n_i-\tilde s_i\,\Omega$, with $\tilde s_i$ the matching channel invariant
$\tilde s,\tilde t,\tilde u$ and $\Omega=(n_s+n_t+n_u)/(\tilde s+\tilde t+\tilde
u)$, well defined precisely because of \eqref{eq:stu}, the same fact that obstructed
naive BCJ. The modified BCJ relation reads
\begin{equation}
	\boxed{\;\tilde s\,A(1234)+(\tilde s+\tilde t)\,A(1324)
		=-\frac{n_u}{\kT}\,\frac{\tilde s+\tilde t+\tilde u}{\tilde u}\;,}
	\label{eq:mbcj}
\end{equation}
whose right-hand side (absent in flat space, where the left-hand side is the vanishing
BCJ combination) measures the curvature obstruction; on the total-energy pole
$\kT\to0$ the obstruction factor $\tilde s+\tilde t+\tilde u$ vanishes and the flat
BCJ relation is recovered. The
KLT kernel deforms in step, acquiring \emph{rank two} at four points, reflecting the two
independent correlators.

\begin{figure}[t]
	\centering
	\begin{tikzpicture}[scale=1.0,every node/.style={font=\footnotesize}]
		\node[wbig,draw=inkcol,align=center] (s) at (0,2.0) {bi-adjoint scalar\\ $\AdS_{d+3}$};
		\node[wbig,draw=inkcol,align=center] (y) at (0,0) {Yang--Mills\\ $\AdS_{d+1}$};
		\node[wbig,draw=inkcol,align=center] (g) at (0,-2.0) {Einstein gravity\\ $\AdS_{d-1}$};
		\draw[warr,draw=inkcol,->] (y)--(s)
		node[midway,right=3pt,align=left,font=\scriptsize]
		{$n_s\to c_s'$ (numerator $\to$ color)\\ \emph{established}};
		\draw[warr,draw=inkcol,->,dashed] (y)--(g)
		node[midway,right=3pt,align=left,font=\scriptsize]
		{$c_s\to n_s$ (square the numerators)\\ \emph{conjectural}};
	\end{tikzpicture}
	\caption{The (A)dS double copy shifts the spacetime dimen\-sion, the dynamical face of
		the web of theories \eqref{eq:seed}. Two dimensions \emph{up} to a bi-adjoint scalar
		(solid, established), two \emph{down} to gravity (dashed, conjectural)
		\cite{Albayrak:2020fyp}. The map is specific to this spectral construction, not a
		universal dimension-shift theorem.}
	\label{fig:dc}
\end{figure}
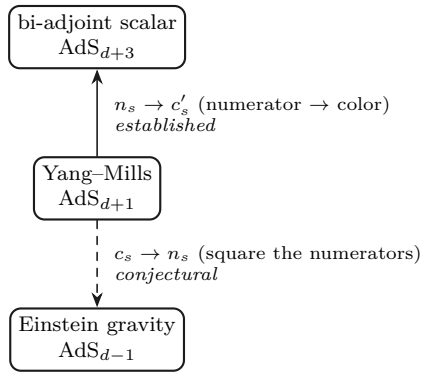

\paragraph{The double copy shifts the dimension.} The most striking output of this
construction is that the double copy can act as a \emph{dimension-shifting}
operation rather than a squaring at fixed $d$ (Figure~\ref{fig:dc}). In the spectral representation of the Yang--Mills
integrand, replacing the numerator $n_s$ by a second color factor $c_s'$ sends
Yang--Mills in $\AdS_{d+1}$ \emph{up} to a bi-adjoint scalar in $\AdS_{d+3}$; this
upward map is established in \cite{Albayrak:2020fyp}. The downward map, squaring the
numerators to reach Einstein gravity in $\AdS_{d-1}$, remains conjectural within this
spectral construction; we return to what obstructs it below, and fixed-dimension
Mellin-momentum constructions realize a four-point gravitational double copy by
other means \cite{Mei:2023jkb}. Modulo that caveat the picture is exactly the web of
theories
\eqref{eq:seed} read dynamically: there the gluon kinematic factor in dimension $d$
equaled the graviton's in $d-2$; here the double copy is the operation that realizes
that equality. The consistency check: $d=3$ in the upward map lands on a $\Delta=3$
scalar in $\AdS_6$, the conformally coupled bi-adjoint scalar, closing the loop.

\paragraph{Where the square breaks, and cosmology.} The maps hold where they have been
checked, but ``gravity $=$ (gauge)$^2$'' is not unconditional in AdS. At three points
the square of the Yang--Mills cubic numerator reproduces the
three-graviton structure of \cite{Maldacena:2011nz}. At four points the naive square produces a candidate graviton
correlator \emph{inconsistent with the AdS cutting rules}
\cite{Meltzer:2020qbr}: it does not factorize
correctly on its partial-energy singularities \cite{Albayrak:2020fyp}. The reason is
structural: the AdS/CFT graviton Ward identities carry extra contact terms that
squaring alone does not enforce. Bulk unitarity is therefore strictly more
constraining than the algebraic double copy, and the two part ways already at four
points; refined constructions can restore the missing data, as the four-graviton
bootstrap of \cite{Mei:2023jkb} demonstrates. Under the
continuation $L_{\dS}=-\ii L_{\AdS}$ the AdS Mandelstam invariants, the modified BCJ
relation \eqref{eq:mbcj}, and the dimension-shifting maps all become statements about
the de~Sitter wavefunction. The double copy is then a candidate organizing principle for
wavefunction coefficients and, through the Born rule, cosmological correlators,
relating gauge and gravitational non-Gaussianities, with the
total-energy pole $\kT\to0$ its flat-space, cosmological-bootstrap singularity, subject
always to the same bulk unitarity conditions that qualify its AdS parent.

\section{Loops, analytic structure, and unitarity}
\label{sec:loops}

The whole apparatus so far is tree level: an operator strips the tensor structure,
and the residue at $\kT\to0$ returns the flat-space amplitude. Loops raise two
obstructions. The loop momentum runs through the internal lines, so the tree
operator no longer commutes past its integral; and the soft limit becomes
\emph{democratic}, with no single propagator on shell and every Witten diagram
contributing at the same order, so factorization requires a curved-space form of
unitarity. One object clears both: the \emph{transition amplitude}, a correlator
with one leg on the spectral shell, what a loop cut produces and what a soft
limit factorizes onto. Following \cite{Albayrak:2020bso,Albayrak:2024ddg}, this
section builds the momentum-space loop integrand, its Appell-$F_4$ analytic
structure, and the cutting rules that reconstruct it, reading the cut throughout
as the bulk face of the cosmological optical theorem.

\subsection{Three layers of integration}
\label{sec:lp-layers}

A one-loop Witten diagram inherits the hybrid character of the Poincar\'e patch:
its momentum-space representation carries three kinds of integration, in nested
layers,
\begin{equation}
	\label{eq:lp-layers}
	\begin{gathered}
		\underbrace{\prod_{f=1}^{u}\int\!\dd^d\bm\ell_f}_{\text{boundary loop momenta}}
		\;\;
		\underbrace{\prod_{b=1}^{n}\int_0^\infty\! p_b\,\dd p_b}_{\text{radial (spectral) momenta}}
		\\[4pt]
		\times\;
		\underbrace{\prod_{i}\int_0^\infty\!\frac{\dd z_i}{z_i^{\,d+1}}}_{\text{bulk-point integrals}}
		\;\big(\cdots\big).
	\end{gathered}
\end{equation}
The $d$ boundary directions stay translation-invariant, so boundary momentum is
conserved at every vertex and an unconstrained loop momentum $\bm\ell$ circulates,
integrated as $\int\!\dd^d\bm\ell$ exactly as in flat space. The radial direction is
different: each internal line, opened by the spectral representation
\eqref{eq:btbb}, carries its own radial momentum $p_b$ with its own integral
$\int p_b\,\dd p_b$, and each vertex contributes a bulk-point integral
$\int\dd z\,z^{-(d+1)}$. As at tree level, the strategy is to do the $z$-integrals
\emph{first} (elementary in $\AdS_4$), leaving the loop and spectral integrals to be
tackled by, respectively, flat-space and hypergeometric technology. For conformally
coupled scalars this is already automatic: the cosmological-polytope rule of
Section~\ref{sec:bypass} extends to loop graphs, performing the bulk-point
integrations combinatorially and delivering the wavefunction loop integrand
$\psiwf$ directly. Two explicit results set the template the spinning case must
match: the unrenormalized $\phi^3$ self-energy in $\AdS_6$ with a hard cutoff $\Lambda$,
\begin{equation}
	\label{eq:lp-selfenergy}
	\mathcal A^{L}(k)=
	-\frac{\ii\pi^2}{72}\Big(\frac{18\Lambda^2}{k}+96\,k\log\tfrac{\Lambda}{k}
	+(39-6\pi^2)\,k-72\,\Lambda\Big),
\end{equation}
whose power-divergent terms are regulator and counterterm dependent (the displayed
formula is not yet a renormalized observable), and the conformally invariant $\phi^4$
bubble in $\AdS_4$, whose scalar factor is the $d=3$ four-point loop integrand
$\psiwf=\lambda_4^2\,\mathcal A^{L}$ \cite{Albayrak:2020isk}.
Each carries a single scalar factor holding all the integration, to be dressed
afterward with whatever tensor structure the theory demands. The whole problem is
to make ``dress afterward'' precise when the tensor structure depends on the loop
momentum.

\subsection{Why the tree operator fails, and the auxiliary vector}
\label{sec:lp-aux}

At tree level (Section~\ref{sec:tree}) the tensor numerator of an internal line was
generated by an operator $\mathcal D^{\bm k}$ acting on a scalar propagator,
Eq.~\eqref{eq:Dop}. Crucially, $\mathcal D^{\bm k}$ depended only on the fixed
external momentum $\bm k$, so it commuted past every integral. At loop level this is
exactly what breaks. The momenta flowing through the two internal lines of a bubble
are $\bm q=\bm k+\bm\ell$ and $\bm q'=\bm\ell$, functions of the loop momentum, so the
operator $\mathcal D^{\bm q}$ that would generate the numerator carries $\bm\ell$ and
cannot be pulled outside $\int\!\dd^d\bm\ell$ \cite{Albayrak:2020bso}. Pulling the tensor
structure out, the whole point of the method, requires a representation of the
numerator in which the operator carries no loop momentum.

The cure is to trade loop-momentum dependence for dependence on an
\emph{auxiliary vector} $\bm v$. Each spinning propagator is rewritten so that its
numerator is generated by $\bm v$-derivatives, while the momentum enters only through
the scalar $(\bm q\cdot\bm v)$; the gluon line becomes
\begin{equation}
	\label{eq:lp-auxprop}
	\begin{gathered}
		\Gbb_{\mu\nu}(\bm q;z,z')=
		\widetilde{\mathcal D}^{\bm v}_{\mu\nu}\;\frac{(\bm q\cdot\bm v)^2}{2\,q^2}\\
		\times\int_0^\infty\! p\,\dd p\,\Big(\tfrac{q^2+p^2}{p^2}\Big)^{a}\,
		\Phi^{d-2}_{d-2}\!\begin{bmatrix}q,p\\ z,z'\end{bmatrix},
		\quad a\in\{0,1\},
	\end{gathered}
\end{equation}
with $\widetilde{\mathcal D}^{\bm v}$ built from the modified transverse/longitudinal
projectors $\widetilde\Pi^{(i)\bm v}$ (pure $\bm v$-derivative operators combined
with soft limits), and the vertices likewise made momentum-free,
$V\to\widetilde V^{\bm v}[\,\tfrac{\ii z^4}{\sqrt2}\sum_i(\bm k_i\cdot\bm v_i)\,]$. Here $\phi^{\mu}_{\nu}(k,z)$ and $\Phi^{\mu}_{\nu}[q,p;z,z']$ denote the normalized
radial profiles of a leg and of an internal line: products of powers of the radial
coordinate(s) with the Bessel functions $K$ and $J$ of Section~\ref{sec:toolkit},
with the upper (lower) label tracking the power (order). For a $K$-leg,
$\phi^{d-2}_{d-2}(k,z)=\sqrt{2/\pi}\,(kz)^{\frac{d-2}{2}}K_{\frac{d-2}{2}}(kz)$,
the bare exponential $e^{-kz}$ in $\AdS_4$; the full set is cataloged
in \cite{Albayrak:2020bso}. The
decisive feature is that $\widetilde{\mathcal D}$ and $\widetilde V$ carry \emph{no}
$\bm\ell$: they depend only on the auxiliary vectors and the fixed external data, so
they commute past the loop, spectral, and bulk integrals alike. For the spinning
vertices and gauges treated in \cite{Albayrak:2020bso}, any loop Witten diagram then
factorizes, as at tree level, into an operator times a scalar factor.
\emph{Loop master formula.} In the auxiliary-vector representation every spinning
loop integrand splits as
\begin{equation}
	\label{eq:auxD}
	W^{\,\mathrm{loop}}=\widetilde{\mathcal D}\,M,
\end{equation}
where $\widetilde{\mathcal D}$ holds all the tensor structure and $M$, the
\emph{scalar factor}, carries every integration in \eqref{eq:lp-layers}. Because
$\widetilde{\mathcal D}$ is built from $\bm v$-derivatives, limits, and contractions,
it is applied \emph{last}, by pure algebra, after the loop is done. The spinning loop
is thereby reduced to a scalar one.
Inside $M$ the loop integral is now an ordinary flat-space tensor integral: powers of
$(\bm\ell\cdot\bm v)$ over the four bubble denominators $\ell^2$, $|\bm k+\bm\ell|^2$,
$\ell^2+p_1^2$, $|\bm k+\bm\ell|^2+p_2^2$, reduced by Feynman parametrization and the
standard rules $\ell^\mu\ell^\nu\to(\ell^2/d)\,\eta^{\mu\nu}$ to scalar master
integrals. General master formulae package this for gluon and graviton diagrams with
arbitrary numbers of legs, lines, vertices, and loops \cite{Albayrak:2020bso}; the loop
momentum disappears, and only the radial spectral integrals remain.

\subsection{Appell \texorpdfstring{$F_4$}{F4} and the odd-\texorpdfstring{$d$}{d} collapse}
\label{sec:lp-F4}

The radial integrals are where the curvature shows itself. Each bulk-point integral of
a bulk-to-boundary Bessel-$K$ leg against two spectral Bessel-$J$ modes is, by a
classical formula \cite{Bailey:1936} (one of a family of Bessel integrals evaluated
in terms of Appell functions by contour methods \cite{Rice:1935}),
\begin{widetext}
	\begin{equation}
		\label{eq:lp-radialF4}
		\int_0^\infty\! z\,\dd z\;\phi^{d-2}_{d-2}(k,z)\,
		J_{\frac{d-2}{2}}(p_1z)\,J_{\frac{d-2}{2}}(p_2z)
		=\frac{2^{\frac{3d-5}{2}}\Gamma(\tfrac{d-1}{2})}{\pi\,k^2}
		\Big(\tfrac{p_1p_2}{k^2}\Big)^{\!\frac{d-2}{2}}
		F_4\!\Big(\tfrac d2,d{-}1;\tfrac d2,\tfrac d2;-\tfrac{p_1^2}{k^2},-\tfrac{p_2^2}{k^2}\Big),
	\end{equation}
\end{widetext}
\noindent
where
$F_4(a,b;c,e;x,y)=\sum_{m,n\ge0}\frac{(a)_{m+n}(b)_{m+n}}{(c)_m(e)_n\,m!\,n!}\,x^m y^n$
is the two-variable hypergeometric \emph{Appell $F_4$} function.
This is the loop-level analogue of the triple-$K$ integral \eqref{eq:tripleK}.
By Bailey's reduction formula,\footnote{The relevant formula here is (7) in \S5.10 of \cite{erdelyi1953higher}. It is interesting to note that the authors of this review were able to derive the algebraic collapse of Appell's function only for odd dimensions at the time \cite{Albayrak:2020bso} was written.} one can derive
\begin{equation}
	\label{eq:F4collapse}
	\boxed{\;
		\begin{gathered}
			F_4\!\Big(\tfrac d2,d{-}1;\tfrac d2,\tfrac d2;
			-\tfrac{p_1^2}{k^2},-\tfrac{p_2^2}{k^2}\Big)\\[2pt]
			=\Bigg(\frac{k^4}{\big(k^2+(p_1{+}p_2)^2\big)\big(k^2+(p_1{-}p_2)^2\big)}\Bigg)^{\!\frac{d-1}{2}}
		\end{gathered}\;}
\end{equation}
which shows that the result collapses to an algebraic function in general and a rational function in \emph{odd} $d$. For
$\AdS_4$ ($d=3$, $n=1$) it is a single rational factor, and the remaining spectral
integrals over $p_1,p_2$ become elementary (up to regularization),
just as the half-integer reductions trivialized the tree-level radial integrals.
In the assembled bubble, a representative gluon term carries the explicit ultraviolet
structure
\begin{equation}
	\label{eq:lp-bubbleresult}
	\begin{split}
		M^{\,\mathrm{gluon}}_{\mathrm{bubble}}\sim\;
		&\frac{\Gamma(-\tfrac d2)\,\Gamma(d+3)}{k^{\,d-6}}
		\int_0^\infty\!\frac{\dd p_1\,\dd p_2}{(p_1p_2)^{3-d}}\\
		&\times\Big[F_4\big(\tfrac d2,d{-}1;\tfrac d2,\tfrac d2;
		-\tfrac{p_1^2}{k^2},-\tfrac{p_2^2}{k^2}\big)\Big]^2
		+\cdots,
	\end{split}
\end{equation}
where the poles of $\Gamma(-\tfrac d2)$, met upon continuation to even $d$, signal the
short-distance (bulk UV) divergence of the loop, and cutoff schemes bring power
divergences demanding local counterterms besides. Their removal by a momentum-space
holographic renormalization, separating these local ambiguities from the nonlocal data,
together with the cutting rules that must organize the accompanying discontinuities, is
a program still in its infancy \cite{Bzowski:2015pba}. In \emph{even} $d$ the collapse is only
algebraic, a half-integer power of the same quartic, and the spectral integrals no
longer reduce to rational algebra: the open frontier. What is guaranteed in all cases is
the flat-space limit: as at tree level (Section~\ref{sec:tree}), the coefficient
of the total-energy pole at $\kT\to0$ is the bulk loop amplitude to all loop orders in dimensional regularization
\cite{Raju:2012zr}, so the loop integrand computed here carries the flat amplitude just
as the tree correlator does.

Among the loop programs that attack the same objects in other
variables,\footnote{Position space: ab-initio $\phi^4$ in $\AdS_4$ to two loops
	\cite{Bertan:2018khc} and its dual anomalous dimensions
	\cite{Bertan:2018afl}; $\AdS_4$ loop Witten diagrams recast as flat-space multi-loop
	Feynman integrals \cite{Heckelbacher:2022fbx}; the spectral representation, where loop kernels factorize into
	principal-series functions of the global-AdS cousin of $p_b$
	\cite{Carmi:2019ocp,Carmi:2021dsn,Carmi:2024tzj,Carmi:2026spv}; analytic loop-level
	momentum-space Witten diagrams from the spectral representation \cite{Chowdhury:2023khl};
	a recursive unitarity-based framework for higher-loop Witten diagrams
	\cite{Herderschee:2021jbi}; CFT crossing, fixing one-loop AdS amplitudes,
	including spinning (supergravity) ones, from tree data up to contact terms, though
	as boundary data rather than an integrand \cite{Aharony:2016dwx}; one-loop gluon
	amplitudes in AdS from the bootstrap \cite{Alday:2021ajh}. de~Sitter: conformally coupled
	one-loop correlators by Weyl transformation to Euclidean AdS
	\cite{Heckelbacher:2020nue}, evaluating to single-valued polylogarithms
	\cite{Heckelbacher:2022hbq}, the position-space shadow of \eqref{eq:F4collapse};
	$\phi^4$ in $\dS_4$ to two loops, where in-in correlators can be simpler than the
	wavefunction but with a subtle flat-space limit \cite{Chowdhury:2023arc}; the four-point bubble by
	partial Mellin--Barnes, its local and nonlocal signals from the factorized part
	\cite{Qin:2024gtr}. Inflationary in-in loops ask instead about secular growth: bounded
	by powers of $\log a$ \cite{Weinberg:2006ac}; infrared logarithms renormalizing
	background quantities \cite{Senatore:2009cf}; Bunch--Davies infrared stability for
	massive fields at one loop \cite{Marolf:2010zp} and all orders \cite{Marolf:2010nz};
	light-field logarithms resummed into a flow reproducing stochastic inflation
	\cite{Starobinsky:1994bd,Baumgart:2019clc}; all-order on-shell factorization of
	cosmological-collider signals \cite{Qin:2023nhv}; a chart of loop-correlator
	singularities \cite{Bhowmick:2025mxh}.} the auxiliary-vector construction is the only
\emph{bulk-integrand} one so far engineered for \emph{spinning} internal lines: it
delivers the gluon and graviton loop integrands at fixed boundary momenta, where the
flat-space limit, the total-energy singularity, and the cosmological cutting rules
are all manifest.\footnote{On the cosmology side the loop literature has grown
	quickly since: systematic leading-loop correlators \cite{Lee:2023jby}; one-loop
	nonanalyticity, factorization, and cutting rules
	\cite{Qin:2023bjk,Liu:2026jzn}; bootstrapping loops from the spectral
	decomposition of the de~Sitter propagator
	\cite{Xianyu:2022jwk,Altshuler:2025qmk}; infrared divergences and their
	resummation in the wavefunction
	\cite{Cespedes:2023aal,Benincasa:2024rfw,Benincasa:2024ptf}; and further loop
	integrands, collider signals at one loop, and evaluation technology
	\cite{Aoki:2026vbc,Chen:2026dqp,Grafe:2026qsm,Zhang:2025nzd,Pimentel:2026kqc}.}

\subsection{Soft limits: democratic scaling and the unitary split}
\label{sec:lp-soft}

The complementary window on the loop's analytic structure is the soft limit, where
unitarity is laid bare. In flat space, sending a leg soft ($\bm q\to0$) makes a single
propagator go on-shell, $1/(2\,\bm q\cdot\bm k_i)$, and the amplitude factorizes onto a
lower-point one: the Weinberg soft theorem \cite{Weinberg:1965nx}.\footnote{Read
	backward, it is a Ward identity of asymptotic symmetry, the viewpoint opened by the
	subleading soft graviton theorem \cite{Cachazo:2014fwa} and systematized
	in \cite{Strominger:2017zoo}.} The mechanism has two ingredients: a single
dominant diagram, and a propagator numerator resolving into physical states,
$\Pi_{\mu\nu}=\sum_h\eps^h_\mu\eps^{h*}_\nu/(k^2-m^2)$. The first fails in AdS. A
bulk-to-bulk line carries its would-be soft singularity only inside the normalized
radial profile $(kz)^{\nu}K_\nu(kz)$, which stays regular as the boundary energy
vanishes, so no denominator in any Witten diagram develops a $1/k_0$ pole and, for the
massless conserved currents considered here, all diagrams scale the same way,
\begin{equation}
	\label{eq:lp-democratic}
	W_{n+1}\;\xrightarrow{\,k_0\to0\,}\;\order{(k_0)^0},
\end{equation}
with none dominant \cite{Albayrak:2024ddg,Chowdhury:2024wwe}: the soft limit in AdS
is \emph{democratic}.\footnote{The same terrain has been reached from the opposite
	direction: the soft $(n{+}1)$-point gluon and graviton correlator as an energy- and
	polarization-derivative acting on the $n$-point one, by Feynman-diagram analysis for
	gluons and a Mellin--momentum bootstrap for gravitons \cite{Chowdhury:2024wwe};
	there too diagrams beyond the Weinberg class contribute at leading order, and the
	soft and flat-space limits do not commute, a derivative organization complementary
	to the Cutkosky-cut one developed below. See also soft limits of the wavefunction
	for scalars with nonlinearly realized symmetries \cite{Bittermann:2022nfh}, and the
	map linking the singularities of cosmological correlators across multiplicities
	\cite{Baumann:2021fxj}; exceptional scalar EFTs in de~Sitter fixed by enhanced soft
	limits \cite{Bonifacio:2021mrf,Armstrong:2022vgl}; loops, recursion, and soft limits
	extended to fermionic correlators in (A)dS \cite{Chowdhury:2024snc}; soft photon,
	gluon, and graviton theorems in (A)dS from conformal invariance \cite{Mei:2025jko};
	constraints on long-range forces in de~Sitter from soft properties of boundary
	correlators \cite{Baumann:2025tkm}; soft theorems from boosts and the other
	broken time symmetries \cite{Hui:2022dnm,Du:2024hol,Du:2024soq}; and Adler-zero
	conditions for inflationary correlators \cite{Green:2022slj}.} A
curved-space organizing principle is needed, and, as in flat space, it is unitarity,
now made manifest in the propagator itself. Resolving the axial-gauge numerator $\eta_{ij}+\bm k_i\bm k_j/p^2$
onto its transverse and longitudinal pieces splits the propagator in two \cite{Albayrak:2024ddg},
\begin{equation}
	\label{eq:softsplit}
	\begin{gathered}
		\Gbb_{ij}(\bm k;z,z')=\Gbb^{\mathrm L}_{ij}(\bm k;z,z')\\
		+\sum_h\int_0^\infty\!\frac{p\,\dd p}{2\,k^{2\nu}}\,
		\frac{\hat A^{h*}_i(\bm k,p,z)\,\hat A^{h}_j(\bm k,p,z')}{k^2+p^2+\ii\eps},
		\quad \nu=\tfrac d2-1,
	\end{gathered}
\end{equation}
with $\hat A^h_i(\bm k,p,z)=\eps^h_i(\bm k)\,k^\nu z^\nu J_\nu(pz)$ an off-shell
normalizable mode. The second term, the \emph{transverse} part
$\Gbb^{\mathrm T}$, is a sum over physical helicities of normalizable-mode bilinears
glued by the flat propagator $1/(k^2+p^2)$: precisely the unitarity sum, now under a
spectral integral. The first term, the \emph{longitudinal} part, has the strikingly
simple closed form\footnote{We note the typo in \cite{Albayrak:2024ddg} where result is recorded as $\frac{\bm k_i\bm k_j}{4\nu\,k^2}\,
	\Big(\frac{\min(z,z')}{\max(z,z')}\Big)^{\!\nu}$, a factor of $(zz')^{\nu}$ off from the true value.}
\begin{equation}
	\label{eq:GL}
	\boxed{\;
		\Gbb^{\mathrm L}_{ij}(\bm k;z,z')
		=\frac{\bm k_i\bm k_j}{4\nu\,k^2}\,\min(z,z')^{2\nu}\;,\;}
\end{equation}
a \emph{crossed} line in the language of Section~\ref{sec:bypass}. Its essential
property is that $\Gbb^{\mathrm L}$ \emph{drops out of the flat-space limit}: it
contributes nothing to the total-energy pole whose coefficient is the amplitude, so
anything built from it is a pure curvature effect, an intrinsically-(A)dS remainder
with no amplitude counterpart. The split \eqref{eq:softsplit} is engineered so that the transverse
part factorizes exactly as unitarity demands, isolating the longitudinal piece as a
separate object.

\subsection{Transition amplitudes and the (A)dS cut}
\label{sec:lp-cut}

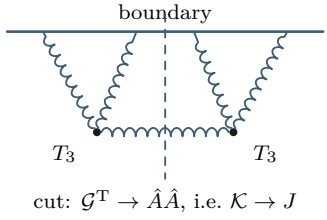
\begin{figure}[t]
	\centering
	\begin{tikzpicture}[scale=0.95]
		\draw[wbndry] (-2.2,1.4)--(2.2,1.4);
		\node[above,font=\footnotesize] at (0,1.4){boundary};
		\node[wdot](vl) at (-0.95,0){}; \node[wdot](vr) at (0.95,0){};
		\draw[wgluon](vl)--(vr);
		\draw[wgluon](vl)--(-1.7,1.4); \draw[wgluon](vl)--(-0.4,1.4);
		\draw[wgluon](vr)--(1.7,1.4); \draw[wgluon](vr)--(0.4,1.4);
		\draw[wcut] (0,1.65)--(0,-0.65);
		\node[font=\footnotesize] at (0,-0.95){cut: $\Gbb^{\mathrm T}\to\hat A\hat A$, i.e.\ $\Kbb\to J$};
		\node[font=\footnotesize] at (-1.4,-0.3){$T_3$}; \node[font=\footnotesize] at (1.4,-0.3){$T_3$};
	\end{tikzpicture}
	\caption{An exchange diagram cut on its internal line into two transition amplitudes
		$T_3$, the (A)dS analogue of a Cutkosky cut. Replacing the transverse propagator
		$\Gbb^{\mathrm T}$ by the normalizable modes it is built from (the operation
		$\Kbb\to J$) severs the diagram into on-shell lower-point pieces sewn by the spectral
		measure.}
	\label{fig:cut}
\end{figure}

The transverse part of \eqref{eq:softsplit} writes the internal line as a sum over physical
states put on-shell by the spectral measure, the curved-space version of the flat
identity $\Pi_{\mu\nu}=\sum_h\eps^h_\mu\eps^{h*}_\nu/(k^2-m^2)$. \emph{Cutting} an
internal line means replacing $\Gbb^{\mathrm T}$ by the normalizable modes $\hat A$ it
is built from (the operation $\Kbb\to J$ that exchanges the bulk-to-boundary kernel
for a bulk mode), and this severs a correlator into pieces joined at on-shell
intermediate states (Figure~\ref{fig:cut}). This is the (A)dS analogue of a Cutkosky
cut, and the objects on either side are \emph{transition amplitudes}: lower-point
correlators in which one leg is a normalizable mode rather than a boundary source,
equivalently correlators between states dual to bulk modes \cite{Raju:2010by,Raju:2011mp}. The
three-point gluon transition amplitude is
\begin{equation}
	\label{eq:lp-T3}
	\begin{split}
		T_3^{\,h_1h_2h}&(\bm k_1,\bm k_2;\bm q,p)\\
		&=V^{ijk}(\bm k_1,\bm k_2,\bm q)\!\int_0^\infty\!\frac{\dd z}{z^{d+1}}\\
		&\quad\times A^{h_1}_i(\bm k_1,z)\,A^{h_2}_j(\bm k_2,z)\,\hat A^{h}_k(\bm q,p,z),
	\end{split}
\end{equation}
with $A^h_i(\bm k,z)=\eps^h_i(\bm k)k^\nu z^\nu K_\nu(kz)$ the bulk-to-boundary leg and
$\hat A$ the cut leg. These are the same objects the on-shell recursion of
Section~\ref{sec:bypass} sews together.
\begin{lesson}
	A loop cut and a soft limit are two faces of the same object. Both replace a
	bulk-to-bulk line by its normalizable modes ($\Kbb\to J$), and both express a
	correlator as transition amplitudes glued by the spectral measure. What flat-space
	unitarity does with on-shell trees, (A)dS unitarity does with transition amplitudes.
\end{lesson}
The claim is made concrete on the four-point $s$-channel gluon diagram, whose soft
limit is computed in three independent ways \cite{Albayrak:2024ddg}: (i) \emph{directly},
as a spectral integral of Appell-$F_4$ radial blocks (the numerically practical form);
(ii) by the \emph{differential operator} \eqref{eq:Dop}, which commutes with the radial
integrals (the compact form); and (iii) by the \emph{unitary decomposition}
\eqref{eq:softsplit}, the most transparent, giving schematically
\begin{widetext}
	\begin{equation}
		\label{eq:lp-4pt}
		\lim_{\bm k_1\to0}W_4=\sum_h\int_0^\infty\!\frac{p\,\dd p}
		{2\,k_2^{2\nu}(k_2^2+p^2+\ii\eps)}\,
		T_3^{\,h_3h_4h}(\bm k_3,\bm k_4;-\bm q,p)\,
		\Big(\lim_{\bm k_1\to0}T_3^{\,h_1h_2h^*}(\bm k_1,\bm k_2;\bm q,p)\Big)
		+W^{\mathrm{soft}}_{\AdS}.
	\end{equation}
\end{widetext}
\noindent
The first term is a unitarity cut (two transition amplitudes sewn across the spectral
integral), and the second is the intrinsically-AdS longitudinal remainder built from
$\Gbb^{\mathrm L}$ of \eqref{eq:GL}, separately
divergent but obeying the clean scaling
$W^{\mathrm{soft}}_{\AdS}(\lambda\bm k)=\lambda^4\,W^{\mathrm{soft}}_{\AdS}(\bm k)$, a
stringent check the explicit $\AdS_4$ result satisfies. That the three routes agree
on the same diagram is a nontrivial test of the transition-amplitude cut.

The seed of the whole construction is the soft limit of the three-point transition
amplitude itself. Evaluating the triple-$K$ scalar factor of a spin-$\ell$ field dual
to a conserved current and sending one leg soft collapses the conformal cases to a
single closed form,
\begin{equation}
	\label{eq:3ptsoft}
	\boxed{\;
		\begin{gathered}
			\lim_{k_1\to0}T_3(k_1,k_2,k_3)
			=(-1)^{n}\,(2n-1)!!\;\frac{k_2^{\,2n+1}-k_3^{\,2n+1}}{k_2^{2}-k_3^{2}}\,,\\
			n=\ell+\tfrac{d-5}{2}\in\mathbb Z_{\ge0}\;.
		\end{gathered}\;}
\end{equation}
Gluons ($\ell=1$) give $n=0$ in $\AdS_4$
and $n=1$ in $\AdS_6$; gravitons ($\ell=2$) give $n=2$ in $\AdS_6$ and $n=3$ in
$\AdS_8$. With momentum conservation ($k_3\to k_2=k$ as $k_1\to0$) the boxed form collapses to
the power law $(-1)^n\,\tfrac{(2n+1)!!}{2}\,k^{2n-1}$, the behavior expected of a
propagator on the cut \cite{Albayrak:2024ddg}.

\subsection{Generalized unitarity and the cosmological optical theorem}
\label{sec:lp-generalized}

The direct auxiliary-vector construction and the unitarity cut are complementary,
exactly as in flat space \cite{Elvang:2013cua}. The direct method delivers the full loop
integrand (rational, cut-free terms included) at the cost of the tensor algebra,
while generalized unitarity reconstructs the cut-constructible part cheaply from
on-shell transition amplitudes, with the tensor structure still carried by
$\widetilde{\mathcal D}$. Cutting the one-loop bubble of Section~\ref{sec:lp-layers} on
its two internal lines turns it, by \eqref{eq:softsplit}, into the spectral integral of two
three-point transition amplitudes (exactly the structure \eqref{eq:lp-4pt} met in the
soft limit). Its discontinuity across a partial-energy channel is therefore fixed by
lower-point on-shell data, and the loop is rebuilt from its cuts by a dispersion
relation in the boundary energies, up to the subtraction polynomials and local
contact terms that renormalization conditions and symmetry must fix. The cut structure itself is supplied in AdS by
Cutkosky rules \cite{Meltzer:2020qbr}, systematized as a
full toolkit of unitarity methods (cutting rules, dispersion relations, and their
CFT avatars) in \cite{Meltzer:2019nbs,Meltzer:2021bmb}, and, on the
cosmology side, by the cosmological cutting rules
\cite{Goodhew:2020hob,Goodhew:2021oqg}, extended to Yang--Mills wavefunction
coefficients in de~Sitter \cite{He:2026bwn}, with correlator discontinuities
obtained directly from scattering amplitudes \cite{Chowdhury:2026upp}.\footnote{The
	same unitarity is now available directly for the in-in correlator: cutting rules
	for in-in observables and their collider application
	\cite{Ema:2024hkj,Ghosh:2024aqd}; factorization of correlators through
	wavefunction-like objects \cite{Stefanyszyn:2024msm}; and further cut,
	dispersive, and open-system formulations
	\cite{Das:2025qsh,Das:2026vfv,Ansari:2026xkm,Lee:2025kgs,Colipi-Marchant:2025oin}.}
The dispersive logic carries over to inflation directly,
reconstructing wavefunction coefficients from analyticity in the energies plus
factorization on their partial-energy singularities \cite{Meltzer:2021zin}. At loop level
the same singularity structure powers the cosmological tree theorem
\cite{AguiSalcedo:2023nds}, which expresses any wavefunction loop integrand as
a sum of tree integrands, the cosmological cousin of flat-space loop--tree duality.

This is where the loop program becomes cosmological. Under the continuation of
Section~\ref{sec:inflation}, holographic correlators map to de~Sitter wavefunction
coefficients $\psiwf$, and the (A)dS cut becomes the \emph{cosmological optical
	theorem} \eqref{eq:cot}: the discontinuity of $\psiwf$ in the boundary energies, taken
with the energy reversal and complex conjugation the theorem prescribes, is a product
of lower-point wavefunction coefficients glued across a cut internal line,
the cutting rules following from unitarity of the bulk time evolution and the
manifestly local test \cite{Jazayeri:2021fvk} from locality; unitarity,
recursion, and soft limits have since been tied together in a single (EA)dS
framework \cite{Ansari:2026sjf}. The transition-amplitude factorization
\eqref{eq:lp-4pt} is the diagram-by-diagram realization of that theorem. And the soft
limit \eqref{eq:lp-democratic}--\eqref{eq:3ptsoft} descends, under the same
continuation, to the squeezed limit of an inflationary correlator: the
single-field consistency relation, fixing the squeezed three-point function in terms of
the tilt of the power spectrum \cite{Maldacena:2002vr}, is the de~Sitter face of the
three-point soft limit \eqref{eq:3ptsoft}, though it carries more than the generic
soft limit alone: its derivation also invokes the adiabatic long mode and its
residual-diffeomorphism Ward identity. A flat-space theorem, deformed by curvature
into a statement about transition amplitudes, reappears as a prediction for the
squeezed non-Gaussianity of the early universe. The open fronts are correspondingly
sharp: in even boundary dimensions the collapse is algebraic rather than rational
and the spectral integrals resist closed form; the
$\Gamma(-\tfrac d2)$ divergences await a systematic momentum-space holographic
renormalization; and, within the auxiliary-vector program, only the bubble is worked out in full, with
the triangle, box, and higher loops charted only in principle by the master formula
$W^{\mathrm{loop}}=\widetilde{\mathcal D}\,M$. These open ends are not mere
bookkeeping: they control the scheme-independent nonlocal data and, after
continuation, its late-time behavior. But the physics the cuts already compute is
cosmological, and we now cross the bridge.

\section{From anti--de~Sitter to inflation}
\label{sec:inflation}

Everything so far was computed, or is most cleanly computed, in Euclidean
anti--de~Sitter space, where the propagators decay, the integrals converge, and the
analytic structure is transparent. This section cashes those results in cosmology.
We carry out the continuation that makes an AdS correlator a de~Sitter wavefunction
coefficient, derive the power spectrum, state the cosmological optical theorem and
\emph{verify} it on the simplest exchange, and show how the same singularity data feed the cosmological
bootstrap and the cosmological collider. The thread is the one announced in
Section~\ref{sec:intro}: one object, two evaluations.

\subsection{The continuation}

The bridge to everything computed earlier is the continuation \eqref{eq:cont} of
Section~\ref{sec:toolkit}, under which a Euclidean AdS Witten diagram becomes a
de~Sitter wavefunction coefficient. The correspondence was established in general
in \cite{Harlow:2011ke}: the Euclidean AdS wavefunction continues to the
de~Sitter one with Euclidean (Bunch--Davies) initial conditions, and the
dictionary is sharpest for conformally coupled scalars, where the convergent
radial integrals of AdS continue to the oscillatory conformal-time integrals of
the wavefunction. Thus the two-site exchange
\begin{equation}
	\label{eq:ds-psi4}
	\psiwf_4=\frac{1}{\kT\,(k_{12}+k_{\ul{12}})\,(k_{34}+k_{\ul{12}})}
\end{equation}
of \eqref{eq:psi2} is at once an AdS scalar correlator and, once the overall
normalization and phase are matched through \eqref{eq:cont}, a de~Sitter
wavefunction coefficient. The continuation is not entirely innocent:
boundary conditions, the $\ii\eps$ prescription, and the distinction between the
wavefunction and the correlator all require care, and massive fields introduce branch
cuts in $\eta$ absent in AdS \cite{DiPietro:2021sjt,Sleight:2019hfp}; for spinning
fields, the in-in correlators admit a shadow formulation \cite{Chowdhury:2025nnk}. But
as an organizing device it
lets every AdS result of this review be read as a statement about inflation.

\subsection{The power spectrum, derived}

The two-point coefficient quoted in Section~\ref{sec:wavefunction} follows from the
free mode \eqref{eq:bd-mode}. The quadratic wavefunction coefficient is the boundary
value of the on-shell action, $\psiwf_2(k)=-\ii\,a^2\phi_k'(\eta)/\phi_k(\eta)$
evaluated at $\eta\to0$. Using $\phi_k'=\tfrac{H}{\sqrt{2k^3}}k^2\eta\,e^{-\ii k\eta}$
and expanding near the boundary,
\begin{equation}
	\label{eq:spectrum}
	\begin{gathered}
		\psiwf_2(k)=-\frac{\ii\,k^2}{H^2\eta}-\frac{k^3}{H^2}+\order{\eta}
		\quad\Longrightarrow\\[3pt]
		\boxed{\;\langle\phi(\vk)\phi(-\vk)\rangle'
			=\frac{-1}{2\,\Rez\,\psiwf_2}=\frac{H^2}{2k^3}\;.}
	\end{gathered}
\end{equation}
The divergent imaginary part $-\ii k^2/H^2\eta$ is a local, scheme-dependent contact
term, the analogue of a boundary counterterm in holographic renormalization. It
drops out of the correlator, which sees only the finite real part
$\Rez\,\psiwf_2=-k^3/H^2$ (negative, as normalizability of $|\Psi|^2$ requires).
The result is scale invariant, the hallmark of inflation. Strictly it is the spectrum
of a massless spectator, a proxy for the curvature perturbation $\zeta$: a mass shifts
the exponent of $\psiwf_2$ off $k^3$, and in single-field slow roll the conversion to
$\zeta$ turns the slow evolution of the background into the observed spectral tilt, as
stated in Section~\ref{sec:wavefunction}.

\subsection{The cosmological optical theorem, verified}

Unitarity of the bulk time evolution, the equality of the wavefunction and in-in
routes of Section~\ref{sec:wavefunction}, constrains the coefficients through the
\emph{cosmological optical theorem} \cite{Goodhew:2020hob}. Continuing all external energies
$k_a\to-k_a$ (which leaves the internal energy $y=k_{\ul{12}}$ fixed, since it
depends on the spatial vectors) and adding the complex conjugate gives a sum over cuts,
\begin{equation}
	\label{eq:cot}
	\begin{split}
		\psiwf_n(\vk)+\psiwf_n^{\ast}(-\vk)
		=-\sum_{\rm cuts}\Ps(y)\,&\big[\psiwf_L+\psiwf_L^{\ast}(-\vk)\big]\\
		\times&\big[\psiwf_R+\psiwf_R^{\ast}(-\vk)\big],
	\end{split}
\end{equation}
with $\Ps(y)$ the power spectrum of the exchanged field, a function of the internal
energy $y$.\footnote{The sign of the cut term tracks the sign convention for
	$\psiwf_n$ in \eqref{eq:Psi}, fixed here by $\psiwf_3=1/\kT$ and \eqref{eq:ds-psi4}
	with plus signs; the opposite convention gives $+\Ps(y)$, as
	in \cite{Goodhew:2020hob}.} Here $\psiwf_n^{\ast}(-\vk)$ is shorthand for complex
conjugation combined with the continuation $k_a\to-k_a$ of the external energies, not
mere evaluation on spatially reversed momenta. The contact coefficient has no internal
line, so for real couplings $\psiwf_3+\psiwf_3^{\ast}(-\vk)=0$; explicitly, $\psiwf_3=1/\kT$ gives
$1/\kT+1/(-\kT)=0$. The four-point exchange \eqref{eq:ds-psi4} is the first nontrivial
test. Writing $x_L=k_{12}$, $x_R=k_{34}$, $y=k_{\ul{12}}$, the left side of
\eqref{eq:cot} is
\begin{equation}
	\label{eq:cot-lhs}
	\begin{split}
		\psiwf_4+\psiwf_4^{\ast}(-\vk)
		&=\frac{1}{\kT(x_L{+}y)(x_R{+}y)}-\frac{1}{\kT(y{-}x_L)(y{-}x_R)}\\
		&=\frac{-2y}{(y^2-x_L^2)(y^2-x_R^2)},
	\end{split}
\end{equation}
while the right side, with the three-point ``half-diagrams'' $\psiwf_L=1/(x_L+y)$,
$\psiwf_R=1/(x_R+y)$ and the conformally coupled internal power spectrum $\Ps(y)=1/2y$,
is
\begin{equation}
	-\frac{1}{2y}\Big[\frac{2y}{y^2-x_L^2}\Big]\Big[\frac{2y}{y^2-x_R^2}\Big]
	=\frac{-2y}{(y^2-x_L^2)(y^2-x_R^2)},
\end{equation}
identical to \eqref{eq:cot-lhs}. The theorem holds on the exchange, and, read
backwards, \emph{fixes} the internal power spectrum to $\Ps(y)=1/2y$: a one-line
derivation of the de~Sitter propagator's normalization from unitarity alone. Iterating
\eqref{eq:cot} gives the cosmological cutting rules \cite{Goodhew:2021oqg} that determine
the discontinuities across all partial-energy singularities. This is the de~Sitter face
of the transition-amplitude cuts of Section~\ref{sec:loops}: cutting an internal line
replaces it by its normalizable modes, turning a coefficient into a product of
lower-point on-shell data. One subtlety: the left side of
\eqref{eq:cot} is not literally a discontinuity, the individual cuts are not uniquely
fixed, and the relation as written does not by itself reduce to the flat-space optical
theorem. The cleanest resolution is geometric: in the \emph{optical polytope} picture
of~\cite{Albayrak:2023hie}, perturbative
unitarity is encoded in a non-convex region of the cosmological polytope, and the
various cutting rules are its different subdivisions.

\subsection{The cosmological bootstrap}

Symmetry and singularities together are strong enough to fix correlators \emph{without}
a bulk computation: the cosmological bootstrap \cite{Arkani-Hamed:2018kmz,Baumann:2022jpr}.
De~Sitter isometries act on the late-time slice as the conformal group of $\mathbb R^d$,
so the wavefunction coefficients obey the momentum-space conformal Ward identities of
Section~\ref{sec:toolkit}. For a massive scalar exchanged in the four-point function,
the dilatation and special-conformal identities reduce to a single second-order
ordinary differential equation in the energy ratio $u=y/(k_1+k_2)$ (with a mirror
equation in $v=y/(k_3+k_4)$ acting on the other pair), with
$y=k_{\ul{12}}$ the internal energy,
\begin{equation}
	\label{eq:boot}
	\begin{gathered}
		\Delta_u\,\psiwf_4=(\text{contact source}),\\[2pt]
		\Delta_u=u^2(1-u^2)\partial_u^2-2u^3\partial_u
		+\big[\Delta_+\Delta_--\cdots\big],
	\end{gathered}
\end{equation}
a hypergeometric equation whose homogeneous solutions carry the exchanged mass; in
the squeezed limit they imprint the non-analytic powers $(k_1/k_3)^{\Delta_\pm}$,
$\Delta_\pm=\tfrac d2\pm\nu$ with
$\nu=\sqrt{\tfrac{d^2}{4}-m^2/H^2}$. One fixes the solution with the two
analytic inputs this review has emphasized, the total-energy residue (the flat-space
amplitude, Section~\ref{sec:tree}) and the partial-energy factorization
\eqref{eq:cot}, supplemented by regularity in the collapsed limit and up to the usual
contact-term ambiguities. Realistic inflation, however, breaks the de~Sitter boosts (the rolling
inflaton selects a preferred clock), so the full conformal group is not available. The
\emph{boostless} bootstrap \cite{Pajer:2020wnj} drops the
boost generators and retains translations, rotations, (approximate) scale
invariance, locality, and unitarity (the
cosmological optical theorem and the manifestly local test) as the defining
constraints; the spinning case is developed in
\cite{Baumann:2020dch}. This weaker, more realistic input still
pins down a great deal, and it places the singularity-based methods of this
review, rather than full conformal symmetry, at the center of the modern
program.\footnote{The boostless program has since been pushed in several
	directions: heavy fields with a small sound speed, where collider signals are
	enhanced \cite{Jazayeri:2022kjy,Pimentel:2022fsc}; graviton and mixed
	graviton--scalar bispectra in the effective theory of inflation
	\cite{Cabass:2022jda,Ghosh:2023agt}; constructive and dispersive solutions of the
	boundary equations \cite{Bissi:2023bhv,Liu:2024xyi}; double massive exchange
	\cite{Aoki:2024uyi}; resonant non-Gaussianity, where scale invariance itself is
	broken \cite{DuasoPueyo:2023kyh}; and bounds and positivity on the resulting
	correlators \cite{deRham:2025mjh,CarrilloGonzalez:2025fqq,Jazayeri:2025vlv}.}

Around these two poles (full conformal symmetry and the boostless axioms), an
ecosystem of unitarity and locality constraints has grown, most of it recognizably
amplitude technology in de~Sitter dress,\footnote{Boundary time-evolution equations
	propagate the massive-exchange correlator itself, bypassing the wavefunction and its
	bulk time integrals \cite{Cespedes:2020xqq}; the boostless inputs distill into a compact
	set of bootstrap rules strong enough to enumerate \emph{every} tree-level single-scalar
	bispectrum up to a fixed number of derivatives \cite{Pajer:2020wxk}; the single-cut
	relation \eqref{eq:cot} extends to systematic cutting rules for arbitrary tree diagrams
	and loop integrands \cite{Melville:2021lst}; consistency constraints rule out many
	(partially-)massless matter couplings in de~Sitter \cite{Sleight:2021iix}; and,
	constructively, a dictionary converts
	any flat-space contact amplitude directly into its de~Sitter correlator
	\cite{Bonifacio:2021azc}. The underlying analytic structure (which
	singularities the wavefunction may have, where it is boundedly analytic, and how
	flat-space analyticity and its violations imprint on cosmology) was charted in
	\cite{Salcedo:2022aal}.} and discrete symmetries
sharpen it further.\footnote{Reality and parity: unitarity and Bunch--Davies initial
	conditions force tree-level parity-even correlators to be real and parity-odd ones to
	factorize \cite{Stefanyszyn:2023qov}, and with scale invariance yield the no-go that
	the parity-odd scalar trispectrum vanishes at tree level for complementary- or
	principal-series exchange, so any detection signals spin, modified initial states, or
	a breakdown of the axioms \cite{Cabass:2022rhr}; CPT: these conditions and the
	cosmological optical theorem share a common origin in a bulk CRT transformation
	\cite{Goodhew:2024eup}; positivity: bounds on the inflationary effective theory
	follow directly from correlators, the de~Sitter face of $S$-matrix positivity
	\cite{Green:2023ids}; in-out: in-in correlators rewritten as amplitude-like in-out
	objects with flat-space Feynman rules, explaining why correlators are often simpler
	than the wavefunction coefficients assembled into them \cite{Donath:2024utn};
	parity-odd signatures and their discrete-symmetry origins are developed further
	in \cite{Goyal:2025hdm}.} Every one of
these constraints consumes the same data this review has been producing: energy poles,
their residues, and the factorization of coefficients on them.

\subsection{The cosmological collider}

The most striking phenomenological target is the \emph{cosmological collider}
\cite{Arkani-Hamed:2015bza,Lee:2016vti}. A massive particle produced during inflation imprints a characteristic
signal in the squeezed limit $k_1\ll k_2\simeq k_3$ of the bispectrum, where the
exchanged field goes on shell. For a field of mass $m$ in the principal series
$m>\tfrac d2 H$, the index $\nu=\sqrt{d^2/4-m^2/H^2}$ is imaginary,
$\nu=\ii\mu$, and the solutions of \eqref{eq:boot} oscillate: the squeezed bispectrum
acquires a non-analytic, oscillatory dependence
\begin{equation}
	\label{eq:collider}
	\frac{\langle\phi\phi\phi\rangle}{\langle\phi\phi\rangle\langle\phi\phi\rangle}
	\ \xrightarrow{\,k_1\to0\,}\
	\Big(\frac{k_1}{k_3}\Big)^{d/2}\Big[\,\mathcal A\,\cos\!\big(\mu\log\tfrac{k_1}{k_3}+\varphi\big)+\cdots\Big],
\end{equation}
whose frequency $\mu$ measures the particle's mass in units of $H$ (for spinning
exchange the mass--frequency relation shifts with the spin); the spin itself is
read off the accompanying angular dependence, a Legendre polynomial in
$\hat{\vk}_1\!\cdot\!\hat{\vk}_3$ multiplying \eqref{eq:collider}. This is an
accelerator operating at inflationary energies,
its particle spectrum recorded in the sky. The squeezed limit that carries the signal
is the soft limit of Section~\ref{sec:loops} continued to de~Sitter, but the
oscillation itself is not a partial-energy pole: it is nonanalyticity in the momentum
ratio, controlled by the late-time dimensions $\Delta_\pm$, and it records genuine
particle production that no local vertex obtained by integrating out the heavy field
can mimic. The cutting rules and the partial-energy factorization \eqref{eq:cot} still
constrain its coefficient; the massive-exchange correlators behind the signal are now
known in closed form \cite{Qin:2023ejc}, and at high energies the correlators are tied
to flat-space amplitudes by a Goldstone equivalence theorem \cite{Green:2024hbw}.

What might this collider record?\footnote{Supersymmetry, the leading mechanism for
	keeping scalars light, generically populates the window $m\sim H$, making the
	squeezed bispectrum a probe of high-scale supersymmetry breaking
	\cite{Baumann:2011nk}; ``heavy-lifting'' couplings to the inflationary background
	drag hidden-sector spectra up toward the Hubble scale, displaying mass ratios and
	spins far beyond any terrestrial accelerator \cite{Kumar:2017ecc}; the Standard
	Model itself sets the irreducible background, its inflationary mass spectrum and
	unavoidable oscillatory signals computed in \cite{Chen:2016uwp}; massive gauge
	bosons, their spin-one angular dependence separating them from scalar exchange, are
	worked out in \cite{Wang:2020ioa}; states with a continuous mass spectrum leave
	their own distinctive signals \cite{Aoki:2023tjm}; analytic formulae extend to
	fields with time-dependent masses \cite{Aoki:2023wdc}; the oscillation phase is
	itself calculable and carries model information
	\cite{Qin:2022lva,Xianyu:2023ytd}; helical signals from a chemical potential
	\cite{Qin:2022fbv}; the effective-theory treatment of when heavy physics
	survives as a signature \cite{Craig:2024qgy}; enhanced, non-Boltzmann-suppressed
	couplings \cite{McCulloch:2024hiz}; and further signal classes and templates
	\cite{Suman:2025tpv,Suman:2025vuf,Wang:2025qww,Cheung:2025dmc,Pimentel:2025rds}.} The signals are small in the minimal setup (with Bunch--Davies initial conditions and
weak couplings, the particle production behind
\eqref{eq:collider} costs a Boltzmann-like factor $e^{-\pi\mu}$, though chemical
potentials, nonadiabatic backgrounds, or direct couplings can lift the suppression),
but they live in the squeezed configurations of Section~\ref{sec:wavefunction}, where
the single-field consistency relation guarantees a nearly signal-free adiabatic floor:
any nonanalytic oscillation resolved above it by a future galaxy or CMB survey is a
discovery. Searches in current data
have already begun \cite{Cabass:2024wob,Philcox:2026tjj}.

\subsection{Holography for cosmology}
\label{sec:holocosmo}

The continuation \eqref{eq:cont} can be taken more seriously than diagram-by-diagram
bookkeeping: it can be read as a proposed duality. In holographic cosmology,
the same continuation maps an inflating four-dimensional background to a
Euclidean domain wall asymptotic to AdS, and the holographic dictionary then trades the
\emph{entire} inflationary computation for one in a dual three-dimensional QFT without
gravity. The late-time observables are read off from momentum-space correlators of the
dual stress tensor: the scalar and tensor power spectra from (the imaginary part of)
the two-point function $\langle TT\rangle(k)$, and the non-Gaussianities from the
three-point function. The bispectrum was computed in \cite{McFadden:2010vh}:
holographic formulae express the inflationary three-point function in terms of
stress-tensor correlators of the dual theory (a super-renormalizable $SU(N)$ gauge
theory of scalars and fermions, \emph{weakly} coupled precisely when the bulk is
strongly quantum and no geometric description exists). A perturbative QFT
computation then predicts a bispectrum of exactly the factorizable equilateral type
with the universal amplitude $f_{\rm NL}^{\rm eq}=5/36$, independent of the details
of the dual within this class of models.
This is a sharp, falsifiable prediction from a non-geometric phase of gravity, and it
has been confronted with data: fitted directly to the Planck
CMB spectra, the holographic models are
competitive with power-law $\Lambda$CDM over the multipoles $\ell\gtrsim30$ where the
dual QFT remains perturbative (and marginally better with the lowest multipoles
excluded), while $\Lambda$CDM is favored globally and the perturbative dual
description breaks down at the largest angular scales \cite{Afshordi:2016dvb}.

For this review the salient point is that the dual computation is not foreign
technology: the stress-tensor correlators holographic cosmology consumes are the
momentum-space boundary correlators of Section~\ref{sec:toolkit}, whose conformal
Ward identities, triple-$K$ representation, and tensorial renormalization
\cite{Bzowski:2013sza,Bzowski:2018fql} were developed in large part \emph{for} this
program, and whose scheme-dependent contact terms are the analogues of the purely
imaginary local term in \eqref{eq:spectrum}, which drops out of $|\Psi|^2$ while
finite real local terms must be tracked observable by observable; the same holographic
technology
systematically renormalizes the IR divergences of de~Sitter in-in correlators
\cite{Bzowski:2023nef}. Whether the three-dimensional
theory is a genuine microscopic dual or bookkeeping for the wavefunction, the
toolkit is shared.

\subsection{What the AdS program teaches inflation}

The momentum-space AdS program feeds this cosmology directly, through the very
objects computed and verified in the preceding sections.
\begin{itemize}
	\item The \emph{total-energy singularity} that organized every tree computation
	(Sections~\ref{sec:tree}--\ref{sec:web}) is the bootstrap's scattering singularity,
	its residue \eqref{eq:flatlim} the flat-space amplitude.
	\item The \emph{cosmological-polytope canonical forms} of Section~\ref{sec:bypass}
	\emph{are} the wavefunction coefficients of the conformal model (more general time
	dependence requires further time integrations); the verified two- and three-site
	wavefunctions are the building blocks of the bootstrap.
	\item The \emph{soft limits} of Section~\ref{sec:loops} continue to de~Sitter and,
	combined with the adiabatic-mode Ward identity, underlie the single-field
	consistency relation \cite{Maldacena:2002vr}; the transition-amplitude cuts of the
	same section are the flat-space face of the cosmological optical theorem
	\eqref{eq:cot} verified above.
	\item The \emph{graviton three-point correlator} of Section~\ref{sec:tree} continues
	to the primordial tensor non-Gaussianity \cite{Maldacena:2011nz}; the
	\emph{double copy} of Section~\ref{sec:bypass} relates scalar, gauge, and
	gravitational non-Gaussianities (exactly on the total-energy residue, with
	computable curvature corrections away from it); and the explicit spinning
	correlators across
	dimensions (Section~\ref{sec:web}) supply templates for the shapes a future survey
	might detect.
\end{itemize}
Holographic and cosmological correlators are thus not merely analogous but, through
\eqref{eq:cont}, two evaluations of one momentum-space object, and the cleanest place
to compute it is often anti--de~Sitter space.

\section{The frontier, and connections to neighboring fields}
\label{sec:frontier}

The program of the preceding sections has reached the point the flat-space program
reached shortly after BCFW: explicit results through several points and one loop, a
recursion, a double copy, a positive geometry, and a unitarity-based reconstruction.
The frontier is the search for the structure \emph{behind} those results (an
amplituhedron for the wavefunction of the universe) and the knitting of momentum
space into the neighboring languages. This section is a map of that frontier, not a
survey; the references are entry points.

\subsection{Differential equations and kinematic flow}

The liveliest direction trades the integral and the sum over tubings for a closed
system of \emph{differential equations} in the external energies
\cite{Arkani-Hamed:2023kig,Arkani-Hamed:2023bsv}. For the conformally coupled
scalars in a power-law cosmology whose wavefunction coefficients are the polytope
canonical forms of Section~\ref{sec:bypass}, one finds a first-order system
\begin{equation}
	\dd\,\vec{\psiwf}=\eps\,\big(\mathbb A\,\dd\log\,\{\text{energy letters}\}\big)\,\vec{\psiwf},
\end{equation}
in which $\vec{\psiwf}$ is a finite vector of ``master'' wavefunction integrals, the
matrix $\mathbb A$ is built from the total- and partial-energy singularities (the
facets of the polytope), $\eps$ is the power-law tilt of the background, and the
``letters'' whose logarithms appear are exactly those energies. Solving the system is a flow in energy space, dubbed \emph{kinematic
	flow}, whose steps generate the transcendental functions (logarithms, dilogarithms,
and their higher-weight relatives) that the time-dependent couplings of a power-law
background introduce: exactly
the transcendentality tiers of Section~\ref{sec:tree}. The geometric origin of the
flow equations has since been identified \cite{Baumann:2025qjx}, and the flow itself
rederived from the splitting of tubings \cite{Ke:2026laa}. The differential-equation
viewpoint makes the singularity structure the primary datum and computes the function
from it, in the same spirit as the modern flat-space method of differential equations
for Feynman integrals; the ``emergence of time'' in the title of \cite{Arkani-Hamed:2023bsv} refers
to the way the bulk time integral is reconstructed from purely boundary,
energy-space data. The singularities that the flow connects had been charted
earlier: an analysis linked the total- and partial-energy singularities
of a correlator to those of its lower-point constituents and took that linkage as the
organizing datum
\cite{Baumann:2021fxj}, the closest antecedent of the thesis this review is built on.

At loop level the kinematic
flow acts on loop \emph{integrands} \cite{Baumann:2024mvm}: the wavefunction
integrand at fixed loop energy obeys a first-order
system generated by the same graphical rules as at tree level, so the combinatorial
growth is tamed before the loop integration is ever attempted. This is the loop-level
analogue of the tree-level statement that the differential equations know the answer
before the time integrals do. The singularities that drive the flow have themselves
been mapped out: the asymptotic structure of cosmological integrals near each of their
singular points can be read directly off the graph combinatorics
\cite{Benincasa:2024lxe}, which fixes which divergences can occur, with what strength,
and how the integral behaves as any subset of energies is scaled (the cosmological
counterpart of the Landau analysis that underpins flat-space integrals). The
finiteness of the system has an explanation from mathematics: the wavefunction
integrals are periods of a
\emph{twisted (relative) cohomology} \cite{De:2023xue}, so the number of master
integrals is a
cohomology dimension, computable before any integration, and the connection matrix
$\mathbb A$ follows from intersection numbers, the same machinery that counts master
Feynman integrals. And the alphabet is not an unstructured list: for path graphs, the singularity
letters of
tree-level correlators in these power-law cosmologies are generated by cluster
algebras of type $A$ \cite{Capuano:2025myy}, the same
structures that organize the symbol alphabets of planar $\mathcal N=4$ amplitudes;
whether a single cluster structure governs arbitrary graphs is open.
The analytic bootstrap that reconstructs six- and seven-point
flat-space amplitudes from their alphabets may then have a cosmological twin
waiting to be run.\footnote{The program has grown quickly. Cohomological and
	period-theoretic foundations: relative twisted cohomology and its intersection
	numbers \cite{De:2024zic,Gasparotto:2024bku}, differential equations from
	Hodge-theoretic and Landau-type data \cite{Grimm:2024tbg,Chen:2024glu,He:2024olr};
	exact evaluations of the same integral family in general power-law backgrounds
	\cite{Fan:2024iek,Liu:2024str}; and further flows, alphabets, and their
	geometry \cite{Ferro:2026oph,Westerdijk:2026msm,Baumann:2026atn}.}

\subsection{Positive geometry: from polytopes to cosmohedra}

The cosmological polytope \cite{Arkani-Hamed:2017fdk,Benincasa:2018ssx}
computes a single graph's wavefunction as a canonical form. But a full correlator is a
\emph{sum} over graphs, and until recently there was no single geometry that delivered
the sum. The \emph{cosmohedron} \cite{Arkani-Hamed:2024jbp} is the proposed
answer for planar $\operatorname{Tr}(\phi^3)$ theory: a polytope obtained by blowing up
the faces of the associahedron that packages the sum over tree amplitudes, from which
the total wavefunction with all diagrams combined is extracted by a prescription that
slightly extends the canonical-form map. Its appearance closes the parallel with the
flat-space story (amplituhedron for the $S$-matrix, cosmohedron for the
wavefunction) and hints at a ``stringy'' completion in which the cosmological
correlator is the analogue of a string amplitude. The construction is already being
made combinatorial and graph-local: the \emph{amplitubes} of
\cite{Glew:2025otn} realize a cosmohedron for an arbitrary individual graph (a ``graph
cosmohedron'' whose faces are labeled by collections of compatible tubes, built
directly on the graph associahedron of Section~\ref{sec:bypass} without reference to
any embedding). The follow-up \cite{Glew:2025arc} shows that the canonical forms
of these objects assemble not just the wavefunction but the \emph{correlator} itself,
the observable that Section~\ref{sec:wavefunction} taught us is quadratic in
$\psiwf$; correlahedra for the same combination are already sketched in
\cite{Arkani-Hamed:2024jbp}. That a single positive geometry can know about the
$\psiwf\,\psiwf^*$ combination is the geometric face of the cosmological optical
theorem. Unitarity's geometry is the optical polytope of Section~\ref{sec:inflation}
\cite{Albayrak:2023hie}, whose non-convexity also delivers the flat-space optical
theorem. A unifying Grassmannian representation has also been proposed for massless spinning
de~Sitter correlators \cite{Arundine:2026fbr}. These constructions are young: the
positive geometries are mostly limited to conformally coupled scalars, the
Grassmannian to massless spinning fields, and extending either to general masses
(gravitons, the cosmological collider) is open and inviting.

\subsection{The structure of de Sitter, and a non-perturbative bootstrap}

The methods of this review are perturbative: they organize the loop expansion of a
weakly coupled bulk theory. Behind them stands the question of what de~Sitter quantum
field theory \emph{is}, whose literature is now converging with the correlator
program. The classic starting point is the analysis of
de~Sitter two-point functions in \cite{Bros:1995js}, which identified
maximal analyticity in the complexified geometry as the correct substitute for the
flat-space spectral condition and derived the thermal (KMS) character of the
Bunch--Davies state from it. The standard orientation to this landscape
remains \cite{Anninos:2012qw}: the static patch, its temperature
and entropy, the unitary representations, and the dS/CFT idea. Two structural results
frame everything the perturbative methods compute. A perturbative $S$-matrix
between the asymptotic past and
future of \emph{global} de~Sitter does exist for massive scalars, and is well defined
order by order \cite{Marolf:2012kh}, so dS has a scattering observable, just
not the one an
inflationary observer measures. And the late-time Bunch--Davies wavefunction was dissected into divergent local
phases and finite universal terms \cite{Anninos:2014lwa}: the anatomy
Section~\ref{sec:wavefunction} performs in momentum space, established as a general
property of the dS wavefunction.

On these foundations a genuinely non-perturbative toolkit is being assembled. The
Hilbert space of quantum
field theory in de~Sitter has been decomposed into unitary irreducible
representations of the dS isometry group (the principal, complementary, and discrete
series that play the role mass and spin play in flat space), explicitly for free
multiparticle states and for conformal field theories in dS \cite{Penedones:2023uqc}. The
K\"all\'en--Lehmann
representation now exists in dS \cite{Loparco:2023rug}: any two-point function of
scalar local operators in the Bunch--Davies
state is an integral over these representations with a \emph{positive} spectral
density, and an inversion formula extracts the density. This is the
non-perturbative skeleton onto which the momentum-space results must fit. The fit is
already being made: a spectral representation of
cosmological correlators themselves \cite{Werth:2024mjg} writes scalar massive-exchange
correlators as
integrals over the principal series of simple ``off-shell'' seed functions, from
which factorization on the partial-energy poles and the cosmological-collider
oscillations of Section~\ref{sec:inflation} are read off as residues. And amputating
the external legs of cosmological correlators defines a de~Sitter $S$-matrix
\cite{Melville:2024ove}, an object that obeys crossing
and unitarity cutting rules like a flat amplitude and reduces to one when the
curvature is switched off. Meanwhile the AdS-to-dS dictionaries of
\cite{Sleight:2020obc,Sleight:2021plv} make the continuation
\eqref{eq:cont} systematic at the level of exchanges: a dS exchange is a specific
linear combination of AdS exchanges with shifted dimensions. The AdS technology of
this review then transports to dS wholesale, boundary terms and contact ambiguities
included.

The de~Sitter analogue of the non-perturbative conformal
bootstrap (constraining correlators by unitarity and analyticity without any bulk
Lagrangian) is being built in parallel. One line of work sets up a de~Sitter
K\"all\'en--Lehmann/spectral representation and the
associated positivity constraints \cite{Hogervorst:2021uvp}; another places
analyticity and unitarity for cosmological correlators with massive states on a firm
footing \cite{DiPietro:2021sjt},
including the analytic continuation of the exchanged-particle spectrum that underlies
the collider signal. These non-perturbative statements are the boundary conditions any
perturbative computation must respect, and the target the momentum-space methods
aim to reproduce and eventually to exceed.\footnote{Nearby developments: an
	interacting theory solved exactly in $\dS_2$ \cite{Anninos:2024fty}; the
	definition and properties of a de~Sitter $S$-matrix
	\cite{Melville:2023kgd,De:2025yls}; unitarity, representation theory and the
	static patch
	\cite{Chakraborty:2023yed,Chakraborty:2025mhh,Chakraborty:2025izq,Hinterbichler:2026xqf};
	and Euclidean/quantum-gravity input on the sphere partition function
	\cite{Collier:2025lux}.}

\subsection{Neighboring fields}

The neighbors divide into two kinds. Some are \emph{parallel languages}: other
representations of the same boundary data, or of closely related observables, each
making a different piece of the structure manifest. Others are \emph{suppliers of technique}: mature subjects whose
machinery this program imports, and whose practitioners already own most of its
tools. 
The five parallel languages, with entry points. \emph{Mellin space} is the
dual-resonance form of conformal correlators, the natural language of holography:
contacts are constants, exchanges are poles at the exchanged twists, and the
large-Mandelstam limit is the Mellin image of the $\kT\to0$ pole; it has yielded
the tree half-BPS $\mathrm{AdS}_5\times S^5$ four-point functions, the one-loop
AdS graviton from CFT dispersion relations, an $n$-gluon AdS recursion, and
inflationary correlators via \eqref{eq:cont}
\cite{Mack:2009mi,Penedones:2010ue,Fitzpatrick:2011ia,Penedones:2016voo,Rastelli:2016nze,Rastelli:2017udc,Alday:2017vkk,Bissi:2022mrs,Chu:2023pea,Sleight:2019hfp,Sleight:2019mgd,Alday:2021odx,Chu:2023kpe,Cao:2024bky,Huang:2025ieo}.
The \emph{position-space conformal bootstrap} works crossing and unitarity in
cross-ratio space, the Lorentzian inversion formula making operator data analytic
in spin: the non-perturbative complement to these results, and the source of the
loop-level CFT data Section~\ref{sec:loops} must reproduce
\cite{Poland:2018epd,Simmons-Duffin:2016gjk,Caron-Huot:2017vep}. The
\emph{$S$-matrix-from-CFT} program charts when a holographic $S$-matrix exists
(large $N$ and a large higher-spin gap imply a local bulk theory) and how to
extract it: the Lorentzian bulk-point singularity, the large-Mandelstam Mellin
limit at loop level, Landau analysis of bulk resonances, smearing kernels for
in/out states
\cite{Heemskerk:2009pn,Gary:2009ae,Fitzpatrick:2011dm,Komatsu:2020sag,Hijano:2019qmi,Gillioz:2020mdd,Gadde:2022ghy,Jain:2023fxc}.
\emph{Scattering equations}: CHY localization of flat amplitudes on algebraic
equations, with an (A)dS boundary-correlator counterpart whose explicit part is
the curved-space double copy of Section~\ref{sec:bypass}
\cite{Cachazo:2013iea,Gomez:2021qfd,Gomez:2021ujt,Armstrong:2022csc}. And
\emph{celestial holography} recasts the flat $S$-matrix as a two-dimensional
celestial-CFT correlator in a boost basis, soft theorems becoming Ward identities
and gravity acquiring $w_{1+\infty}$
\cite{Raclariu:2021zjz,Pasterski:2016qvg,Pasterski:2021rjz,Pasterski:2021raf,deGioia:2023cbd,Arkani-Hamed:2020gyp,Albayrak:2020saa,Donnay:2023mrd,Iacobacci:2022yjo,Sleight:2023ojm}.

The four suppliers of technique. \emph{Feynman-integral technology}: kinematic
flow is the differential-equation method reaching cosmology, the canonical
$\dd\log$ form solved by iterated integrals, together with the symbol and
polylogarithm technology that once turned a seventeen-page two-loop answer into a
few lines; the same apparatus, but $\eps$ deforms the cosmology rather than the
dimension, and the letters are energies rather than Mandelstams
\cite{Kotikov:1990kg,Henn:2013pwa,Henn:2014qga,Goncharov:2010jf,Duhr:2014woa,Weinzierl:2022eaz,Chen:2023iix,Hoefnagels:2025cnt}.
\emph{Positive geometry as mathematics}: canonical forms and the graph
associahedra of Section~\ref{sec:bypass} are algebraic combinatorics predating
their physics use; cosmohedra, amplitubes, and the cluster-algebraic energy
alphabets (so far established for path graphs) repay the debt with new polytope
families
\cite{Arkani-Hamed:2017tmz,CarrDevadoss:2006,Arkani-Hamed:2024jbp,Glew:2025otn,Capuano:2025myy}.
\emph{Nonequilibrium field theory}: the in-in contour of
Section~\ref{sec:wavefunction} is the Schwinger--Keldysh closed-time path of
many-body physics, $\pm$ vertices its diagrammatics; in return,
influence-functional and open-system methods suit de~Sitter infrared effects and
decoherence
\cite{Schwinger:1960qe,Keldysh:1964ud,Kamenev:2011,Calzetta:2008iqa,Chen:2017ryl,Loganayagam:2023pfb,Salcedo:2024smn,Colas:2025app}.
And \emph{collider physics proper}, the oldest neighbor: a better recursion,
double copy, unitarity method, or LHC integration algorithm becomes, through the
$\kT\to0$ limit and its curved-space corrections, a statement about
early-universe correlators \cite{Dixon:1996wi,Elvang:2013cua,Cheung:2017pzi}.
Three of these neighbors are more than parallels. The total-energy pole is not \emph{analogous}
to the flat-space limit of a CFT; it belongs to that family: the bulk-point residue,
the large-Mandelstam Mellin amplitude and the residue at $\kT\to0$ extract one object
three ways, and Section~\ref{sec:tree} performs the third. Celestial holography runs
the machine backwards, packaging the flat $S$-matrix \emph{into} a conformal
correlator rather than exposing it inside one; the two meet in the flat-space
limit of a holographic CFT, where the celestial sector sits inside an ordinary CFT in
the very limit that produces the total-energy pole. And because that residue
\emph{is} an amplitude, not an analogue of one, collider physics is the third:
its practitioners and cosmologists are increasingly computing the same functions.

All of these frontiers are powered by the two facts that powered the explicit
computations: the correlator is
controlled by its energy singularities, and its leading singularity is the flat-space
amplitude. The differential equations flow between those singularities; the cosmohedron
is bounded by them; the non-perturbative bootstrap constrains them; the neighboring
languages relocate them; the neighboring fields supply the technology to exploit
them. It is one structure, seen from many sides.

\section{Outlook}
\label{sec:outlook}

\begin{center}
	\begin{minipage}{0.88\columnwidth}
		\centering\small\itshape\color{slatedeep}
		Who truly knows, and who can here declare it,\\
		whence it was born, and whence comes this creation?\\
		The gods are later than this world's production.\\
		Who knows, then, whence it first came into being?\\[5pt]
		{\upshape\scriptsize\color{slatesoft}%
			N\=asad\=iya S\=ukta, \d{R}g Veda 10.129, trans.\ R.~T.~H. Griffith}
	\end{minipage}
\end{center}
\vspace{8pt}

We have followed a single momentum-space object (the boundary correlator, equivalently
the wavefunction coefficient; the equal-time correlator then follows by the Born rule)
from its definition to its cosmological payoff. The many
techniques are one method: compute the scalar skeleton once, restore the spin and the
curvature by algebra, remove the bulk integral by combinatorics or on-shell data, and
read the answer, through the single continuation \eqref{eq:cont}, as either a
holographic correlator or a cosmological observable. Table~\ref{tab:dict} collects the
dictionary that has organized the review.

\begin{table*}[t]
	\centering\small
	\renewcommand{\arraystretch}{1.2}
	\caption{The dictionary. Each flat-space structure has an anti--de~Sitter avatar and a
		cosmological reading, all organized by the total energy $\kT$; entries on a row are
		related by continuation, factorization or a limit, not by literal identity.}
	\begin{tabular}{@{}l@{\hspace{2.4em}}l@{\hspace{2.4em}}l@{}}
		\toprule
		\textbf{flat space} & \textbf{anti--de~Sitter} & \textbf{de~Sitter / inflation}\\
		\midrule
		$S$-matrix $A_n$ & boundary correlator $\langle JJ\cdots\rangle$ & wavefunction coeff.\ $\psiwf_n$ ($\langle\zeta\cdots\rangle$ via $|\Psi|^2$)\\
		on-shell external leg & bulk-to-boundary $\Kbb_\nu$ ($K_\nu$) & Bunch--Davies mode\\
		internal/cut line & bulk-to-bulk $\Gbb_\nu$ ($J_\nu$, spectral $p$) & internal mode; cuts via normalizable modes\\
		Mandelstam $s,t,u$ & partial/total energies $k_{\ul{12}},\kT$ & energies; folded/squeezed limits\\
		$s+t+u=0$ & $\tilde s+\tilde t+\tilde u\propto\kT\neq0$ & (broken boosts; flat identity as $\kT\to0$)\\
		amplitude pole & total-energy residue $=A_n$ & scattering (flat-space) singularity\\
		BCFW recursion & all-line shift $+$ transition amplitudes & bootstrap ODE\\
		double copy & dimension-shifting double copy & scalar/tensor non-Gaussianity relations\\
		unitarity cut & $K\to J$ transition amplitude & cosmological optical theorem\\
		amplituhedron & cosmological polytope / tubings & cosmohedron\\
		Feynman integral DEs & (energy) differential equations & kinematic flow\\
		\bottomrule
	\end{tabular}
	\label{tab:dict}
\end{table*}

Underlying the whole program is one conviction: the primordial correlators are not
an intractable tangle of nested time integrals but a controlled, amplitude-like
structure, computable by importing (and sharpening) the technology of the flat-space
$S$-matrix. The sky is, to the precision of the coming surveys, a set of correlation
functions frozen out of a bulk quantum evolution; the methods assembled here compute
their primordial input from first principles, and read in it the physics of the first
instants of the universe. The remaining steps to data (the conversion to $\zeta$ or
$\gamma$, late-time transfer functions, survey likelihoods) are standard cosmology; this
review supplies the part that is universal.

\subsection{Ten open problems}
\label{sec:tenproblems}

The open problems are sharp. We list ten, drawn from the review's own threads, each
stated so that a solution would be recognizable.

\begin{enumerate}
	\item \emph{Close the one-loop bubble in even dimension.} The collapse
	\eqref{eq:F4collapse} is algebraic in every $d$ but rational only in odd $d$; in even
	$d$, $\AdS_5$ chief among them, the half-integer power leaves square-root structure and
	the spectral integral of \eqref{eq:lp-bubbleresult} (Section~\ref{sec:loops}) stands.
	Position space says the answer is polylogarithmic
	\cite{Heckelbacher:2020nue,Heckelbacher:2022hbq}; identify that function class, its
	branch prescription and its alphabet directly in momentum space, at one loop and beyond.
	
	\item \emph{Momentum-space holographic renormalization at one loop.} Tree-level
	three-point divergences are renormalized completely: counterterms, beta functions,
	anomalies \cite{Bzowski:2013sza,Bzowski:2015pba}. The divergent spectral integrals of
	Section~\ref{sec:lp-layers}, such as the self-energy \eqref{eq:lp-selfenergy}
	\cite{Carmi:2019ocp,Carmi:2021dsn}, have no such scheme: what is the general counterterm
	structure adapted to the boundary-momentum/spectral representation, including spinning
	fields, and which CFT data does it renormalize?
	
	\item \emph{A double copy consistent with unitarity.} The naive four-point graviton
	square fails the AdS cutting rules \cite{Meltzer:2020qbr,Albayrak:2020fyp}
	(Section~\ref{sec:bypass-dc}), yet the graviton trispectrum \emph{can} be assembled from gluons
	with correction terms \cite{Armstrong:2023phb}, and a Mellin-momentum bootstrap
	assembles complete four-graviton data \cite{Mei:2023jkb}. Locate where ``gravity $=$
	gauge$^2$'' breaks in curved space, determine which contact terms are forced by Ward
	identities and cuts, and find the replacement for BCJ beyond
	\eqref{eq:mbcj} for which squaring and cutting commute, plausibly in the
	differential \cite{Herderschee:2022ntr,Cheung:2022pdk} or Mellin-space
	\cite{Zhou:2021gnu} representations.
	
	\item \emph{Positive geometry with spin and mass.} The polytopes and tubings
	\eqref{eq:tubing}, the cosmohedra \cite{Arkani-Hamed:2024jbp,Glew:2025otn} and the optical
	polytope \cite{Albayrak:2023hie} all live at conformal coupling. Extending them to
	gravitons and massive exchange (the light-mass expansion \cite{Benincasa:2019vqr} is the
	first step) would place the collider signal \eqref{eq:collider} inside a geometry,
	and could connect to the orthogonal-Grassmannian description of massless spinning
	correlators \cite{Arundine:2026fbr}.
	
	\item \emph{The alphabet of kinematic flow.} The differential equations
	\cite{Arkani-Hamed:2023kig,Arkani-Hamed:2023bsv} generate the tiers of
	Figure~\ref{fig:tiers} letter by letter and extend to loop integrands
	\cite{Baumann:2024mvm}, with cohomological master counting \cite{De:2023xue} and mapped
	asymptotics \cite{Benincasa:2024lxe}, but graph by graph. Missing is the analogue of the
	symbol \cite{Goncharov:2010jf,Duhr:2014woa}: an all-order characterization of which
	functions cosmology can produce, with the type-$A$ cluster structure of path graphs
	\cite{Capuano:2025myy} as a first entry, not yet a classification for arbitrary trees
	and loops.
	
	\item \emph{The auxiliary vector at higher points and loops.} The representation
	\eqref{eq:auxD} reduces any spinning loop to a scalar one, with master formulae
	(Section~\ref{sec:lp-aux}) at arbitrary multiplicity and loop order
	\cite{Albayrak:2020bso}, but only low-lying cases are evaluated. Carry five-point,
	triangle, box and two-loop spinning scalar factors to closed renormalized form,
	determine when the tensor reduction commutes with regularization, and mesh the results
	with the general-$d$ tree formulae \cite{Albayrak:2023kfk}.
	
	\item \emph{A genuine de~Sitter $S$-matrix.} Amputating correlators defines a
	boost-invariant late-time $S$-matrix \cite{Melville:2024ove}; the global construction
	\cite{Marolf:2012kh}, the in-out formalism \cite{Donath:2024utn} and the Hilbert-space
	analysis \cite{Penedones:2023uqc} define related but differently scoped objects.
	Establish their domains and their maps to the wavefunction and to in-in correlators: do
	they agree where they overlap, and how is each assembled from the transition amplitudes
	of Section~\ref{sec:lp-cut}?
	
	\item \emph{Branch cuts of massive exchange from AdS.} Massive dS exchange requires
	continuing \eqref{eq:cont} in the spectral parameter, producing branch cuts and
	particle production \cite{Sleight:2020obc,Sleight:2021plv,Werth:2024mjg} (secular
	growth is a distinct, loop-level late-time effect); at
	loop level the nonanalytic part factorizes on shell to all orders \cite{Qin:2023nhv}.
	Derive the analytic structure behind the collider oscillation \eqref{eq:collider} from
	the AdS side and match the non-perturbative constraints \cite{DiPietro:2021sjt}.
	
	\item \emph{From closed forms to templates.} Planck bounds \cite{Planck:2019kim} will
	tighten substantially with CMB-S4 and with SPHEREx's large-scale-structure survey
	\cite{Abazajian:2019eic,SPHEREx:2014bgr,Meerburg:2019qqi}, and estimators need
	oscillatory and spinning shapes evaluated fast across the whole kinematic domain.
	Turning \eqref{eq:4pt-full} and \eqref{eq:ds-psi4} into a validated numerical pipeline
	for general masses and spins, propagated through late-time transfer functions and
	survey windows, is the shortest path from formalism to data.
	
	\item \emph{One geometry.} Read column-wise, Table~\ref{tab:dict} gives diagrams,
	unitarity and time evolution each a separate geometric avatar: cosmological polytope
	\cite{Arkani-Hamed:2017fdk}, optical polytope \cite{Albayrak:2023hie}, cosmohedron
	\cite{Arkani-Hamed:2024jbp}, kinematic flow \cite{Arkani-Hamed:2023bsv}. The amplituhedron
	lesson (one object, locality and unitarity as outputs \cite{Arkani-Hamed:2013jha}) suggests
	these are facets of a single geometry for the wavefunction of the universe. A solution
	should say whether they are related by projections, blow-ups or different canonical
	forms on a common space, and why the wavefunction and the Born-rule correlator require
	different geometric data.
\end{enumerate}

\begin{acknowledgments}
It is a pleasure to thank our collaborators and friends: Santiago Agui Salcedo,
Daniel Baumann, Paolo Benincasa, Jinwei Chu, Chandramouli Chowdhury, Carlos Duaso Pueyo,
Austin Joyce, Hayden Lee, David Meltzer, Xinkang Wang, and the many other colleagues for  conversation over several years. We would like to thank Jinwei Chu for carefully reading the manuscript and providing valuable feedback.
S.A. was supported by the BAGEP Award of the Science Academy in Turkey
and by the TÜBİTAK (The Scientific and Technological Research Council of
Turkey)\footnote{Please see ROR ID:
\href{https://ror.org/04w9kkr77}{https://ror.org/04w9kkr77}} 2232-b
International Fellowship for Early Stage Researchers program, project number
122c153. S.K.\ thanks the Department of Physics at the University of Chicago for research funding
and the late Peter Littlewood for his friendship and inspiration.
\end{acknowledgments}

\appendix
\section{Conventions and special functions}
\label{app:conv}

\paragraph{Metrics and continuation.} We use the Poincar\'e patch of Euclidean
$\AdS_{d+1}$, $\dd s^2=z^{-2}(\dd z^2+\dd\bm x^2)$, with the conformal boundary at
$z\to0$ and the $d$ boundary directions $\bm x$ Fourier-transformed. De~Sitter
results follow, where applicable, by the analytic continuation $\eta=\ii z$,
$R_{\dS}=-\ii R_{\AdS}$ of Section~\ref{sec:tk-dict} (with the subtleties spelled
out in Section~\ref{sec:inflation}), with $\eta\in(-\infty,0)$ conformal time and scale
factor $a=-1/(H\eta)$. Boundary momenta are $\bm k_a$; $k_a\equiv|\bm k_a|$ is the
``energy'' of a leg; the total energy is $\kT=\sum_a k_a$; a partial energy is a sum
of magnitudes over a subset, $k_S=\sum_{a\in S}k_a$; and an underline denotes the
magnitude of a vector sum, $k_{\ul S}\equiv|\sum_{a\in S}\bm k_a|$, the energy
flowing on an internal line. For a channel $S$, the associated channel energy is
$E_S\equiv k_S+k_{\ul S}$, written $k_{12\ul{12}}$ and the like in the main text;
this is the combination whose analytic continuation to zero produces a
factorization singularity. Correlators carry a stripped momentum-conserving delta
function, indicated by the prime $\langle\cdots\rangle'$.

\paragraph{Objects.} We write $W_n$ for a Euclidean-AdS boundary correlator (with
color ordering or tensor labels when needed), $\psiwf_n$ for a de~Sitter
wavefunction coefficient, $\langle\cdots\rangle'$ for an equal-time correlator, and
$\mathcal A_n$ for a flat-space scattering amplitude. Analytic continuation relates
$W_n$ to $\psiwf_n$ after convention matching; the Born rule relates $\psiwf_n$ to
equal-time correlators; and the leading $\kT\to0$ coefficient extracts
$\mathcal A_n$.

\paragraph{Bessel functions.} The bulk-to-boundary propagator uses the modified
Bessel function $K_\nu$ (the solution decaying into the bulk), the bulk-to-bulk mode
the oscillatory $J_\nu$. The half-integer cases that make $\AdS_4$ ($d=3$) elementary
are
\begin{gather}
	K_{1/2}(x)=\sqrt{\tfrac{\pi}{2x}}\,e^{-x},\qquad
	K_{3/2}(x)=\sqrt{\tfrac{\pi}{2x}}\,e^{-x}\Big(1+\tfrac1x\Big),\notag\\
	J_{1/2}(x)=\sqrt{\tfrac{2}{\pi x}}\sin x,\qquad
	J_{-1/2}(x)=\sqrt{\tfrac{2}{\pi x}}\cos x.\notag
\end{gather}
The Bessel completeness relation
$\int_0^\infty p\,\dd p\,J_\nu(pz)J_\nu(pz')=\delta(z-z')/z$, combined with the
radial powers appropriate to the bulk field, is what builds the spectral
bulk-to-bulk propagator \eqref{eq:btbb}.

\paragraph{Triple-\texorpdfstring{$K$}{K} integral.} The conformal three-point seed of momentum-space
CFT is
\begin{equation}
	I_{\alpha\{\nu_1\nu_2\nu_3\}}
	=\int_0^\infty\!\dd z\;z^{\alpha}\prod_{a=1}^3 k_a^{\nu_a}K_{\nu_a}(k_a z),
\end{equation}
convergent for $\alpha+1>\sum_a|\nu_a|$ and otherwise defined by analytic
continuation; its singular cases produce the conformal anomalies and beta functions
catalogued in \cite{Bzowski:2015pba}. Its degenerations at $\nu=\half$ are the rational
building blocks of Section~\ref{sec:toolkit}.

\paragraph{Hypergeometric and Appell functions.} Tree-level non-conformal exchanges
evaluate to ${}_2F_1$ of energy ratios; the dilogarithm of Section~\ref{sec:tree} is
the $\phi^3$ $\AdS_4$ case. One-loop radial integrals give the Appell function
$F_4(a,b;c,c';x,y)=\sum_{m,n}\tfrac{(a)_{m+n}(b)_{m+n}}{(c)_m(c')_n\,m!\,n!}x^my^n$;
at the bubble parameters it collapses to a closed form in every dimension, rational
in odd $d$ (Section~\ref{sec:loops}).

\paragraph{Three-dimensional spinor helicity.} A Euclidean boundary ($d=3$) momentum $\vk$ is
uplifted to the complex null vector $\hat{\vk}=(\ii k,\vk)$; upon complexification
(equivalently, in $(2,1)$ Lorentzian boundary signature, where the spinors are
real) it factorizes as
$\hat k_{\alpha\beta}=\lambda_{\alpha}\bar\lambda_{\beta}$; the brackets are
$\langle ij\rangle=\epsilon^{\alpha\beta}\lambda_{i\alpha}\lambda_{j\beta}$,
$\langle\bar\imath\bar\jmath\rangle=\epsilon^{\alpha\beta}\bar\lambda_{i\alpha}\bar\lambda_{j\beta}$,
$[i\bar\jmath]=\lambda_i^\alpha\bar\lambda_{j\alpha}$, with energy
$k_i=-\tfrac{\ii}{2}[i\bar\imath]$, in the conventions of
Section~\ref{sec:bypass-recursion}. The Schouten identity
$\langle ij\rangle\lambda_k+\langle jk\rangle\lambda_i+\langle ki\rangle\lambda_j=0$
is a spinor-algebra identity used together with momentum conservation; the
generalized version that supplies the momentum-conservation analogue appears in
Section~\ref{sec:bypass-recursion}. The same three-dimensional variables are used
systematically for momentum-space spinning and higher-spin CFT correlators in
\cite{Jain:2020rmw,Jain:2021vrv,Maldacena:2011jn}.

\paragraph{The wavefunction and the Born rule.} The late-time Hartle--Hawking
wavefunction \cite{Hartle:1983ai}, with Bunch--Davies initial
conditions \cite{Bunch:1978yq}, is
$\Psi[\phi]=\exp[\sum_{n\ge2}\tfrac1{n!}\int\psiwf_n\,\phi^n]$ with $\psiwf_n$ the
wavefunction coefficients; the equal-time correlators follow from $|\Psi|^2$, the
leading connected two- and three-point moments being
$\langle\phi\phi\rangle'=-1/(2\Rez\,\psiwf_2)$ (with
$\Rez\,\psiwf_2<0$ by normalizability) and
$\langle\phi\phi\phi\rangle'=-2\Rez\,\psiwf_3/\prod_a 2\Rez\,\psiwf_2(k_a)$. We index
$\psiwf_n$ by the number of external points.

\bibliographystyle{utphysModified}
\bibliography{refs}

\end{document}